\documentclass{jpp}
\usepackage{epsfig}
\usepackage{graphicx}
\usepackage{epstopdf} 

\usepackage{amsmath}
\usepackage{amssymb}
\usepackage{mathabx}

\usepackage{natbib}
\usepackage{subcaption}
\usepackage{float} 
\usepackage{here} 

\usepackage{xcolor} 

\usepackage{tikz}
\usetikzlibrary{patterns}
\usetikzlibrary{decorations.pathmorphing}

\usepackage{color}

\newcommand\beq{\begin{equation}}
\newcommand\eeq{\end{equation}}

\newcommand{\ben}{\begin{eqnarray}}
\newcommand{\een}{\end{eqnarray}}
\newcommand{\benn}{\begin{eqnarray*}}
\newcommand{\eenn}{\end{eqnarray*}}

\newcommand{\apar}{ A_{\parallel}}

\newcommand{\pa}{\partial}

\newcommand{\lapp}{\Delta_\perp}

\newcommand{\gpar}{\nabla_{\parallel}}

\newcommand{\teb}{\textcolor{blue}}

\allowdisplaybreaks

\makeatletter
\tikzset{
        hatch distance/.store in=\hatchdistance,
        hatch distance=5pt,
        hatch thickness/.store in=\hatchthickness,
        hatch thickness=5pt
        }
\pgfdeclarepatternformonly[\hatchdistance,\hatchthickness]{north east hatch}
    {\pgfqpoint{-1pt}{-1pt}}
    {\pgfqpoint{\hatchdistance}{\hatchdistance}}
    {\pgfpoint{\hatchdistance-1pt}{\hatchdistance-1pt}}%
    {
        \pgfsetcolor{\tikz@pattern@color}
        \pgfsetlinewidth{\hatchthickness}
        \pgfpathmoveto{\pgfqpoint{0pt}{0pt}}
        \pgfpathlineto{\pgfqpoint{\hatchdistance}{\hatchdistance}}
        \pgfusepath{stroke}
    }
\makeatother

\title{Inverse cascade and magnetic vortices in  kinetic Alfv\'en-wave turbulence}

\author{G. Miloshevich, D. Laveder, T. Passot,  P.L. Sulem}

\affiliation{Universit\'e C\^ote d'Azur, CNRS, Observatoire de la C\^ote d'Azur, Laboratoire J.L. Lagrange, Boulevard de l'Observatoire, CS  34229, 06304 Nice Cedex 4, France}

\begin{document}

\maketitle

\begin{abstract}
A Hamiltonian two-field  gyrofluid model for kinetic Alfv\'en waves (KAWs) in a magnetized electron-proton plasma, retaining ion finite-Larmor-radius corrections and  parallel magnetic field fluctuations, is used to study the inverse cascades that develop when turbulence is randomly driven at sub-ion scales. In the directions perpendicular to the ambient field, the dynamics of the cascade turns out to be nonlocal and the ratio  $\chi_f$ of the wave period to the characteristic nonlinear time at the driving scale affect some  of its properties. For example, at small values of $\chi_f$, parametric decay instability of the modes driven by the forcing can develop, enhancing for a while inverse transfers. The balanced state, obtained at early time when the two counter-propagating waves are equally driven,  also becomes  unstable at small $\chi_f$, leading to an inverse cascade. For $\beta_e$ smaller than a few units, the cascade slows down when reaching the low-dispersion spectral range. For higher $\beta_e$, the ratio of the KAW to the Alfv\'en frequencies displays a local minimum. At the corresponding transverse wavenumber, a condensate is formed, and the cascade towards larger scales is then  inhibited. Depending on the parameters, a parallel inverse cascade can develop, enhancing the elongation of the ion-scale magnetic vortices that generically form.
\end{abstract}


\section{Introduction}

Large-scale magnetic structures are commonly observed in astrophysical media and have also been shown to stabilize the H mode in tokamaks~\citep{solano10}. The question then arises as to what are the physical processes contributing to  their formation and prescribing their characteristic size. In astrophysics, among the potential mechanisms, turbulent inverse cascades have been suggested. In this context, special attention was paid in the literature to the cascade of magnetic helicity in the framework of  incompressible magnetohydrodynamics (MHD) in the absence of an ambient field \citep{Frisch75, Pouquet76, Meneguzzi81, Alexakis06, Muller12, Linkmann16, Linkmann17, Pouquet19}. Extension to compressible MHD with Mach number that goes up to unity  \citep{Balsara99, Brandenburg01}, as well as the case of incompressible Hall-MHD \citep{Pouquet20} were also studied. The cascade is however prevented by the presence of an ambient field, due to the non-conservation of the magnetic helicity based on the fluctuations, and due to the lack of gauge-invariance of the conserved  generalized magnetic helicity that can be defined in this regime \citep{Matthaeus82, Stribling94, Brandenburg04}. Conservation of the magnetic helicity  together with the existence of an inverse cascade \citep{KimCho15} are nevertheless recovered in  electron magnetohydrodynamics (EMHD) that describes whistler waves  at electron scales where the ion motion is negligible and the dispersion associated with the Hall effect is significant.   One is thus naturally led to wonder about the possibility of the existence of an inverse cascade for kinetic Alfv\'en waves (KAWs) at sub-ion scales.

A question arises concerning the possible drivers for such inverse cascades. Bypassing the classical local direct cascade, energy could be directly injected at small scales via nonlocal interactions  mediated by magnetic reconnection occurring in thin ion-scale current sheets. It has recently been proposed that these reconnection events could generate an inverse flux
toward larger scales, as well as  starting a transfer of
energy toward smaller scales \citep{Franci17}. 
Another example of possible mechanism {was} discussed in the context of Alfvenic turbulence in the distant ion foreshock region where observation of an inverse cascade was reported, related to the existence of nonlinear parametric instabilities generated by upstream accelerated protons reflected on the bow shock  \citep{He19}.

The dynamics of a strongly magnetized plasma characterized by small perturbations of a homogeneous equilibrium
is appropriately described by the gyrokinetic formalism, from which reduced gyrofluid models of various complexity can be derived. Such models can provide a uniform description covering a spectral range extending from  MHD  to electron scales. They  capture  the transition from Alfv\'en waves  to  KAWs, 
which are known to play a dominant role in the solar wind at MHD \citep{Belcher71,Reville20} and sub-ion  \citep{Alexandrova09, Sahraoui10, Salem12, Podesta13} scales, respectively. Their nonlinear dynamics is  isolated in the description provided by a
reduced two-field gyrofluid model discussed in \citet{PST18,PS19,MPS19}. When neglecting electron inertia, it extends to finite beta parameters the model considered in \cite{Zocco11} taken in the isothermal limit,  retaining the coupling to the parallel magnetic fluctuations.

Kinetic effects, such as Landau damping, cyclotron resonance or micro-instabilities are definitely not taken into account in the two-fluid gyrofluid model. The coupling to magnetosonic waves, which could play a role at larger scales (e.g. they permit the decay instability of Alfvén waves at MHD scales) is also not retained. Nevertheless, the 
two-field gyrofluid model  enables  the study (for a broad range of plasma parameters) of imbalanced turbulence, characterized by an excess of the energy carried by one of the two types of counter-propagating waves. Imbalanced Alfv\'enic turbulence is 
ubiquitous in the solar wind \citep{Tu89, Lucek98, Wicks13}, with the degree of imbalance dependent on the type of wind \citep{Tu90,Bruno14,Bruno17,DAmicis19} and of the distance from the Sun \citep{Roberts87, Marsch-Tu90,Chen20}. In the framework of the two-field model, imbalance is easily characterized by the generalized cross-helicity (GCH) which is an ideal quadratic invariant  that  reduces to the  negative  cross-helicity at the MHD scales and to the magnetic helicity at the sub-ion scales. When turbulence is driven by injection of energy and of GCH, these quantities are expected to cascade forward to the smaller scales and/or backward to the larger ones, depending in particular on the injection scale compared to the ion Larmor radius (or the sonic Larmor radius). Indeed, no inverse cascade can take place at the MHD scales in the presence of a strong ambient field, while an inverse cascade of magnetic helicity was predicted in the (dispersive) sub-ion range, by analogy with EMHD \citep{Schekochihin09}, and also on the basis of absolute equilibrium arguments \citep{PST18}. Based on these observations, one expects that as the inverse cascade approaches the MHD scales, its properties will be significantly affected.

In addition to providing a mechanism for the formation of large-scale structures in fluids and plasmas, inverse cascades can reveal various interesting phenomena, like critical transitions that are observed in split-cascade configurations when the relevant dimensionless parameter is varied~\citep{alexakis18}. This includes transitions from inverse to forward cascade of energy in thin-layer turbulence
~\citep{benavides17} and transitions from MHD to fluid turbulence, when the relative strength of the magnetic forcing  parameter is varied~\citep{seshasayanan14}.  Furthermore, there are examples of such criticality in rotating and stratified flows where helicity conservation can be broken when  dynamical parameters such as Rossby and Froude numbers are varied~\citep{Marino_2013}.

The present paper addresses the existence and the properties of the inverse cascades which can develop in the two-field gyrofluid model when energy and GCH are injected. We will vary the injection rate, the driving scale and the ratio $\beta_e$ of equilibrium electron thermal pressure  to the magnetic pressure due to the ambient field,  and suggest that the ratio $\chi_f$ of the wave period to the characteristic nonlinear time at the driving scale affects some properties of the inverse cascade in the transverse spectral plane (e.g. its "degree of self-similarity"), at least at early time. At late times,  the properties of the cascade appear to be rather dependent on the driving scale and $\beta_e$. 
Section \ref{2fluids} provides a description of the model,  together with a brief discussion of the  Fj{\o}rtoft argument  often used for predicting the existence of an inverse cascade. Section \ref{numerics} specifies the numerical set up and  the conditions of the simulations. Section \ref{sec:global} discusses the global properties of the cascade dynamics in the context of our fiducial simulation where turbulence is driven far into the sub-ion range. Section \ref{sec:turb-strength} addresses the effect of the turbulence strength on the early dynamics, and in particular the transition from a self-similar spectrum to a propagating spectral bump, the emergence of the  parametric decay instability or  the instability of  balanced turbulence, as the amplitude of the magnetic fluctuations is reduced.  Section \ref{smallkf} addresses the situation where the cascade reaches the weakly dispersive range.   Section \ref{arrest} is concerned with the arrest of the cascade and the generation of a finite-scale condensate, when $\beta_e$ is large enough.  The coherent structures, in the form of magnetic vortices that are generated in  physical space as consequences of the inverse cascades
and of their arrest, are described in Section \ref{vortices}.   Section \ref{conclu} is the Conclusion. The Appendices include a brief description of the decay instability in the context of the present model and the derivation of the shell-to-shell transfers.

\section{The two-field gyrofluid model} \label{2fluids}

\subsection{Equations and conservation laws}

 A  description  of the Alfv\'en wave dynamics from  the MHD  to  the electron scales, in an electron-proton plasma, is provided by the two-field gyrofluid model which involves the gyrokinetic scaling corresponding to a strong spectral  anisotropy and weak nonlinearity. In the absence of dissipation and driving, it involves two  equations for the electron gyrocenter number density $N_e$ and the parallel\footnote{At the order at which the equations are considered, parallel or longitudinal refers to the direction of the ambient field, except in the case of the parallel derivative $\nabla_\|$ where the derivative is taken in the direction of the local magnetic field.} component of the magnetic potential $A_\|$, in the form\footnote{In Eq. (\ref{eq:A}), the term $[B_z, (2\delta^2/\beta_e) \Delta A_\|]$ is subdominant within the asymptotics leading to the model, but as mentioned in \citet{PST18}, this term is retained in order to preserve the Hamiltonian structure, which is an important property of the model in the sense that it ensures the absence of uncontrolled dissipation. The second term in the rhs of Eq. (\ref{L3}) is also subdominant but has been retained as it becomes relevant when $\beta_i$ becomes of order unity. This term is needed to recover the system of equations governing inertial kinetic Alfv\'en waves with $\tau \gg 1$ (see also \citet{Chen-Bold17}).
 The effect of subdominant terms will only be significant on a time much longer than the time interval for which the asymptotic model is valid.}
 \begin{eqnarray} 
&&\partial_t N_e +[\varphi,N_e]-[B_z,N_e]+\frac{2}{\beta_e}\nabla_\| \Delta_\perp A_\|=0\label{eq:gyro-2fields-Ne} \label{eq:Ne}\\
  &&\partial_t  L_e A_\| -\left [\varphi,\frac{2\delta^2}{\beta_e}\Delta_\perp A_\| \right ]
  + \left [B_z,\frac{2\delta^2}{\beta_e}\Delta_\perp A_\|\right ] 
 + \nabla_\| \left (\varphi-N_e-B_z \right )=0\label{eq:gyro-2fields-A}.\label{eq:A}
\end{eqnarray}
In the above equations,  the parallel magnetic fluctuations $B_z$ and the electron gyrocenter number density $N_e$ are related to the electric potential $\varphi$ by $B_z= M_1 \varphi$ and $N_e= -M_2 \varphi$, where $M_1$ and $M_2$ are Fourier multiplier operators written as $M_1 = L_1^{-1}L_2$ and $M_2= L_3 + L_4 L_1^{-1}L_2$ (here, $\delta^2 = m_e/m_i$ denotes the electron to proton mass ratio), the operators $L_i$ being defined as
\begin{eqnarray}
&&L_1 = \frac{2}{\beta_e}  +(1+2\tau)(\Gamma_0-\Gamma_1)   \label{L1}\\
&&L_2 = 1 +\frac{1-\Gamma_0}{\tau} - \Gamma_0 +\Gamma_1 \\
&&L_3 = \frac{1-\Gamma_0}{\tau} -\delta^2\Delta_\perp \label{L3}\\
&&L_4 = 1-\Gamma_0+\Gamma_1\\
&&L_e = 1-\frac{2 \delta^2}{\beta_e} \lapp.
\end{eqnarray}
Here, $\Delta_\perp = \partial_{xx} + \partial_{yy}$ is the Laplacian in the plane transverse to the ambient field and  $[f,g]= \partial_x f \partial_y g-\partial_y f \partial_x g$  the canonical bracket of two scalar functions $f$ and $g$.
Furthermore, $\Gamma_n$ denotes the (nonlocal) operator 
$\Gamma_n(-\tau \Delta_\perp)$ associated with  the Fourier multiplier  $\Gamma_n(\tau k_\perp^2)$, defined by  $\Gamma_n(x) = I_n(x) e^{-x}$ where $I_n$ is the modified Bessel function of first type of order. Furthermore,$\tau$ denotes  the  ratio  of  the  proton to  the  electron temperatures at equilibrium. 
For a scalar function $f$, the parallel gradient operator $\gpar$ is defined by
\beq
\gpar f=-[\apar , f]+\frac{\pa f}{\pa z}.
\eeq
One recovers the fluctuating magnetic field from the expression of $B_z$ given above and ${\boldsymbol B}_\perp =\nabla_\perp\times (A_\|\widehat{{\boldsymbol z}})$. 

Characterizing the plasma equilibrium state by the number density $n_0$, the temperatures $T_{i0}$ and $T_{e0}$, and the ambient field $B_0$, the model is written above in a non-dimensional form, using the following units: time is normalized to the inverse ion gyrofrequency $\Omega_i^{-1} = (eB_0/(m_ic))^{1/2}$ (where $e$ is the proton charge and $c$ the speed of light), lengths by the sonic Larmor radius $\rho_s= c_s/\Omega_i$ where $c_s=\sqrt{T_{e0}/m_i}$ is the sound speed. Thus wavenumbers and wavevector components are measured in units of $\rho_s^{-1}$. The other normalization factors include $B_0$ for the parallel magnetic fluctuations $B_z$, $B_0\rho_s$ for the parallel magnetic potential $A_\|$,   the equilibrium number density $n_0$ for the electron gyrocenter density $N_e$ and $T_e/e$ for the electric potential $\varphi$. Furthermore, the parameter $\beta_e= 8 \pi n_0 T_{e0} /B_0^2$ is the ratio of the equilibrium electron pressure to the magnetic pressure due to the ambient field.
 
\begin{figure}
    \centering
      \includegraphics[width=.5\linewidth]{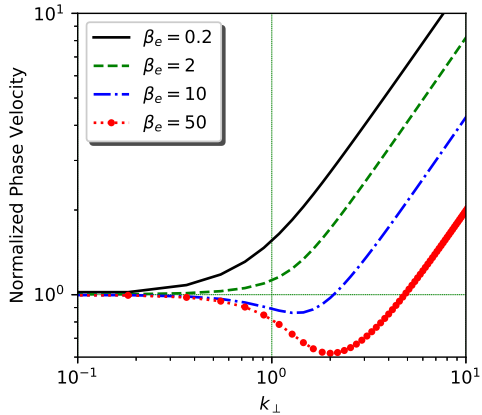} 
   \caption{Normalized parallel-phase velocity $v_{ph} \sqrt\frac{\beta_e}{2}$ (or equivalently, ratio of the KAW frequency $v_{ph} k_\|$ to the Alfv\'en frequency $ v_A k_\|$), versus the transverse wavenumber $k_\perp$ for various values of $\beta_e$.}
    \label{fig:phasevelocities}
\end{figure}

At the level of the linear approximation, the KAW parallel-phase velocity $v_{ph}= \omega/k_z$  is given by 
\begin{equation}
v_{ph}^2\equiv\left(\frac{\omega}{k_z}\right)^2 =  s^2\frac{k_\perp^2}{1 + \frac{2\delta^2 k_\perp^2}{\beta_e}} \frac{1 -{\widehat M}_1 + {\widehat M}_2} {{\widehat M}_2}, \label{eq:vph}
\end{equation}
where the hat refers to the Fourier symbol of the operator and $s= (2/\beta_e)^{1/2}$ is the  Alfv\'en velocity $v_A$ measured in sound speed units ($s=v_A/c_s$).
We  plot in Fig. \ref{fig:phasevelocities} $v_{ph}/s$, versus the transverse wavenumber $k_\perp$, when electron inertia is neglected.  For all values of $\beta_e$, the ratio $v_{ph}/s$, that can also be viewed as the wave frequency normalized by the Alfv\'en wave frequency $v_A k_\|$, is asymptotically equal to unity at large scale (dispersionless limit). On the other hand,
while for small ($\beta_e = 0.2$) or moderate ($\beta_e=2$) values of the electron beta parameter, $v_{ph}/s$  involves a monotonic transition to a $k_\perp$-scaling at sub-ion scales, for  larger $\beta_e$ (e.g. 10 or 50), it displays a local minimum at a wavenumber that (in inverse sonic Larmor radius units) increases with $\beta_e$ at fixed $\tau$ and only depends on $\beta_i$ when measured in units of the ion Larmor radius $\rho_i$. A similar behaviour is observed in Fig. 3 ($\tau= 100$) and 5 ($\tau =1$) of \cite{Howes06} where this quantity,  is computed  from the gyrokinetic theory and also in the framework of the full linear kinetic theory taken in the quasi-transverse limit. 
As discussed below, it turns out that the sensitivity of the dispersion to the 
transverse wavenumber, as measured by the parallel-phase velocity $v_{ph}$, has an important effect on the nonlinear dynamics of the cascade. In particular, it has an impact on the generation of a finite-size condensate and on the formation of large-scale coherent structures. 

In physical space, the  eigenmodes, which can be referred to as Elsasser potentials, are given by 
\begin{equation}
\mu^\pm = \Lambda \varphi \pm s A_\|, \label{mupmequation}
\end{equation} 
where
\begin{equation}
\Lambda = D_e^{-1} (1+M_2-M_1)^{1/2} M_2^{1/2} ,
\end{equation}
with $D_e^2 = (-\lapp) L_e$. 	They obey 
\begin{eqnarray}
&&\partial_t \mu^\pm \pm V_{ph} \partial_z \mu^\pm \nonumber \\
&& +\frac{1}{4} \Lambda^{-1} D_e^{-2} M_3\Big \{ [\Lambda^{-1} M_3 (\mu^++\mu^-) , \Lambda^{-1}M_2 (\mu^+ + \mu^-)]
+[(\mu^+-\mu^-), \lapp (\mu^+-\mu^-)] \Big \} \nonumber \\
&&\pm \frac{1}{4} D_e^{-2}\lapp \Big \{
[(\mu^+ -\mu^-), \Lambda^{-1} M_2 (\mu^++\mu^-)]
+ [L_e (\mu^+ -\mu^-), \Lambda^{-1} (1-M_1)(\mu^+ +\mu^-)]
\Big \}=0.\nonumber \\ 
&&\label{Dtmu}
\end{eqnarray}

The systems (\ref{eq:Ne})-(\ref{eq:A}) or (\ref{Dtmu})
preserve the energy ${\mathcal E}$ and the generalized cross-helicity ${\mathcal C}$, defined as 
\begin{eqnarray}
{\mathcal E} &=& \frac{1}{2} \int \Big ( \frac{2}{\beta_e} |\nabla_\perp A_\| |^2 
+ \frac{4\delta^2}{\beta_e^2}|\Delta_\perp A_\| |^2 
- N_e(\varphi -N_e-B_z) \Big ) d^3 {x} \nonumber \\
&=&\frac{1}{4} \int \left \{ (D_e \mu^+)^2 + (D_e \mu^-)^2 \right \} d^3x.
\label{energy}\\
{\mathcal C} &=&-\int N_e \Big ( 1 - \frac{2\delta^2}{\beta_e}\Delta_\perp \Big) A_\|d^3 {x}.\label{defC} \nonumber \\
&=& \frac{1}{4} \int \left \{\left ( V_{ph}^{-1/2}D_e \mu^+\right)^2 - \left ( V_{ph}^{-1/2}D_e \mu^-\right)^2 \right\} d^3x,
\end{eqnarray}
where $V_{ph}$ is the operator in physical space, which in Fourier space corresponds to the multiplication by $v_{ph}$.
Note that ${\mathcal C}$ was defined with an opposite sign in \citet{PST18}.

\subsection{Normal-field formulation}

As is common with noncanonical Hamiltonian systems, the associated Poisson bracket  possesses Casimir invariants, corresponding to $C_\pm = \int G^\pm d^3x$,
where $G^\pm = L_e A_\| \pm \delta N_e$ are referred to as normal fields. In terms of these fields, the two-field gyrofluid model 
can be rewritten in the form
\begin{equation}
\partial_t G^\pm + [\varphi^\pm, G^\pm] + \partial_z
\left ( \varphi^\pm \mp \frac{1}{\delta} G^\pm \right) =0,\label{Gpmequation}
\end{equation}
where $\varphi^\pm =\varphi -B_z \pm \frac{1}{\delta} A_\|$. In terms of these variables, GCH simply reads
\begin{equation}
{\mathcal C} = -\frac{1}{4\delta}\int  \left \{(G^+)^2 - (G^-)^2 \right \} \, d^3x.  \end{equation}
Similarly,
\begin{equation}
{\cal E} = \frac{1}{2} \int \left \{\frac{1}{2\beta_e} L_e^{-1} |\nabla_\perp (G^+ + G^-)|^2 + \frac{1}{4\delta^2}M_3 M_2^{-1} (G^+ -G^-)^2  \right \}\, d^3x.
\end{equation}

Note that the system  of equations (\ref{Gpmequation}) for $G^\pm$ becomes degenerate in the limit $\delta \to 0$, as the two equations become identical and  reproduce the equation for $A_\|$.   The equation for $N_e$ corresponds to the first order in a development in $\delta^2$. As seen below, this formulation nevertheless has a major interest for estimating the energy and GCH spectral transfers estimated in  Appendix \ref{App:transfers}. 

\subsection{Fj{\o}rtoft argument for an inverse cascade}

\teb{The} \citet{Fjortoft53} argument originally refers to a simple method to predict the existence of an inverse cascade by noting the impossibility of simultaneous direct cascades of two ideal invariants whose spectra differ by a power of the wavenumber. Other versions of the argument can be found in \citet{Nazarenko11} (and references therin). We will
first revisit the use of the \citet{Fjortoft53} argument to support the existence of an inverse GCH cascade at sub-ion scales \citep{Schekochihin09}. In this range, the 
two-field model simplifies and  reduces, when neglecting electron inertia, to
electron reduced magnetohydrodynamics (ERMHD) \citep{Schekochihin09, BHXP13}, under the conditions $\tau k_\perp^2 \gg 1$, $\tau\sim 1$,
	\begin{align}
	\partial_t A_\| + \left(1+\frac{1}{\tau}\right) \nabla_\|  \varphi=0  \label{eq:ERMHD1}\\
	\partial_t \varphi - \frac{\frac{2\tau}{\beta_e}}{1+\frac{\beta_e}{2}(1+\tau)} \nabla_\|\Delta_\perp A_\| =0. \label{eq:ERMHD2}
	\end{align}
In this regime,  the invariants read 
\begin{eqnarray}
&&{\mathcal E} =  \frac{1}{\beta_e}\int \left (|{\boldsymbol B}|^2 + \frac{2}{\beta_e} \frac{1}{1+\tau} B_z^2 \right )d^3x \\
&&{\mathcal C} 
=  \left ( 1 + \frac{2}{\beta_e}\frac{1}{1+\tau} \right ) \int A_\|B_z d^3x. \label{C-magn-hel}
\end{eqnarray}
In the incompressible limit where the beta parameter tends to infinity, the GCH invariant reduces,in the quasi-transverse limit,  to the  generalized magnetic helicities of EMHD \citep{Biskamp99}, or of extended MHD (XMHD) when the ion velocity and electron inertia are taken to zero (Eq. (35) of \citet{Abdelhamid16} and Eq. (29) of \citet{Miloshevich17}).

The magnetic field ${\boldsymbol B}$ can be written ${\boldsymbol B} = \nabla_\perp \times(A_\| {\widehat {\boldsymbol z}}) + B_z {\widehat {\boldsymbol z}}$.
Introducing  ${\boldsymbol A}_\perp$ such that $B_z = {\widehat {\boldsymbol z}}\cdot (\nabla_\perp \times {\boldsymbol A}_\perp)$, one has $\int A_\| B_z d^3x = (1/2)\int  {\boldsymbol A}\cdot {\boldsymbol B} d^3x$  with ${\boldsymbol A} = {\boldsymbol A}_\perp + A_\| {\widehat {\boldsymbol z}}$ (see e.g. Eq. (F6) of \citet{Schekochihin09}), together with $\int |{\boldsymbol B}|^2 d^3x = \int |\nabla_\perp \times {\boldsymbol A}|^2 d^3x$. 

In the large $\beta_e$ limit,  ${\mathcal E}=  \frac{1}{\beta_e}\int |{\boldsymbol B}|^2 dx $ and, like in incompressible MHD without ambient field (see e.g. \citet{Pouquet19} for a recent review), assuming that the helicity is of a given sign and that it is maximal, one gets $E_c(k_\perp)=  (1/k_\perp) E(k_\perp)$, which leads to  conjecture the existence of an inverse cascade of magnetic helicity, by generalizing the argument developed by \citet{Fjortoft53} for two-dimensional incompressible turbulence. 

In the case of finite $\beta_e$, the energy and GCH spectra cannot be directly related to each other. Nevertheless, when turbulence is not too strong, the first and third terms in the energy given by the first line of  Eq. (\ref{energy}) are approximatively in equipartition (this property was checked to be accurately satisfied in the present simulations). In the ERMHD regime, where the coefficients $M_1$ and $M_2$ are scale-independent, it is easily seen that in this case ${\mathcal E} \sim \int |{\boldsymbol B}|^2 d^3x$, which then permits using the above argument.  In summary, the phenomenological argument for the existence of an inverse GCH cascade at finite $\beta_e$ requires both equipartition between total magnetic and internal energies and single-sign maximal helicity, while in the large $\beta_e$ limit the first condition is not necessary.

In contrast, when considering the Hall reduced magnetohydrodynamc  (HRMHD) equations for dispersive Alfv\'en waves (Eqs. (E19)-(E20) of  \citet{Schekochihin09} or Eqs.(2.24)-(2.25) of \citet{PS19}), which correspond to the regime $\tau \ll 1$ and $\tau k_\perp^2 \ll 1$, $N_e ({\boldsymbol k})\approx - k_\perp^2 \varphi({\boldsymbol k})$ and $B_z({\boldsymbol k}) \approx (1+ 2/\beta_e)^{-1} k_\perp^2\varphi{(\boldsymbol k})$, one has
\begin{eqnarray}
   &&{\mathcal E} = \frac{1}{2} \int \left [ \frac{2}{\beta_e}|\nabla_\perp A_\| |^2 + \left (1 + \frac{k_\perp^2}{1+ \frac{\beta_e}{2}}\right) \left(1+ \frac{2}{\beta_e}\right)^2 \frac{1}{k_\perp^2} B_z^2 \right ] d^3x \\
   && {\mathcal C} = \left( 1 + \frac{2}{\beta_e}\right) \int A_\| B_z d^3 x =\frac{1}{2}\left( 1 + \frac{2}{\beta_e}\right)  \int {\boldsymbol A}\cdot {\boldsymbol B}d^3x.
\end{eqnarray}
Two regimes are then to be distinguished. When formally taking the limit $k_\perp \to \infty$ (which in the HRMHD model requires en extremely small value of $\tau$) and assuming equipartition of the magnetic and internal energies, one writes the total energy in the form
\begin{equation}
    {\mathcal E}= \frac{1}{\beta_e}\int|{\boldsymbol B}_\perp|^2 d^3x = \left ( \frac{1 + \frac{2}{\beta_e}}{2+ \frac{2}{\beta_e}}\right)\frac{2}{\beta_e} \int |{\boldsymbol B}|^2 d^3x.
\end{equation}
The same argument as in the case of ERMHD can then be used to conjecture the existence of an inverse GCH cascade. As expected, this argument does not apply in the RMHD regime corresponding to the limit $k_\perp \to 0$. Using in this case the strongest assumption of local-in-scale energy balance, one gets
\begin{equation}
    B_z({\boldsymbol k}) = \sqrt{\frac{2}{\beta_e}} \frac{1}{1+ \frac{2}{\beta_e}}k_\perp |{\boldsymbol B}_\perp({\boldsymbol k})|,
\end{equation}
and thus 
\begin{equation}
    {\mathcal C} = \sqrt{\frac{2}{\beta_e}} \int |{\boldsymbol B}_\perp|^2 d^3 x.
\end{equation}
In this case, ${\mathcal E}$ and ${\mathcal C}$ are proportional and no energetic condition prevents the existence of a simultaneous direct cascade of the two invariants. From the above argument, we can expect that an inverse cascade of GCH develops when turbulence is driven at sub-ion scales and that 
this cascade progressively slows  down when approaching the non-dispersive scales.

\section{Numerical setup and conditions of the simulations} \label{numerics}

\begin{table}
\begin{center}
\def~{\hphantom{0}}
\begin{tabular}{cccccccccccc}
  &  &   &   &   &   &   &  & &\\
  & $\beta_e$ & $k_f$ & $\epsilon_E$ & 
  $\epsilon_E/\epsilon_C$ & $L/2\pi$ & 
     $N$ & $\nu$ & $\delta^2$ & ${\rm min}(\Delta t)$ & $\chi_f^t$ & $|B_\perp(k_f)|$\\
       
\hline
\hline

\hspace{0.5cm} $R_{1.3}^{0.5}$ \hspace{0.2cm} & $0.2$ & $1.3$ & $0.37$ & $30.1$ & $27.5$ & $360$ & $7.8\times 10^{-5}$ & $ 10^{-5}$ & $8.0\times 10^{-3}$ & 0.17 & $0.40$ \\ 

\hline 

\hspace{0.5cm} $R_{1.3b}^{0.2}$ \hspace{0.5cm} & $0.2$ & $1.3$ & $0.37$ & $\infty$ & $27.5$ & $360$ & $7.8\times 10^{-5}$ & $ 10^{-5}$ & $1.0\times 10^{-2}$ & 0.17 & $0.38$ \\ 

\hline 

\hspace{0.5cm} $R_{1.3sb}^{0.2}$ \hspace{0.5cm} & $0.2$ & $1.3$ & $5.95$ & $\infty$ & $27.5$ & $360$ & $7.8\times 10^{-5}$ & $ 10^{-5}$ & $1.0\times 10^{-2}$ & 0.43 & $0.70$ \\ 

\hline
\hline

\hspace{0.5cm} $R_{1.3}^{2}$ \hspace{0.5cm} & $2$ & $1.3$ & $0.37$ & $6.34$ & $27.5$ & $360$ & $7.8\times 10^{-5}$ & $ 10^{-5}$ & $1.0\times 10^{-2}$ & 0.61 & $1.0$ \\

\hline

\hspace{0.5cm} $R_{1.3b}^{2}$ \hspace{0.5cm} & $2$ & $1.3$ & $0.37$ & $\infty$ & $27.5$ & $360$ & $7.8\times 10^{-5}$ & $ 10^{-5}$ & $1.0\times 10^{-2}$ & 0.61 & $0.85$ \\

\hline

\hspace{0.5cm} $R_{6.5}^2$ \hspace{0.5cm} & $2$ & $6.5$ & $0.37$ & $27.2$ & $5.5$ & $240$ & $4.0\times 10^{-8}$ & $ 10^{-6}$ & $4.0\times 10^{-3}$ & 0.22 & $0.27$ \\

\hline

\hspace{0.5cm} $R_{13}^{2}$ \hspace{0.5cm} & $2$ & $13$ & $0.37$ & $54.2$ & $2.75$ & $240$ & $2.5\times 10^{-11}$ & $ 10^{-6}$ & $5.0\times 10^{-4}$ & 0.14 & $0.25$ \\

\hline

\hspace{0.5cm} $R_{13s}^{2}$ \hspace{0.5cm} & $2$ & $13$ & $95.2$ & $54.2$ & $2.75$ & $240$ & $1.0\times 10^{-10}$ & $ 10^{-6}$ & $5.0\times 10^{-4}$ & 0.88 & $1.65$ \\

\hline

\hspace{0.5cm} $R_{13w}^{2}$ \hspace{0.5cm} & $2$ & $13$ & $0.0058$ & $54.2$ & $2.75$ & 240 & $2.5\times 10^{-11}$ & $ 10^{-6}$ & $5.0\times 10^{-4}$ & 0.035 & $0.085$\\

\hline

\hspace{0.5cm} $R_{13b}^{2}$ \hspace{0.5cm} & $2$ & $13$ & $0.37$ & $\infty$ & $2.75$ & $240$ & $2.5\times 10^{-11}$ & $ 10^{-6}$ & $5.0\times 10^{-4}$ & 0.14 & $0.26$ \\

\hline

\hspace{0.5cm} $R_{13sb}^{2}$ \hspace{0.5cm} & $2$ & $13$ & $95.2$ & $\infty$ & $2.75$ & $240$ & $1.0\times 10^{-10}$ & $ 10^{-6}$ & $5.0\times 10^{-4}$ & 0.88 & $1.68$ \\

\hline

\hspace{0.5cm} $R_{36}^{2}$ \hspace{0.5cm} & $2$ & $36.0$ & $0.37$ & $149$ & $1.0$ & $240$ & $5.0\times 10^{-14}$ & $10^{-6}$ & $5.0\times 10^{-4}$ & 0.071 & $0.14$ \\

\hline

\hspace{0.5cm} $R_{36b}^{2}$ \hspace{0.5cm} & $2$ & $36.0$ & $0.37$ & $\infty$ & $1.0$ & $240$ & $5.0\times 10^{-14}$ & $ 10^{-6}$ & $5.0\times 10^{-4}$ & 0.071 & $0.16$ \\

\hline
\hline

\hspace{0.5cm} $R_{0.89}^{10}$ \hspace{0.5cm} & $10$ & $0.89$ & $0.37$ & $2.05$ & $20.3$ & $360$ & $1.0\times 10^{-5}$ & $10^{-5}$ & $4.0\times 10^{-3}$ & 1.74 & $2.3$ \\

\hline

\hspace{0.5cm} $R_{1.3}^{10}$ \hspace{0.5cm} & $10$ & $1.3$ & $0.37$ & $1.94$ & $13.5$ & $360$ & $4.0\times 10^{-7}$ & $10^{-5}$ & $3.125\times 10^{-3}$ & 1.55 & $3.5$ \\

\hline

\hspace{0.5cm} $R_2^{10}$ \hspace{0.5cm} & $10$ & $2.0$ & $0.37$ & $2.20$ & $9.05$ & $360$ & $1.5\times 10^{-8}$ & $ 10^{-5}$ & $2.0\times 10^{-3}$ & 1.29 & $2.6$\\

\hline

\hspace{0.5cm} $R_{6.5}^{10}$ \hspace{0.5cm} & $10$ & $6.5$ & $0.37$ & $6.34$ & $5.5$ & $240$ & $4.0\times 10^{-8}$ & $10^{-6}$ & $2.0\times 10^{-2}$ & 0.61 & $0.50$ \\

\hline

\hspace{0.5cm} $R_{13}^{10}$ \hspace{0.5cm} & $10$ & $13$ & $0.37$ & $12.6$ & $2.75$ & $240$ & $2.5\times 10^{-11}$ & $ 10^{-6}$ & $5.0\times 10^{-4}$ & 0.39 & $0.58$ \\

\hline
\hline 

\hspace{0.5cm} $R_{13}^{50}$ \hspace{0.5cm} & $50$ & $13$ & $0.37$ & $2.63$ & $2.75$ & $240$ & $2.5\times 10^{-11}$ & $ 10^{-6}$ & $5.0\times 10^{-4}$ & 1.11 & $1.45$ \\

\hline

\hspace{0.5cm} $R_{36}^{50}$ \hspace{0.5cm} & $50$ & $36$ & $0.37$ & $7.21$ & $1.0$ & $240$ & $5.0\times 10^{-14}$ & $ 10^{-6}$ & $5.0\times 10^{-4}$ & 0.57  & $1.1$ \\

\hline

\end{tabular}
\captionof{table}{Parameters of the runs, together with the nonlinear parameter $\chi_f^t$ at the driving scale. Runs are named after the  perpendicular wavenumber of the driving and the value of $\beta_e$, appearing respectively as subscript and superscript. Additional subscripts $s$ and $w$ refer to energy injection rates respectively stronger or weaker than the usual value $\epsilon_E = 0.372$. In all the simulations, $\epsilon_{E^+} / \epsilon_{E^-} = 1.5$, except in runs $R_{1.3b}^{0.2}$, $R_{1.3sb}^{0.2}$, $R_{1.3b}^{2}$, $R_{13b}^{2}$ and $R_{13sb}^{2}$, where  $\epsilon_{E^+} / \epsilon_{E^-} = 1$ (balanced driving).}\label{Tab:Tcr}
\end{center}
\end{table}

We performed three-dimensional numerical simulations of the two-field gyrofluid, in the form given by Eq. \eqref{Gpmequation}, supplemented with injection at a wavenumber $k_f$  and small-scale dissipation.
Usually, when retaining electron inertia in Eqs (\ref{eq:Ne})-(\ref{eq:A}), $\beta_e$ must be small enough in order for electron FLR terms to be negligible compared to electron inertia. In order for both electron FLRs and inertia terms to be negligible at the forcing wavenumber (we here concentrate on the inverse cascade), one needs to have (in dimensional units) both $k_f\rho_e=2^{1/2}\delta (k_f \rho_s)\ll 1$ and $k_f d_e=(2/\beta_e)^{1/2}\delta (k_f\rho_s)\ll 1$. In all the simulations listed in Table \ref{Tab:Tcr} (except the one with $k_f\rho_s=36$), one has $k_f\rho_e<1$. Electron inertia will thus be even smaller if $\beta_e\gtrsim 2$.
In order to make the effects associated with electron inertia completely negligible at all the scales, we prescribed an electron-to-proton mass ratio $\delta^2$ smaller than the physical one \footnote{After the paper has been submitted, a new version of the code integrating Eqs. (\ref{eq:Ne}) and (\ref{eq:A}) with $\delta=0$ has been developed and it was checked that, for the presented simulations, the results are indeed  indistinguishable.}.

The integration domain is assumed to be periodic of size $L$ in the three directions, and  a Fourier pseudo-spectral method was used, with aliasing  suppressed by spectral truncation at $2/3$ of the maximal wavenumber. 
Time stepping was performed using a third-order Runge-Kutta scheme. The prescribed parameters of the simulations presented in the following are described in Table \ref{Tab:Tcr}, together with the typical level $B_\perp(k_f)$ at the injection wavenumber and a phenomenological  estimate $\chi_f^t$ of the nonlinear parameter (defined below). They involve resolutions of $N=240$ or $N=360$ grid points in each direction and a time step $\Delta t$ that has usually to be reduced as the simulation proceeds.

All the simulations are driven by an additive random forcing, white-noise in time in such a way that the energy injection rate $\epsilon _E^\pm$ of each type of counter-propagating waves is prescribed. This is done by a forcing terms in the equations for ${\widehat{ D\mu^\pm}}$, in the form of $f_{\boldsymbol k}^\pm =  \frac{1}{\sqrt{\Delta t}} A^{\pm} \exp \left[ {-\frac{(k_\|-k_{0\|})^2 }{\sigma_{0\|}^2}} \right] \exp \left[ {-\frac{(k_\perp-k_{0\perp}^2)}{\sigma_{0\perp}^2}} \right]$, truncated so that only  wavenumbers $k_\perp$ and $k_\|$  such that $|k_{\|,\perp}-k_{0\|,\perp}| \le 3 \sigma_{0\|,\perp}$ are driven (in practice, only 3 Fourier modes are forced in each direction) and multiplied by a factor $a + ib$ where $a$ and $b$ are Gaussian real random  variables of zero mean value and variance unity, drawn independently for each of the fields at each time step. In Eqs. (\ref{Gpmequation}) for $G^\pm$ that are integrated numerically in the following sections, this corresponds to driving terms  of the form $\displaystyle{\frac{1}{2}\left [\frac{L_e}{s} \mp \delta M_2 \Lambda^{-1} \right]  D_e^{-1}f^+ -\frac{1}{2} \left [\frac{L_e}{s} \pm \delta M_2 \Lambda^{-1}\right]  D_e^{-1}f^-}$. In addition,  dissipation terms  $\nu (-\Delta)^4 G^\pm$ are included.

When not otherwise specified, we choose $k_{0\|} = k_{0_\perp} = k_f$ and $\sigma_{0\|} = \sigma_{0\perp} = \sigma_f$. Due to the ordering underlying the derivation of the gyrofluid model (where longitudinal gradient balances transverse nonlinearity), this choice corresponds to a quasi-transverse driving in the primitive physical variables. 
The energy injection rate,  $\epsilon_E\equiv \epsilon_E^+ + \epsilon_E^-$, is chosen to be equal to $0.372$ for most of the runs, while the GCH injection rate $\epsilon_C\approx (\epsilon_E^+ - \epsilon_E^-)/v_{ph}(k_f)$  varies with $k_f$ and $\beta_e$. 
Since the goal of the simulations was to study the large-scale dynamics and because of constraints on computational resources we had to introduce hyperviscosity coefficients $\nu$ that often do not permit the  development of a small-scale inertial range.  As a consequence there is no real  scale separation between injections and dissipation.

\section{Global properties of the GCH inverse cascade}\label{sec:global}

In the small-scale limit, the model  reduces to ERMHD.  When $\beta_e \to \infty$, this latter model can be derived in the strong anisotropic limit from EMHD (see e.g. \citet{Galtier15}) for which direct numerical simulations in the case of  imbalanced driving display an inverse cascade of magnetic helicity \citep{KimCho15}. Therefore, it is of interest to study  the  dynamics when the driving takes place at scales comparable to or moderately smaller than the sonic Larmor radius (taken as the length unit), for finite values of $\beta_e$.  

It is first useful to define the various spectra and fluxes that will be used to illustrate the cascade properties.
We denote by  $E(k_\perp)$ and $E_C(k_\perp)$ (hereafter, transverse spectra)  the spectral density in the transverse plane of the energy and GCH, respectively. Similar quantities  $E(k_\|)$ and $E_C(k_\|)$ are defined for the parallel spectra\footnote{{Note that the parallel spectra here refer to spectra along the direction of the ambient magnetic field. Especially when the nonlinearity parameter of the simulation is of order unity, these quantities can differ from the parallel spectra associated with second-order structure functions tied to the local magnetic field lines, whose definition requires the reintroduction of a large but finite value of the ambient magnetic field (see e.g. \citet{Maron01} in the MHD case). In this case, a given field line can indeed wander throughout the whole perpendicular domain as it extends along the parallel direction, even when the field line distortion is small. This could potentially affect the interpretation of the parallel spectra of Fig. \ref{fig:R1}.}}. Integration of these spectra 
with respect to the transverse ($k_\perp$) or parallel ($k_\|$) wavenumbers, reproduces the corresponding ideal quadratic invariants. It is also of interest to consider the Elsasser
energy and GCH transverse spectra  $E^\pm(k_\perp)$ and $E_C^\pm(k_\perp)$ related to the  energy  and GCH  by the relations $E(k_\perp) = \frac{1}{2} (E^+(k_\perp) + E^-(k_\perp))$ and $E_C(k_\perp) = \frac{1}{2} (E_C^+(k_\perp) - E^-_C(k_\perp))$ with $E^\pm_C(k_\perp) = E^\pm(k_\perp)/v_{ph}(k_\perp)$. Here, $E^\pm(k_\perp)$ can be viewed as the spectrum of the field $D_e \mu^\pm$ and $E_C^\pm(k_\perp)$ of the field $V_{ph}^{-1/2}D_e \mu^\pm$.
Furthermore, the perpendicular and parallel fluxes of the energy and GCH  are respectively defined as the negative nonlinear contributions to  $\partial_t \int_0^{k_\perp} E(k'_\perp) dk'_\perp$ or $\partial_t \int_0^{k_\perp} E_C(k'_\perp) dk'_\perp$(for the perpendicular fluxes) or to 
$\partial_t \int_0^{k_\|} E(k'_\|) dk'_\|$ or $\partial_t \int_0^{k_\|} E_C(k'_\|) dk'_\|$ (for the parallel fluxes). They are explicitly calculated using Eq. (\ref{eq:fluxes}) together with Eqs. (\ref{CTransfer})-(\ref{ETransfer}) of Appendix \ref{App:transfers}.

After presenting our fiducial run, which will serve to illustrate the existence and main properties of the GCH inverse cascade, we will address the effect of a variation of the energy injection rate $\epsilon_E$, and thus of  the turbulence strength.

\begin{figure}
     \begin{subfigure}{.48\textwidth}
        \centering
        \includegraphics[width=1\textwidth]
        {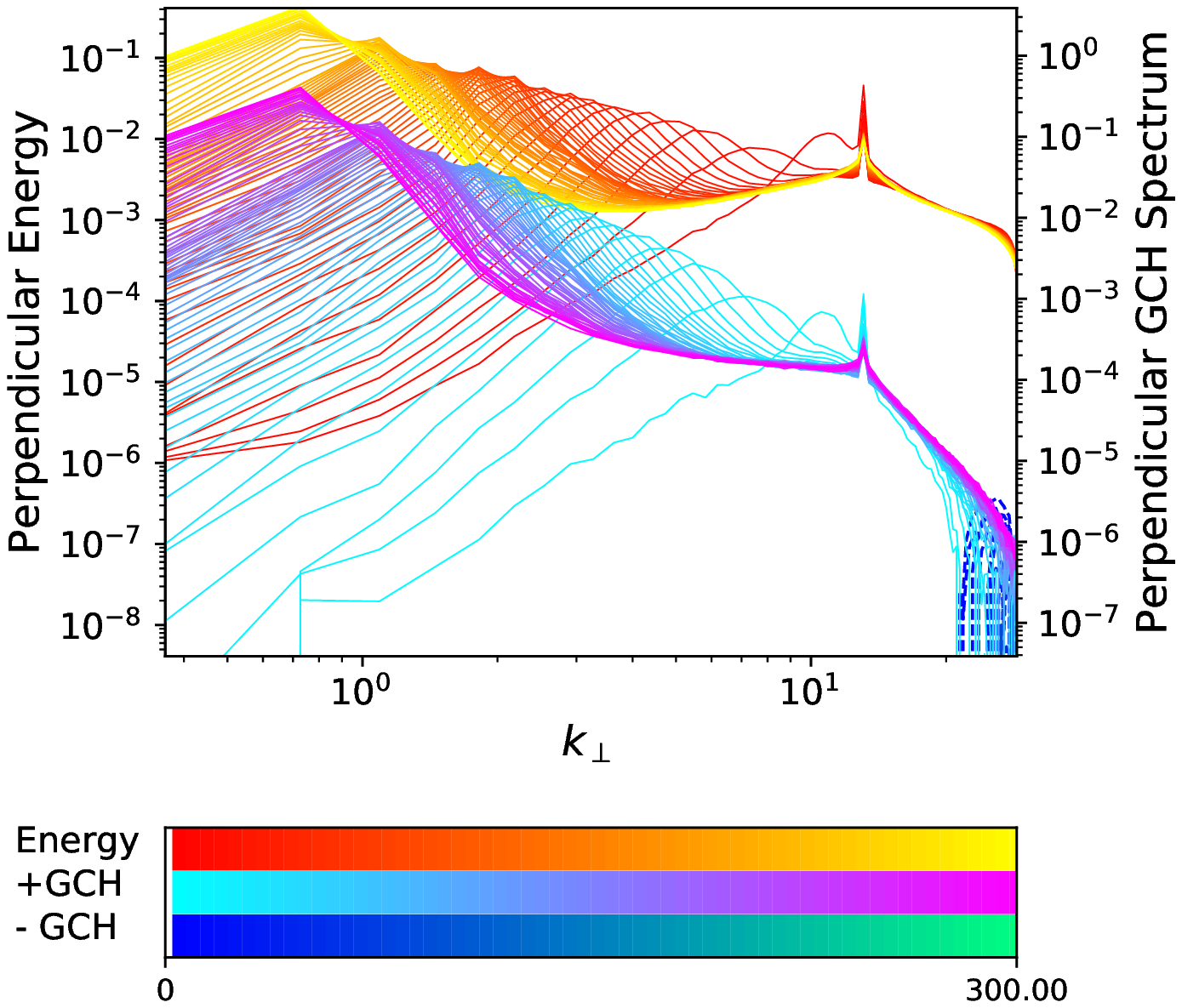} 
        
    \end{subfigure}
     \begin{subfigure}{.48\textwidth}
        \centering
        \includegraphics[width=1\textwidth]
        {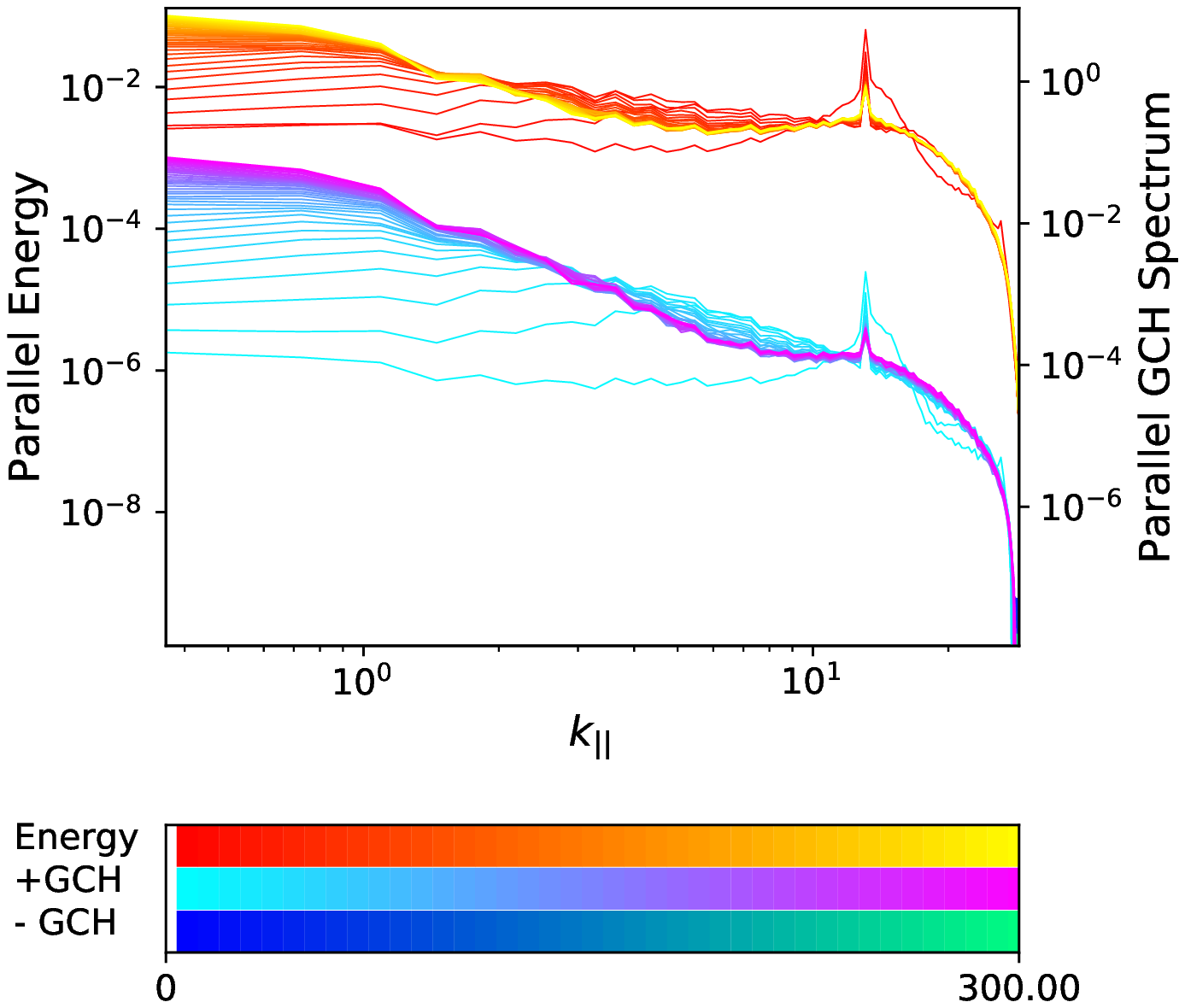} %
        
    \end{subfigure}
     \begin{subfigure}{.48\textwidth}
      \centering
      \includegraphics[width=1\linewidth]
      {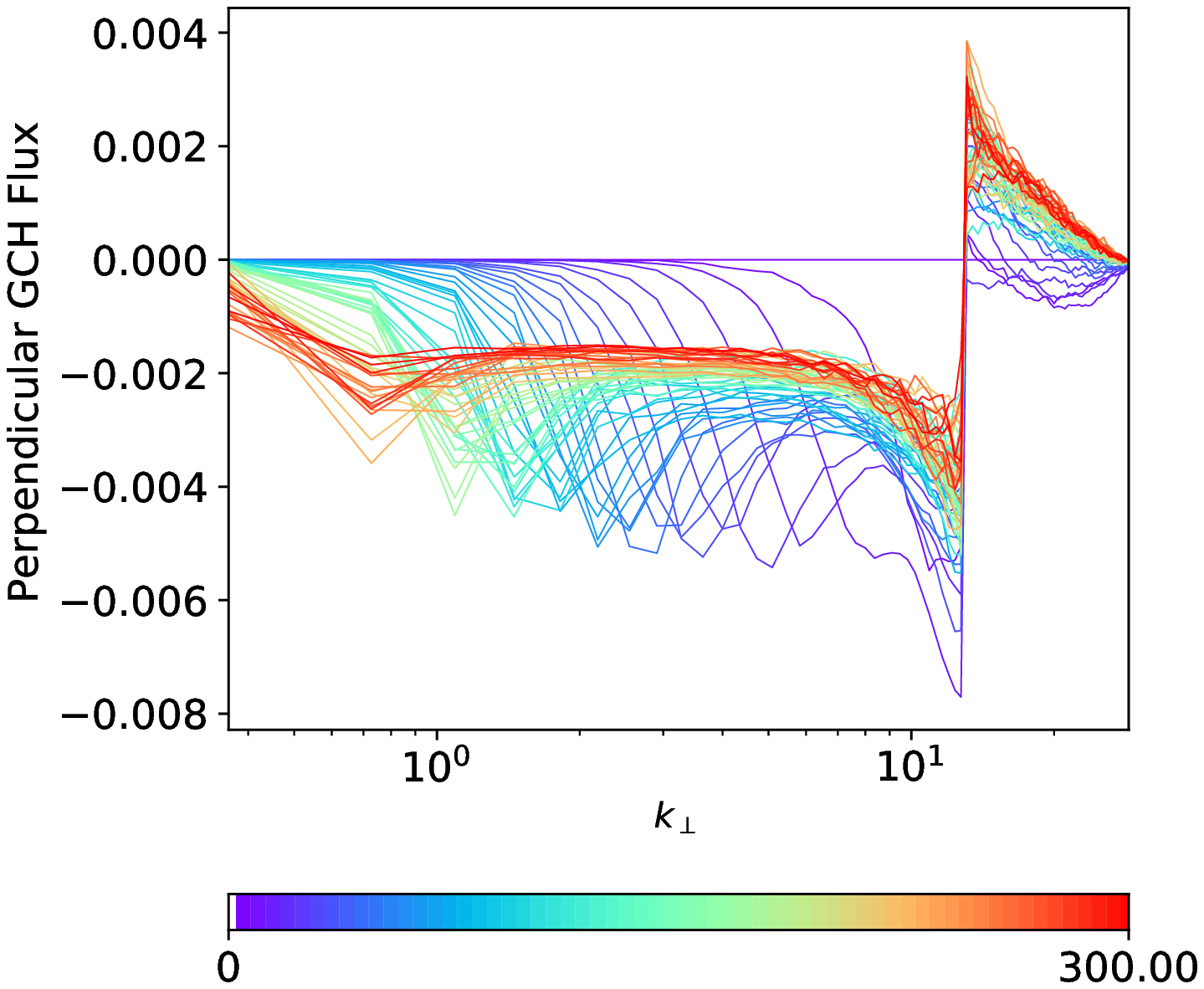} 
    \end{subfigure}
   \begin{subfigure}{.48\textwidth}
      \centering
      \includegraphics[width=1\linewidth]
      {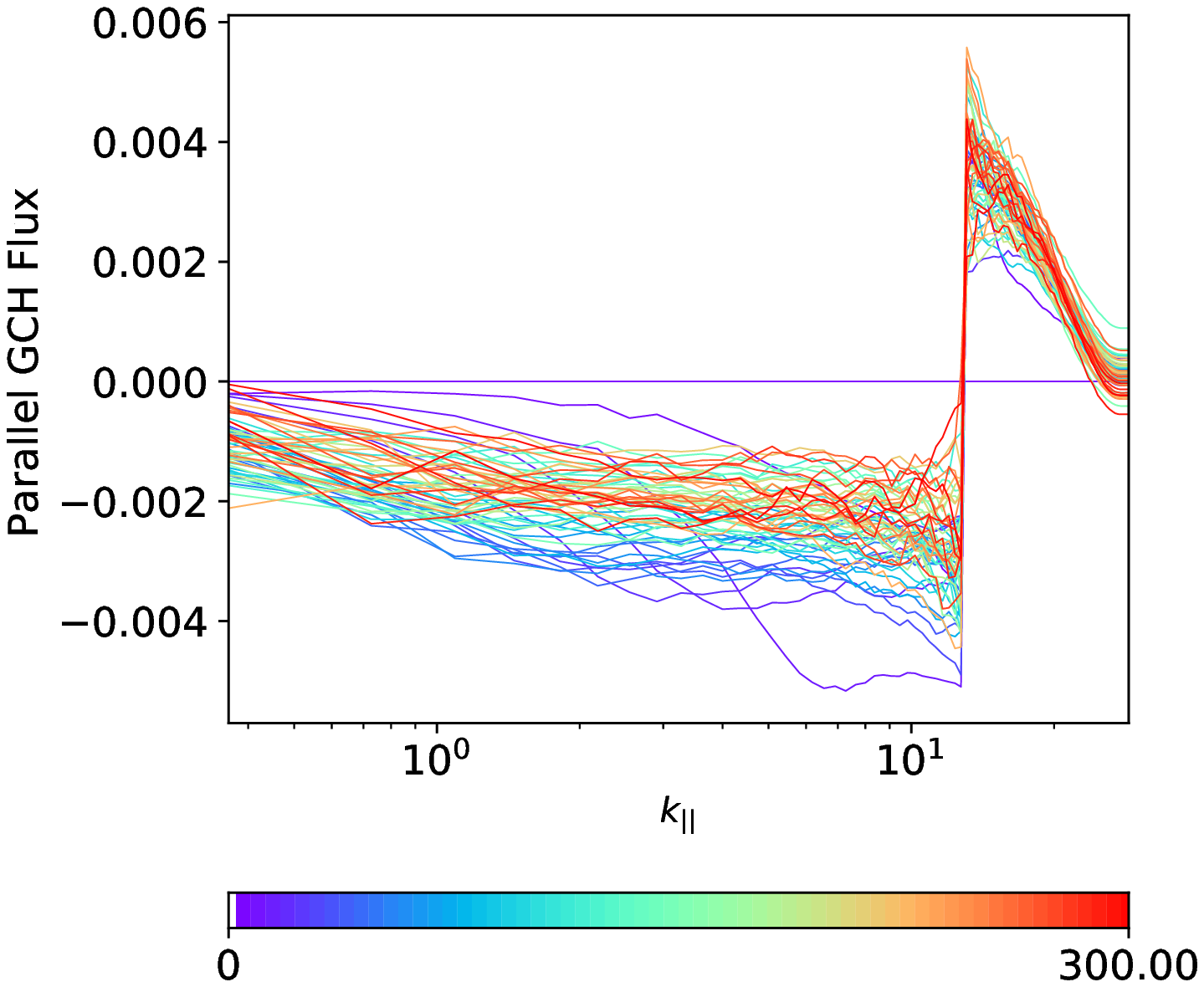} %
    \end{subfigure}
      \caption{Perpendicular (left) and parallel (right) energy and GCH spectra (top) and GCH fluxes (bottom) for Run $R_{13}^2$. The integration time that extends up to $t=300$ (in $\Omega_i^{-1}$ units) is indicated in the the color bars, where +GCH refers to positive value and -GCH to the modulus of negative values (dashed lines visible at the smallest scales) of the GCH spectra. For clarity, the axes of the energy and GCH spectra have been shifted. Same convention on forthcoming graphs.
     }
     \label{fig:R14SpecandFlux}
\end{figure}

\begin{figure}
      \begin{subfigure}{.48\textwidth}
        \centering
        \includegraphics[width=1\textwidth]
        {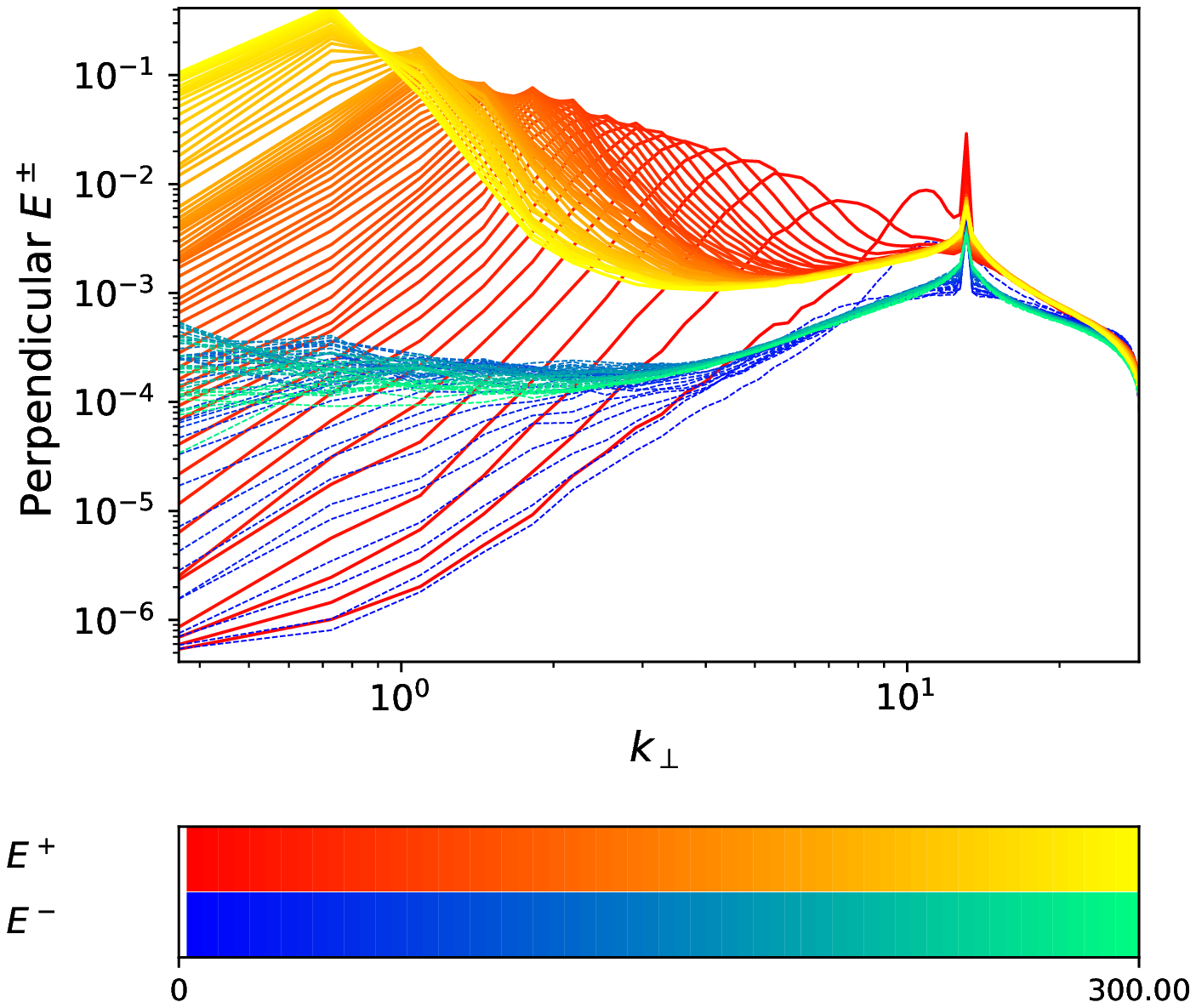} %
    \end{subfigure}
    \begin{subfigure}{.48\textwidth}
      \centering
      \includegraphics[width=1\linewidth]
      {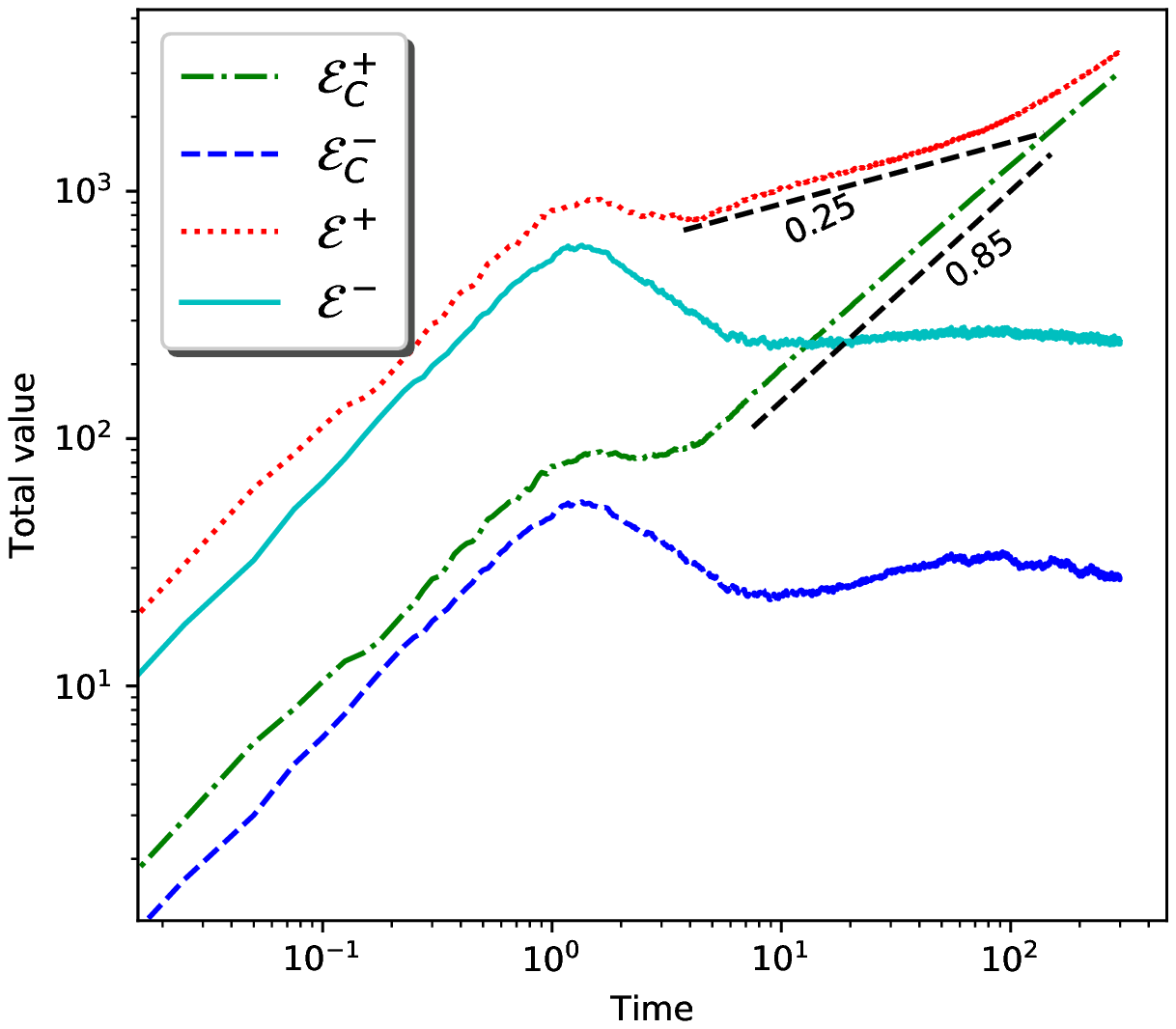} 
    \end{subfigure}
     \caption{Perpendicular Elsasser spectra  (left) and time evolution of the total Elsasser energies ${\cal E}^\pm$  and GCHs ${\cal E}_C^\pm$ (right) for Run $R_{13}^2$.
     }
     \label{fig:R14elsasser}
\end{figure}

\subsection{Spectra and fluxes}
When choosing $\beta_e=2$ and $k_f = 13$, an inverse cascade of GCH quickly develops, giving rise to  a spectral bump propagating  towards large transverse scales. This is illustrated in Fig. \ref{fig:R14SpecandFlux} (top, left). In this case, the inverse cascade extends over more than one decade before reaching the weakly dispersive scales. The observed dynamics is qualitatively similar to that obtained in numerical simulations of EMHD  \citep{KimCho15} and also of incompressible MHD in the absence of an ambient field \citep{Muller12, Linkmann17}.
At the final time of the simulation, the energy has almost reached the lowest modes and a separate simulation is needed to address the situation where the cascade crosses $k_\perp=1$ and penetrates into the weakly dispersive range. It will be discussed in Section \ref{smallkf} with run $R_{1.3}^2$.

The  perpendicular GCH flux displayed in the bottom left panel shows at late times a remarkably constant range, associated with an inverse cascade, in the spectral region between the peak and the forcing. Interestingly, a self-similar inverse cascade develops in the parallel direction as well, as shown in Fig. \ref{fig:R14SpecandFlux} (top, right), also associated 
with a negative and constant GCH flux (bottom, right). The existence of this parallel inverse cascade has important consequences on the types of structures that form in physical space, as discussed in Section \ref{vortices}.
When considering energy fluxes (not shown), we note that the perpendicular one is significant at wavenumbers larger than $k_f$ but very small for $k_\perp<k_f$. Energy is transferred to large scales together with the cross-helicity, but only the latter undergoes a true inverse cascade. In the parallel direction, the energy flux, although not displaying a constant range, is decaying more slowly as $k_\perp$ decreases away from $k_f$.

In Fig. \ref{fig:R14elsasser}, we show the Elsasser spectra $E^\pm(k_\perp)$ (left) and the time evolution of the total Elsasser energies ${\mathcal{E}^\pm =\sum_{k_\perp} E^\pm}(k_\perp)$   and of the positive/negative GCHs ${\mathcal E}_C^\pm =\sum_{k_\perp} E_C^{\pm}(k_\perp)$ (right).  The inverse GCH cascade is only associated with an inverse transfer of $E^+$. This behaviour is specific to cases where the ratio $\epsilon^+/\epsilon^-$ is small or moderate, as also seen in EMHD simulations by \citet{KimCho15}. In this run,  a non-negligible transfer of $E^-$ is nevertheless visible, which might develop into a different type of inverse cascade involving longer spatial and temporal scales.  Simulations with a spectral range extending to much larger scales and spanning a longer time interval would be necessary to characterize this effect more precisely. Both $\mathcal{E}^+$ and  $\mathcal{E}_C^+$ grow at long time as power laws with exponents similar but not identical to those reported in \citet{KimCho15} in the case of EMHD. Furthermore, we  verified that, while keeping the parameters of run $R_{13}^{2}$, but injecting only one type of wave (infinitely imbalanced driving), the other type of wave is generated by interactions of driven modes. While the energy of the dominant mode is transferred to the large scales through the propagation of a spectral bump whose envelope obeys a $k_\perp^{-3/2}$ scaling law,  the subdominant wave undergoes a self-similar cascade with a $k_\perp^{-1}$ spectrum (not shown), indicating a behavior of both spectra similar to that observed in EMHD simulations with maximal helicity injection \citep{KimCho15}.

\subsection{Shell-to-shell transfers} \label{transfer}

\begin{figure}
    \begin{subfigure}{.5\textwidth}
      \centering
      \includegraphics[width=1\linewidth]
      {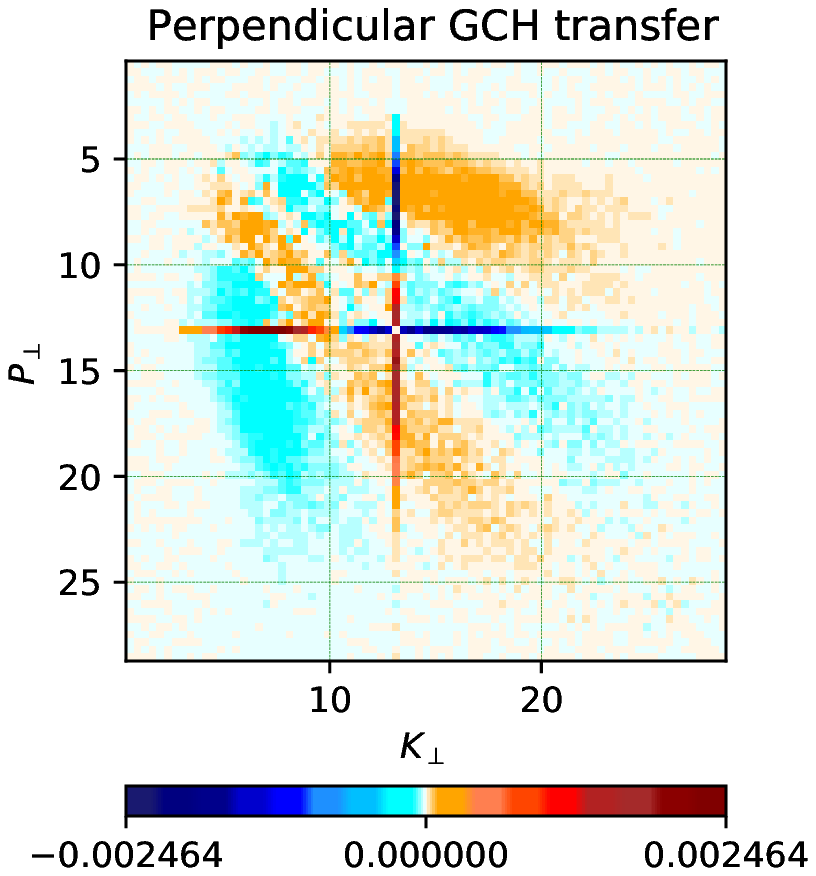} %
     \end{subfigure}
    \begin{subfigure}{.5\textwidth}
      \centering
      \includegraphics[width=1\linewidth]
      {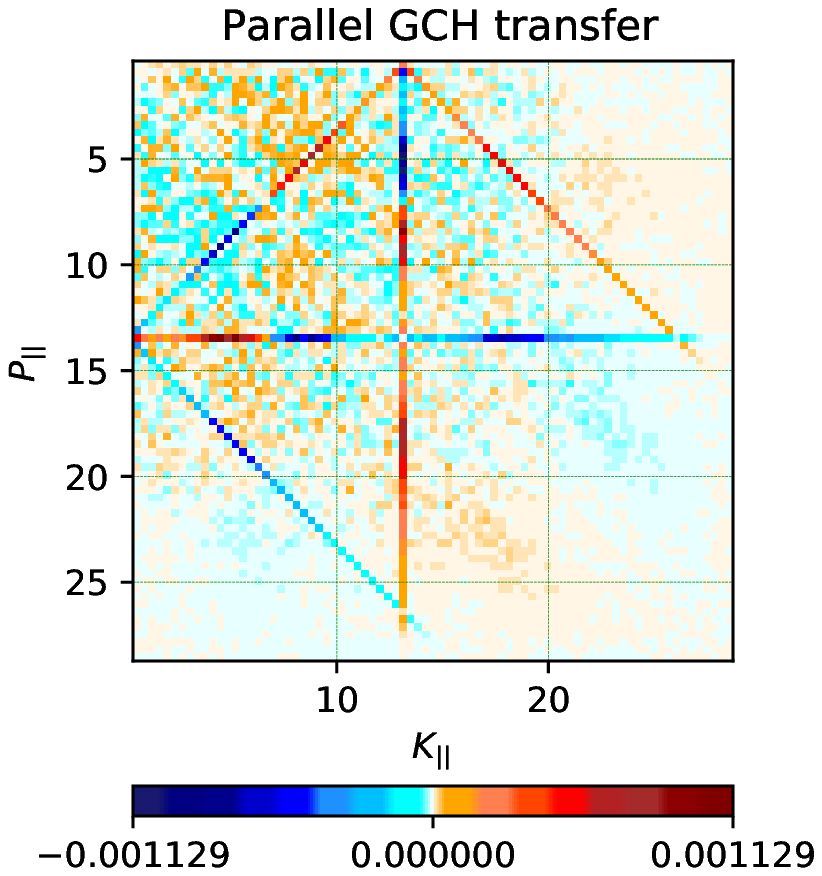} %
    \end{subfigure}
     \begin{subfigure}{.5\textwidth}
      \centering
      \includegraphics[width=1\linewidth]
      {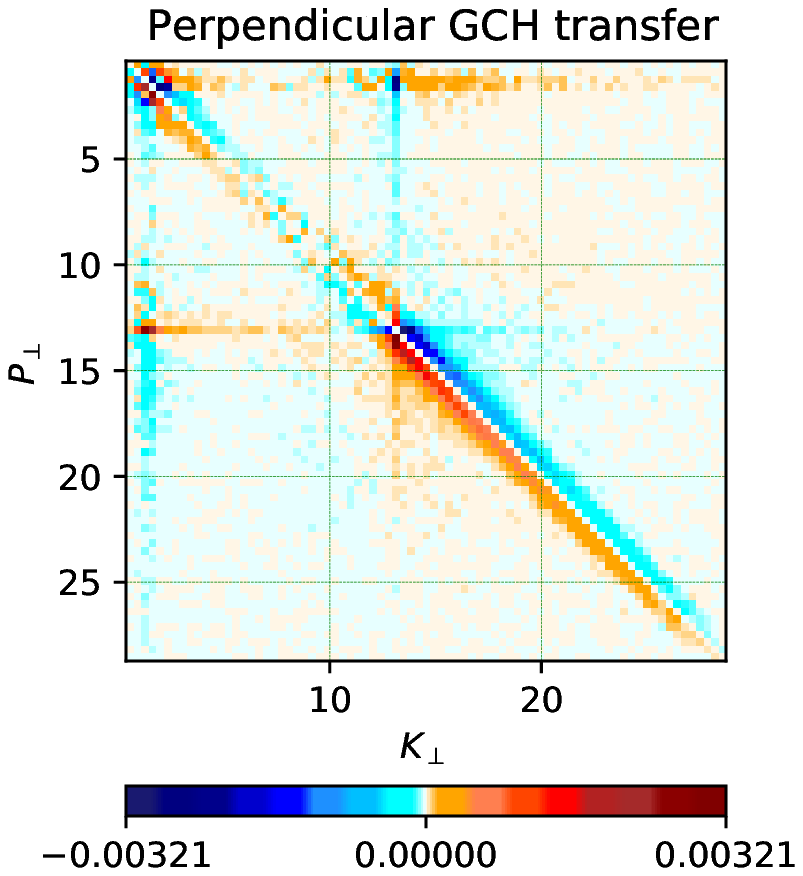} %
    \end{subfigure}
    \begin{subfigure}{.5\textwidth}
      \centering
      \includegraphics[width=1\linewidth]
      {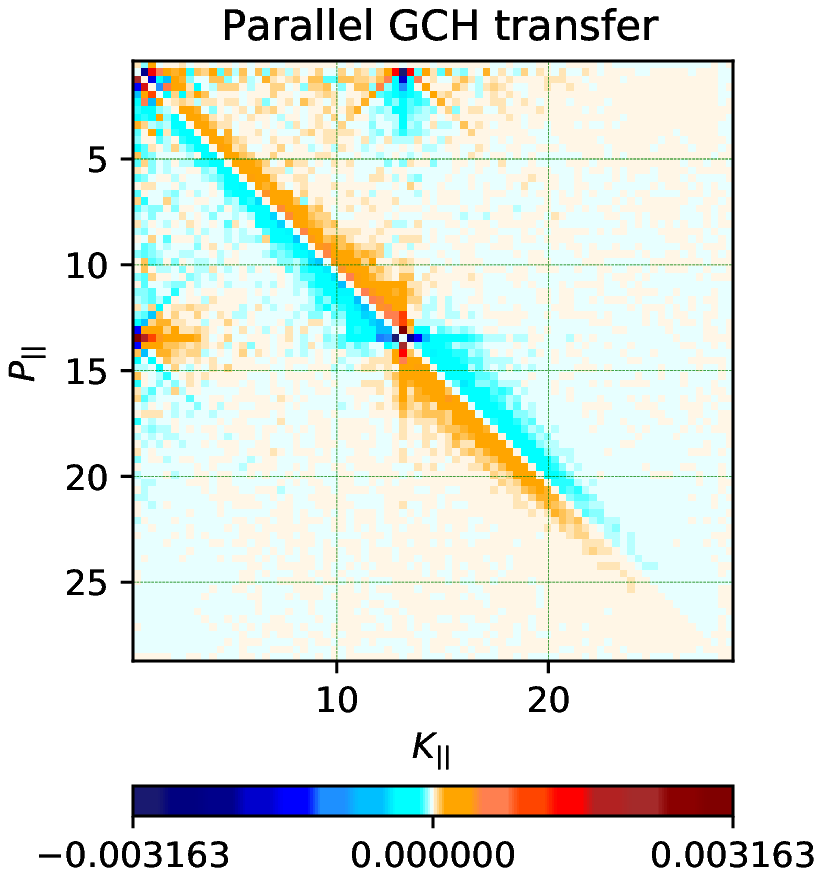} %
    \end{subfigure}
     \caption{
    Perpendicular (left) and parallel (right) transfers of GCH for run $R_{13}^2$ at an early ($t=5$, top) or late ($t=100$, bottom) time of the simulation.
   }
    \label{fig:R14-GCH-Transfers}
\end{figure}

In the following, we analyze the shell-to-shell transfers, i.e. the amount of energy and GCH transferred per unit time from one spectral shell in the transverse plane to another, after integration on the parallel wavenumbers. The transfer between parallel wavenumbers, after integration on the transverse wavenumbers is also considered. These quantities provide useful information concerning the degree of (spectral) locality of the nonlinear interactions. {Their expressions are derived in Appendix \ref{App:transfers} and explicitly given in Eqs. (\ref{CTransfer}) and (\ref{ETransfer})}. It is to be noted that in this context $R_{13}^2$ is generic and the statements about the non-local nature of transfers apply to the other simulations. However, as shown in the following subsection, other features may differ.

Figure \ref{fig:R14-GCH-Transfers} shows the GCH shell-to-shell transfers both in the transverse plane (left) and in the longitudinal direction (right), at an early time ($t=5$, top), and at a time close to the end of the simulation ($t=100$, bottom).
Transfers in the perpendicular directions are computed from  Eqs. (\ref{CTransfer}) and (\ref{ETransfer}) of Appendix \ref{App:transfers}. Similar expressions are used for the longitudinal transfers, where cylindrical shells are replaced by slabs. 

At early time, 
inverse transfers of GCH appear to be strongly nonlocal, i.e. off-diagonal which indicates transfer between non-neighboring shells.
We observe the characteristic cross-signature which demonstrates the interactions of the forcing with other shells. The modes located on the upper vertical and left horizontal axes that meet at the forcing wavenumber are associated with an inverse transfer close to the forcing.  On the contrary, the nonlocal transfers between the forcing and the wavenumbers that are much larger, are consistently associated with direct transfers. 

The inverse GCH cascade is mainly driven by the nonlocal interactions between the forcing and the propagating bump, that can be seen as the orange (cyan) areas located in the regions where $k_\perp$ (respectively $p_\perp$) is slightly smaller and larger than $k_f$ with  $p_\perp$ (respectively $k_\perp$) between $5$ and $10$.

In the parallel direction, there is no significant inverse transfer but interacting triads involving the forcing and its first harmonic are clearly visible (see also the parallel spectrum at the first output in the top right panel of Fig. \ref{fig:R14SpecandFlux}). The oblique lines ($k_\| + p_\| = k_f$ and circular permutations) and those parallel to the axes ($k_\| = k_f$ or $p_\|=k_f$)  correspond to the interactions between triads including the driving mode. No other interaction is significant at this early time.

The transfers  at late times display more local features, both in the parallel and perpendicular directions. The perpendicular inverse transfer of GCH still involves nonlocal interactions between the spectral bump and the forcing, while energy (not shown) has predominantly local features. It is noteworthy to mention that early bump propagation is always dominated by the nonlocal cross-type interaction between the forcing and the bump. The parallel GCH transfers associated to the direct cascade towards scales smaller than the forcing is local, while the inverse parallel cascade involves both strongly local and non-local features.

\begin{figure}
    \begin{subfigure}{.5\textwidth}
      \centering
      \includegraphics[width=1\linewidth]
      {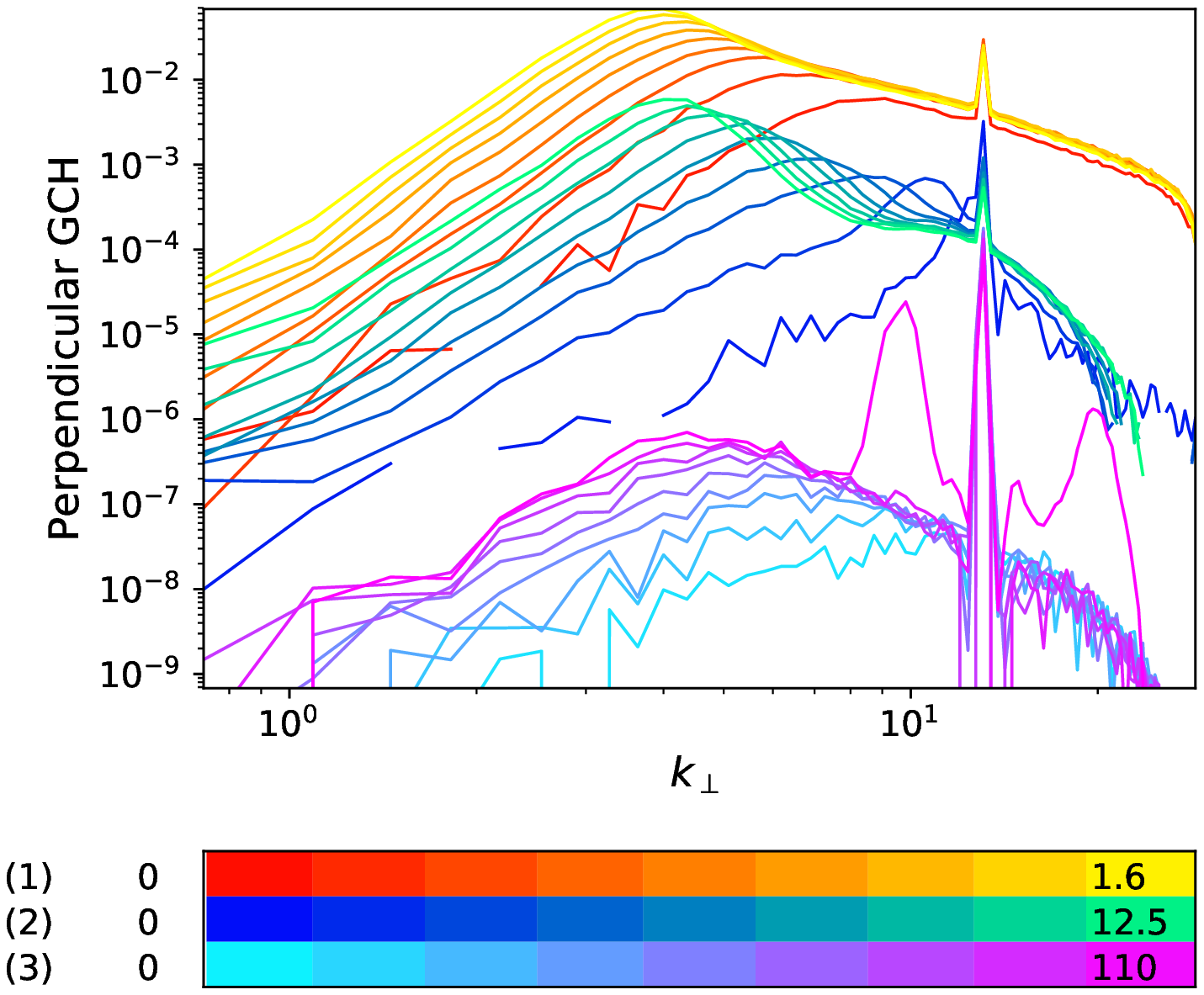} %
      \label{fig:R1Early}
    \end{subfigure}
    \begin{subfigure}{.5\textwidth}
      \centering
      \includegraphics[width=1\linewidth]
      {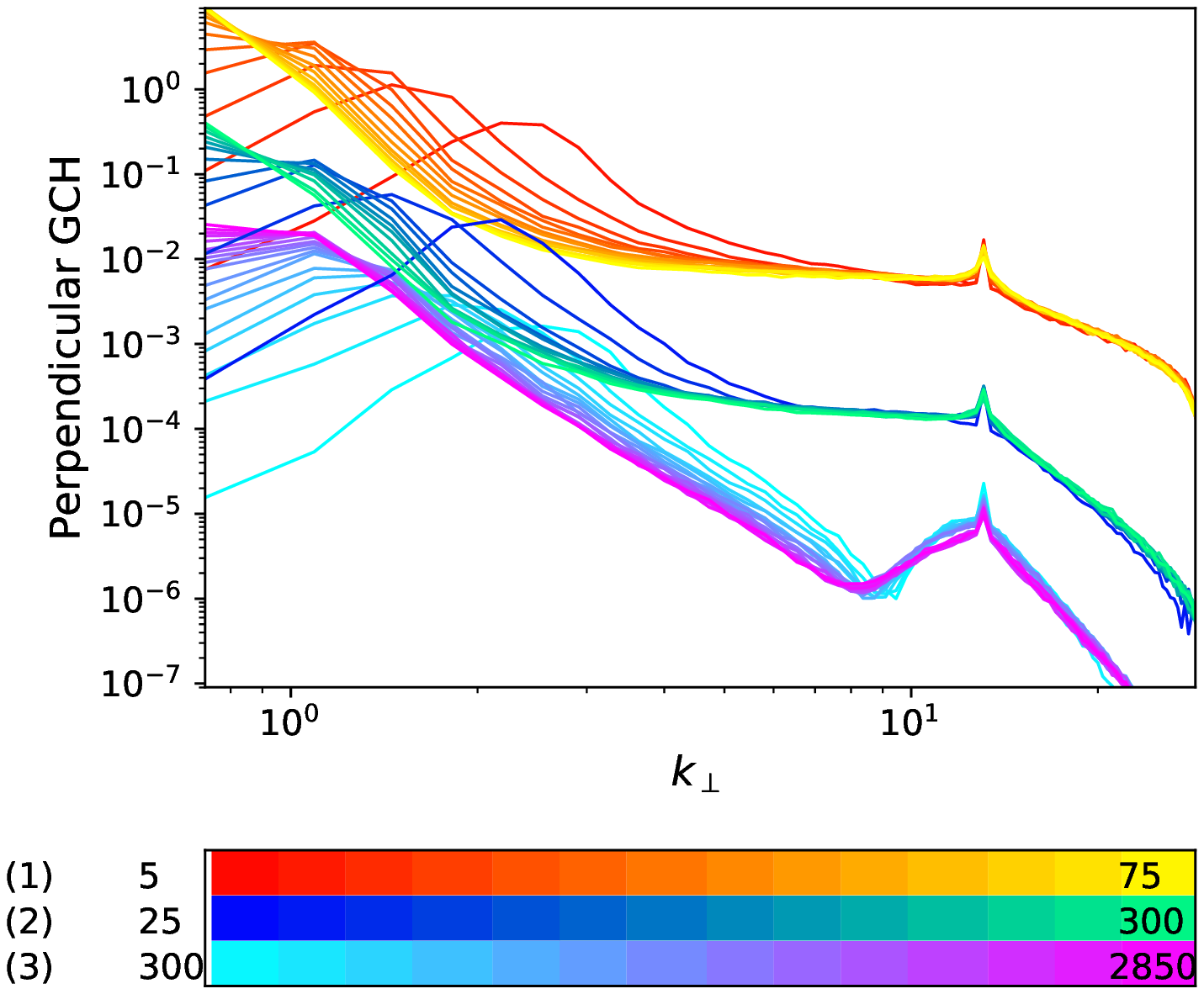} %
      \label{fig:R1Late}
    \end{subfigure}
    \caption{Snapshots at short (left) and long (right) times of the GCH spectra for runs $R_{13s}^2$ (1), $R_{13}^2$ (2) and $R_{13w}^2$ (3) corresponding  to strong, intermediate and weak injection rates respectively. Final times displayed in the left panel are $t=1.6$ (1), $t=12.5$ (2), $t=110$ (3), and in the right panel $t=15$ (1), $t=300$ (2), $t=2850$ (3).}
    \label{fig:R1Spec}
    \label{fig:vary-epsilon}
\end{figure}

\section{Influence of the turbulence strength on the early-time dynamics} \label{sec:turb-strength}

Since the dynamics involves both waves and the nonlinear dynamics of the purely perpendicular component, it is natural to investigate the impact of changing the amplitude of the perpendicular magnetic fluctuations. This is easily done by changing the injection rate $\epsilon_E$.

\subsection{Varying the amplitude of the fluctuations}
When varying the energy injection rate, the nature of the turbulent cascade that develops at relatively early time changes significantly (see Fig. \ref{fig:vary-epsilon}, left). At large amplitude, a self-similar  cascade  starts developing (as clearly seen in run $R_{13s}^2$) but the spectrum rapidly distorts. When the driving is decreased (run $R_{13}^2$), the dynamics is that of a propagating spectral bump, as described in the previous subsection. Further decreasing the injection rate, the early dynamics is  dominated  by  the  development  of  a parametric  decay  instability of the pump waves driven by the random forcing (conspicuous in Fig. \ref{fig:vary-epsilon} (left) where  two bumps emerge at time $t\approx 110$). Indeed, in contrast with  Alfv\'en waves in ideal MHD, three KAWs can interact resonantly, producing a decay instability \citep{Voitenko98a} (see Appendix \ref{App:decay}  for a brief description of  this linear instability in the context of the two-field model, in the absence of injection and dissipation). This instability is  competing with the turbulence dynamics (see \citet{Shi17,Fu18} for a study in the MHD context). Even though the growth rate becomes larger when the level of the magnetic fluctuations is increased, the transfer to neighboring modes due to turbulence is more efficient, thus hiding or preventing the instability. When $\epsilon_E$ is small enough, as it is the case for run $R_{13w}^2$, the growth rate is smaller, but the turbulence is also less efficient, making the appearance of the decay possible.
A more detailed study of the resonance conditions will be discussed in the next subsection.

Decreasing $\epsilon_E$ only affects the early-time dynamics. Indeed, as seen in the right panel of Fig. \ref{fig:vary-epsilon}, the late evolution of the two runs with strong and intermediate driving displays similar features associated with  a  strongly  bent  spectrum  with  a bump at large scale, separated from the forcing by a spectral domain where the dynamics tends to settle to an absolute equilibrium\footnote{We indeed observe that the GCH flux (which is flat in the spectral range between the driving and the spectral bump) slowly decreases as time evolves (see Fig. \ref{fig:R14SpecandFlux} for run $R_{13}^2$).}.  The case of a weak driving has the same tendency but the evolution is much slower. Similar simulations that display decay at early time but with a larger value of $\chi_f$ show in a more conspicuous way an evolution towards a flat spectrum in the range between the driving and the maximum of the spectrum.

Even if the inverse cascades displayed by the simulations previously discussed may suggest the existence of qualitatively different kinds of cascades, a more likely scenario is that there is a continuous transition between a self-similar cascade and a spectral bump propagating to large scales.
Inspection of the time evolution of individual  energy and GCH transverse Fourier modes  indicates that when the inverse cascade reaches a given wavenumber, the corresponding modes first grow and then decrease with a characteristic decay time that becomes shorter when wave effects are enhanced. The relative effect of the strength of the turbulence compared to that of the waves, measured by the nonlinearity parameter at the driving scale $\chi_f$,  is then likely to characterize the cascade properties.

The nonlinear parameter $\chi({\boldsymbol k})$ at  the wavenumber ${\boldsymbol k}$ can be estimated as follows.
From Eq.(\ref{eq:A}) with $\delta=0$, we estimate  the non-linear time $\tau_{NL}(k_\perp)$ 
(defining ${\widehat M}_3 = 1 -{\widehat M}_1 + {\widehat M}_2$) by
\begin{equation}
    \tau_{NL}({\boldsymbol k})^{-1} = k_\perp^2 {\widehat M}_3 |\varphi({\boldsymbol k})|.
\end{equation}
The nonlinear parameter $\chi({\boldsymbol k}) =  \tau_{NL}({\boldsymbol k})^{-1} / (v_{ph} k_\|)$ reads (using Eq. (\ref{eq:vph})), 
\begin{equation}
    \chi({\boldsymbol k}) = \frac{1}{s} k_\perp {\widehat M}_2^{1/2} {\widehat M}_3^{1/2} \frac{1}{k_\|}|\varphi({\boldsymbol k})|.
\end{equation}
At least when the nonlinearities are not too strong, one can expect equipartition of the magnetic  and potential energy spectra (see e.g. \citet{Slepyan15} and references therein), a property which is observed to hold with a good accuracy in the most of the simulations  considered in this paper (in particular, the ratio of the two spectra at $k_f$ is typically $1\pm 0.03$, the error reaching $4\%$ at large $\beta_e$). This  leads to a phenomenological estimate that near the injection scale
\begin{equation}
    |\varphi({\boldsymbol k})| \approx s {\widehat M}_2^{-1/2} {\widehat M}_3^{-1/2} |{\boldsymbol B}_\perp({\boldsymbol k})|,
\end{equation}
which allows one 
to relate the nonlinear parameter $\chi_f$ at the injection wavevector ${\boldsymbol k}_f$ (whose perpendicular and parallel components are denoted $k_{\perp f}$ and $k_{\| f}$)  to the magnitude of magnetic fluctuations $B_\perp (k_{\perp f})= \sqrt{k_\perp  E_{B_\perp} (k_\perp)}$  (where  $E_{B_\perp} (k_\perp)$ holds for the one-dimensional transverse magnetic spectrum), by
\begin{equation}
\chi_f = \frac{k_{\perp f}}{k_{\| f}} |B_\perp(k_{\perp f})|.\label{chif-B}
\end{equation}

In order to have a qualitative relation between 
the level of magnetic field fluctuation $|B_\perp(k_{\perp f})|$ at the driving wavenumber and $\epsilon_E$, it is nevertheless possible
to use a simple  strong-turbulence phenomenology and write, still assuming energy equipartition,
\begin{equation}
    \epsilon_E\approx\frac{2}{\beta_e}|B_\perp({k_{\perp f}})|^2 \tau_{NL}(k_{\perp f})^{-1}\approx\frac{2}{\beta_e}|B_\perp({k_{\perp f}})|^3 k_{\perp f} v_{ph}(k_{\perp f}).
\end{equation}
This prompts the definition of the nonlinearity parameter based on turbulent phenomenology by, up to a numerical constant,
\begin{equation}
    \chi_f^t= \frac{k_{\perp f}}{k_{\| f}}\left ( \frac {\beta_e}{2} \frac{\epsilon_E}{k_{\perp f}v_{ph}(k_{\perp f})} \right )^{1/3}. \label{chif}
\end{equation}
Assuming that $\tau$ is of order unity, we find, using Table 1 in  \citet{PS19}, that at large scales $v_{ph}\sim s$,  and at small scales  $v_{ph}\sim s k_\perp$ when $\beta_e\lesssim 1$  and $v_{ph}\sim s^2 k_\perp$ when $\beta_e$ is large. For a driving acting at sub-ion scales, $\chi_f^t\sim\epsilon_E^{1/3}k_{\perp f}^{1/3}k_{\| f}^{-1}\beta_e^{1/2}$ for small or moderate $\beta_e$ and $\chi_f^t\sim\epsilon_E^{1/3}k_{\perp f}^{1/3}k_{\| f}^{-1}\beta_e^{2/3}$ for large $\beta_e$. 

The expression of the $\chi_f$ parameter suggests that the dynamics will be sensitive to the angle of the injected KAWs with respect to the ambient magnetic field. Indeed, in simulations for which $k_{\| f}\ll k_{\perp f}$, which makes  $\chi_f$ very large, the dynamics is quasi-2D. The waves have very long time scales and the inverse cascade is more self-similar. Moreover, in these conditions, the dynamics is influenced by the quasi-conservation of the squared magnetic potential, which is observed to undergo a self-similar inverse cascade (not shown).

Note that for small $\beta_e$,  where $v_{ph} \sqrt{\frac{\beta_e}{2}}$ has a finite limit, the parameter $\chi_f^s$ becomes independent of $\beta_e$ when velocities are measured in units of Alfv\'en speed instead of sound speed and the length unit is kept unchanged. Indeed, in this  limit, after a proper 
rescaling of the dependent and independent variables, the equations
become independent of $\beta_e$ and reproduce the model of \cite{Zocco11}, taken in the isothermal limit. Interestingly, the small-$\beta_e$ limit is already accurately approached when $\beta_e=0.2$. In contrast, at large $\beta_e$, the parameter $\chi_f^t$ increases with  $\beta_e$, even when using the Alfv\'en velocity unit. 

The values of $\chi_f^t$ for the different runs considered in this paper are given in Table 1, together with $|B_\perp(k_{\perp f})|$, which is measured immediately after the transient phase, at the beginning of the simulation. Note that, for some runs, $|B_\perp(k_{\perp f})|$ varies significantly during the time evolution.  The ratio $\chi_f/\chi_f^t$ has a mean value of $1.78$ with a standard deviation of $0.43$. Both quantities vary in the same way from one run to the other at moderate values of $\beta_e$, but not necessarily at larger values. Although relatively large, such 
 uncertainty nevertheless permits using $\chi_f^t$ as a crude prediction of the nonlinear parameter $\chi_f$ which can be only measured a posteriori. It is natural to attempt to isolate the parameter $\chi_f$ to see if it solely controls the onset of self-similar - bump transitions etc. However, since it is not possible to precisely assign $\chi_f$ when initializing the simulations, we would have to initiate multiple runs to be able to select the appropriate desired values. Such ensemble would drain computational resources. Nevertheless,  this  conjecture  that $\chi_f$ is the governing parameter appears to  be  consistent  with  the  various  simulations presented in this section.
 
\subsection{Analysis of a decay event} \label{decay-event}
\begin{figure}
    \begin{subfigure}{.5\textwidth}
      \centering
      \includegraphics[width=1\linewidth]
      {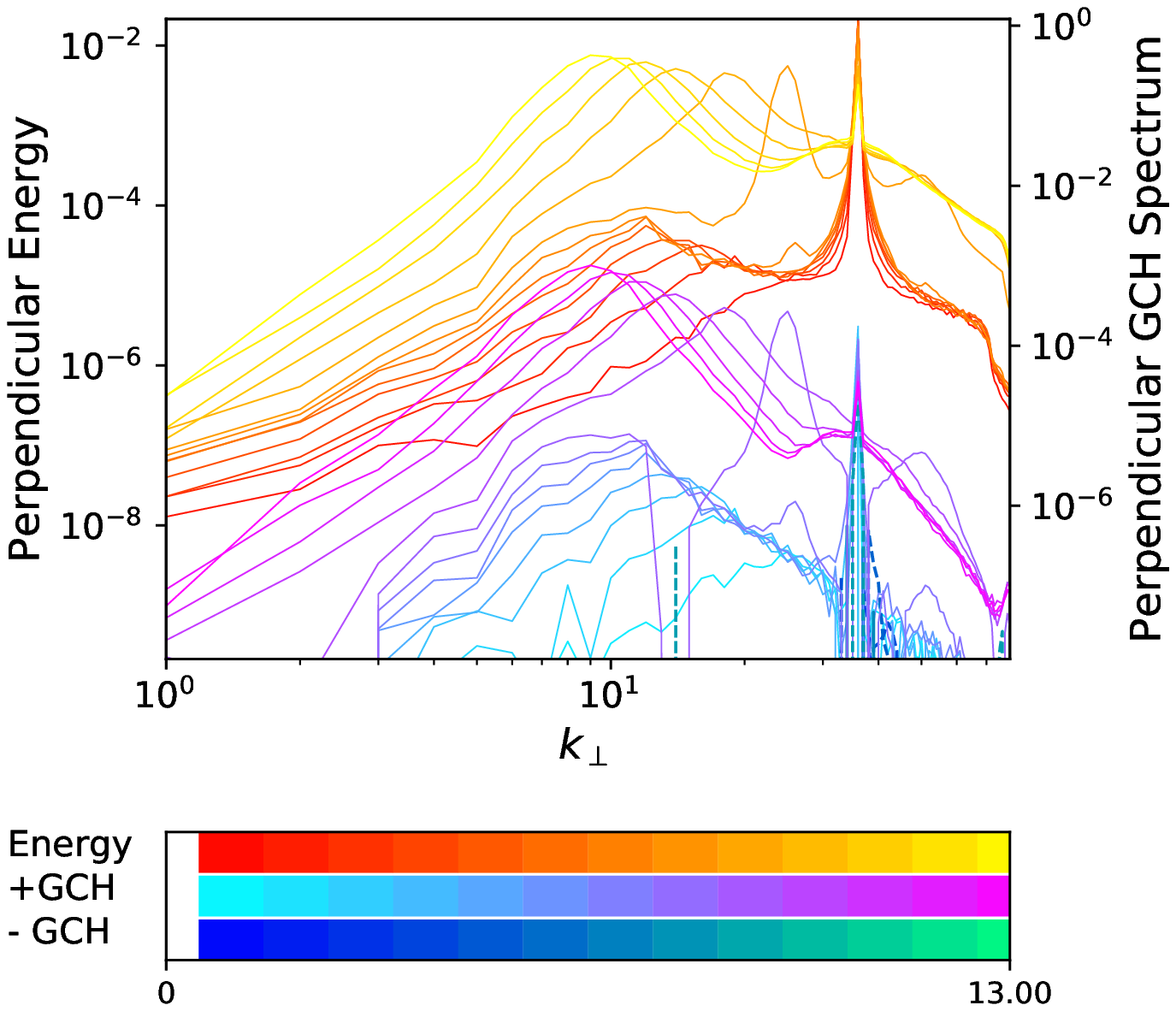} %
    \end{subfigure}
    \begin{subfigure}{.5\textwidth}
      \centering
      \includegraphics[width=1\linewidth]
      {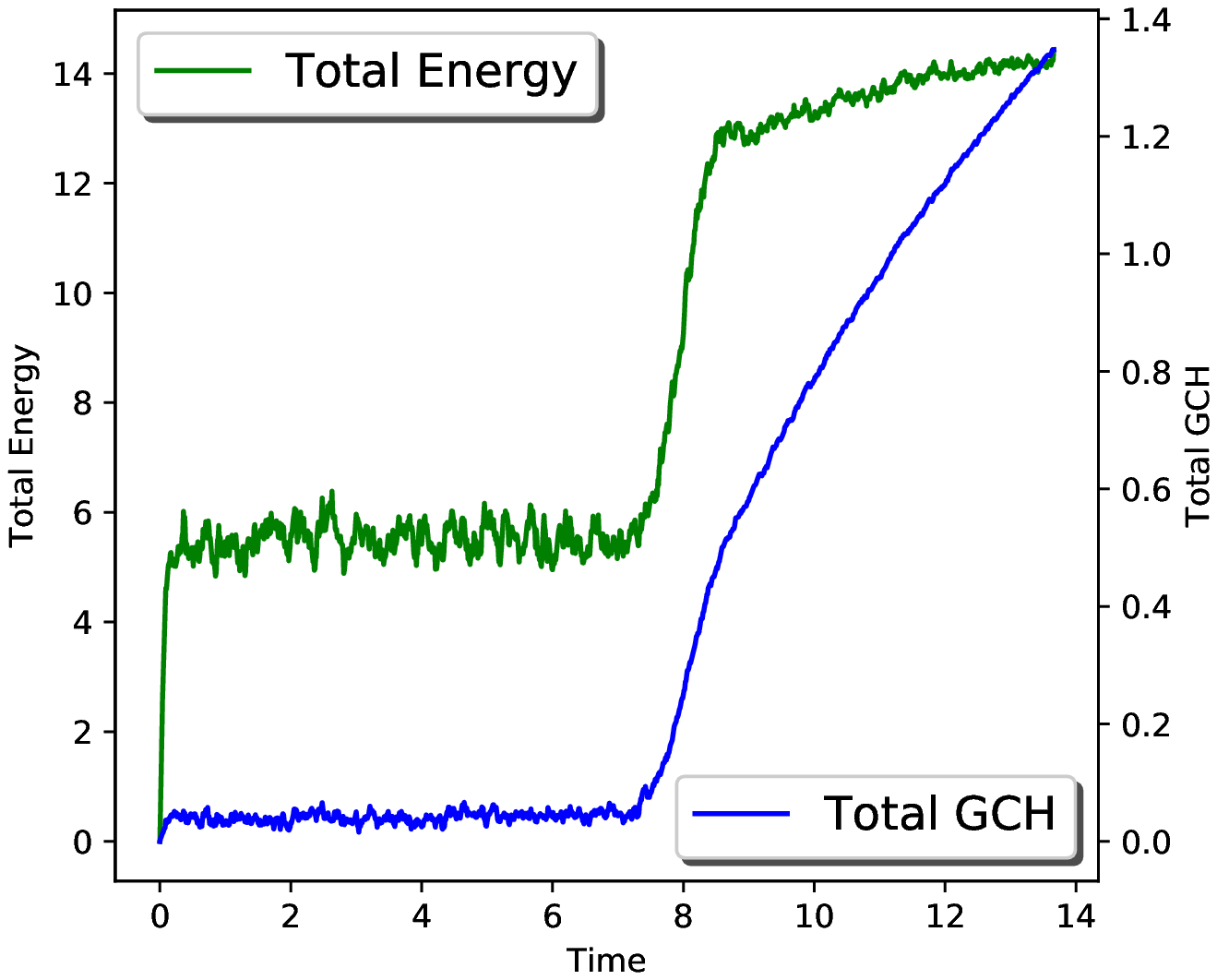} %
      \label{fig:R10bPerpSpec}
    \end{subfigure}
    \caption{Transverse energy and GCH spectra (left) and time variation of the total energy and GCH (right) for run $R_{36}^{2}$.}
    \label{fig:R10andbSpec}
\end{figure}
\begin{figure}
     \begin{subfigure}{.5\textwidth}
   \centering
    \includegraphics[width=1\textwidth]{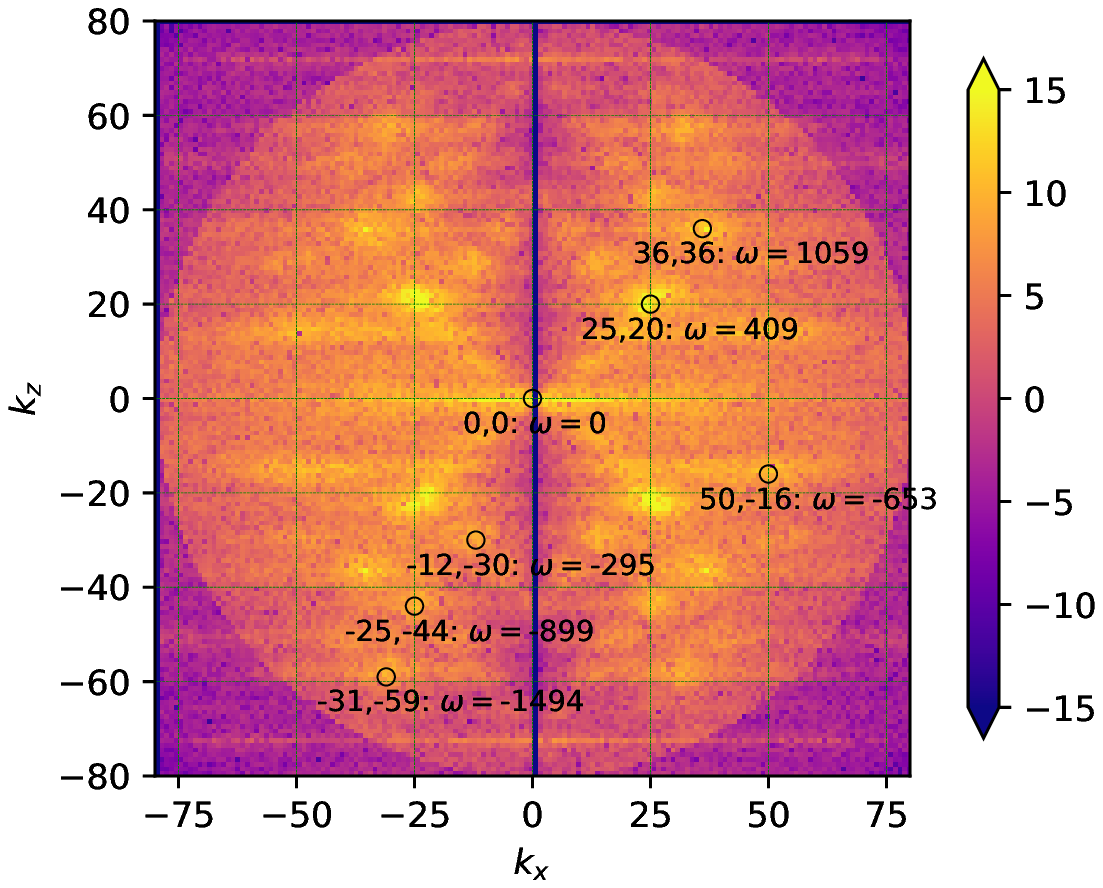}
    \end{subfigure}
    \begin{subfigure}{.5\textwidth}
       \includegraphics[width=.75\linewidth,angle=-90]
      {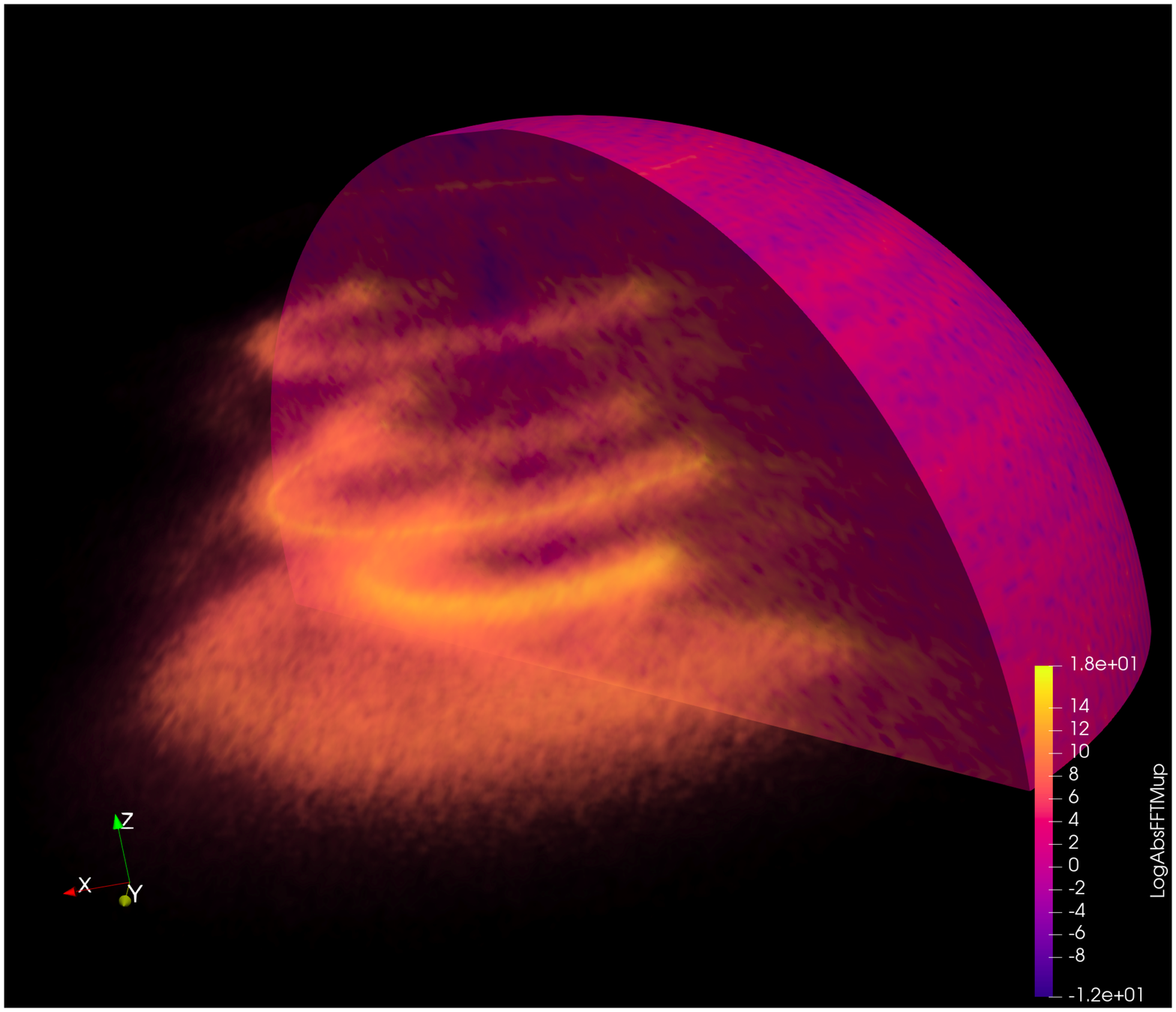} 
    \end{subfigure}
\caption{Color scale plot of $\ln|{\widehat \mu}^+|^2$ for run $R_{36}^2$ showing decay instability. Left: Cross-section  in a  $(k_x, k_z)$-plane. Because the forcing is statistically isotropic in the $(k_x, k_y)$-plane, this plot is sufficient to characterize the full $(k_x, k_y, k_z)$ space (see text). Among the visible peaks, some of them, such as (50,-16), (25,20) and (36,36) (which corresponds to the forcing) are associated with resonant triads. Right: Three-dimensional view.  The dominant peaks visible as dots in the $(k_x, k_y)$-plane appear as circles symmetric around the $k_z$ axis. Some circles are not visible since a transparency mask was applied to clearly visualize the most intensive peaks. 
}
\label{fig:R10FFT}
\end{figure}
Another example of the decay instability is encountered when, taking the same injection rate and $\beta_e$ as for run $R_{13}^2$, the  driving wavenumber $k_{\perp f}$ is increased to $36$. Figure \ref{fig:R10andbSpec}(left)
clearly shows that, after the development of self-similar spectra (with exponents -2 for energy and -3 for GCH, together with a very weak negative GCH flux), a transient regime develops, corresponding to the onset of unstable modes, an effect that does not persist after turbulence has become sufficiently developed.  
This suggests that, if the decay instability can be a leading mechanism in the regime of weak turbulence \citep{Voitenko98b}, its effect  cannot be more than a transient in the strong turbulence regime. When present, the decay instability nevertheless significantly affects the global dynamics. As seen in Fig. \ref{fig:R10andbSpec} (right),  a sharp  increase of the total energy and GCH takes place when the decay instability is acting. The instability can be identified at a time close to $t=7$. Later on, energy and GCH  are still growing, but at lower rates, consistent with the existence of an inverse cascade. As expected, increasing $\beta_e$ for example to $\beta_e=50$ (not shown), the other parameters being fixed, $\chi_f$ increases to 1.1  and the decay instability is suppressed. It is also interesting to note that the cascade that develops at early time in this case is self-similar.

In order to highlight the presence of resonant interactions in run $R_{36}^{2}$, we display in Fig. \ref{fig:R10FFT} (left) the color scale plot of the two-dimensional spectrum  $|{\widehat \mu}^+({\boldsymbol k})|^2$ in the plane $(k_x, k_z)$, showing decay instability at work (at $t \approx 8$).  Since the forcing is statistically isotropic in the $(k_x, k_y)$-plane, this plot is sufficient to demonstrate the features of the full $k_x, k_y$ and $k_z$ dependence. We see numerous peaks, some of them associated with resonant triads. In fact, due to axisymmetry, each peak is essentially a circle penetrating the plane (see Fig. \ref{fig:R10FFT} (right)). This is important when evaluating the resonance condition since it provides more freedom in choosing the corresponding $k_x$ and $k_y$ wavevector components. Of particular interest are triads with coordinates in the $(k_x,k_z)$-plane given by (50,-16), (25,20) and (36,36) (which corresponds to the forcing). We indeed see that, for this triad, both $k_{1z} + k_{2z} = k_{fz}$ and $\omega_1 + \omega_2 = \omega_f$. 
In this context, what appears as $k_x$ should be viewed as $k_\perp$.
Resonance conditions in the transverse plane then consist of 5 equations (3 norms of transverse wavevectors and 2 resonance conditions) for 6 unknowns. Choosing arbitrarily one of them, we can easily construct a resonant triad. There is thus an infinite number of such resonant triads which can be viewed as defining a resonant manifold. 

\subsection{Instability of the balanced state}\label{sec:instability-balanced}

\begin{figure}
    \begin{subfigure}{.5\textwidth}
      \centering
      \includegraphics[width=1\linewidth]
      {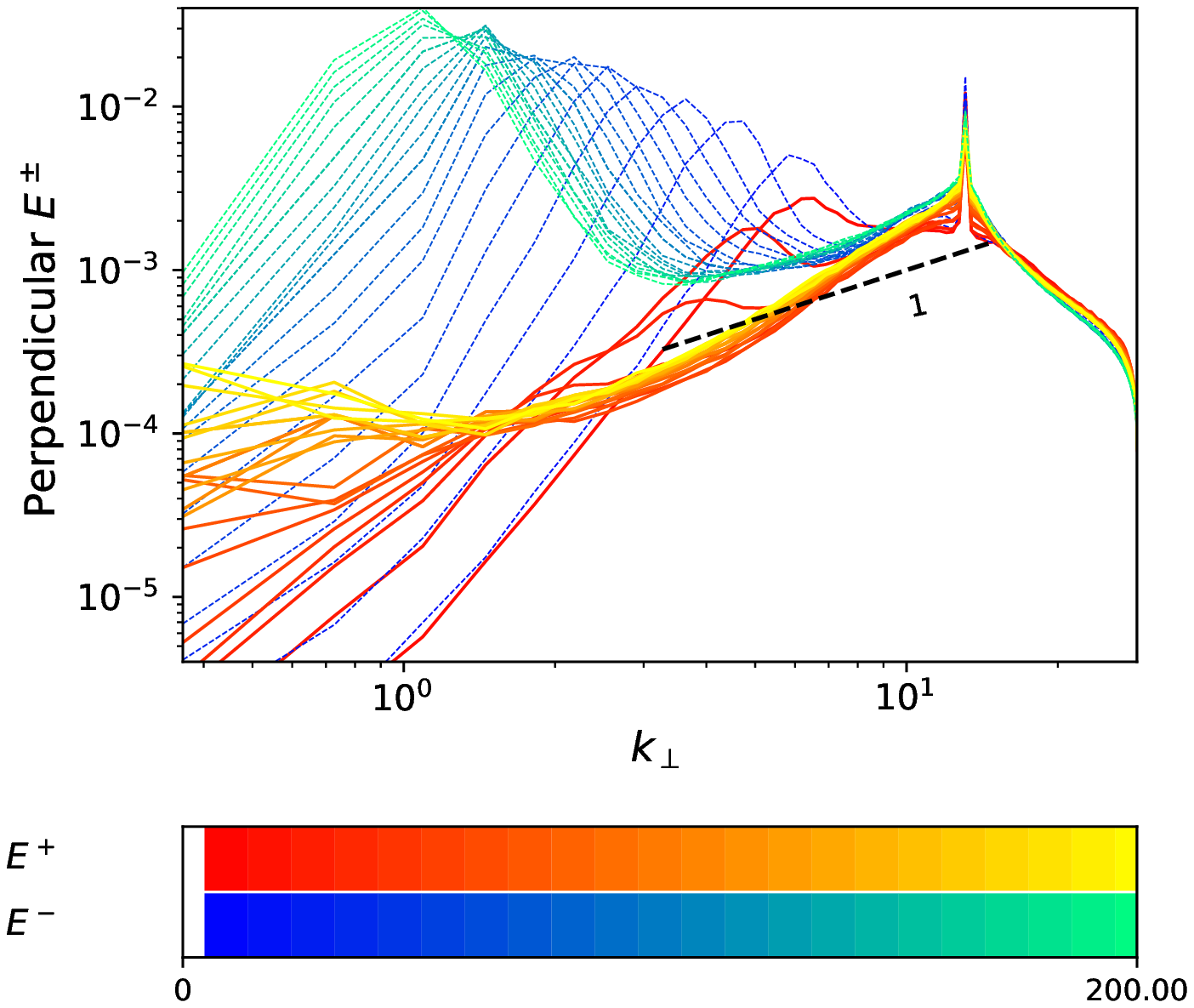} %
    \end{subfigure}
    \begin{subfigure}{.5\textwidth}
      \centering
      \includegraphics[width=1\linewidth]
      {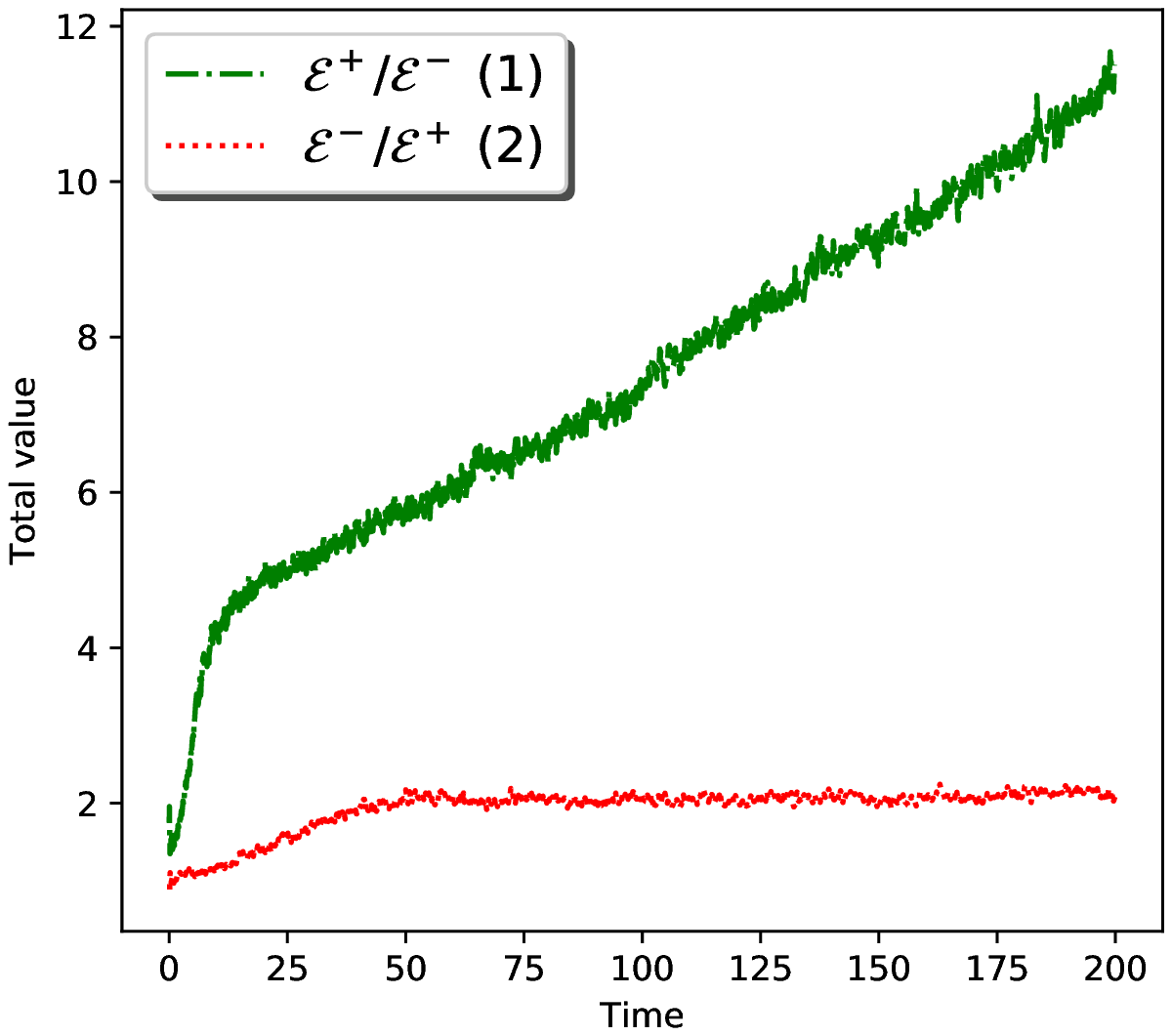} %
    \end{subfigure}
    \caption{Left: $E^\pm(k_\perp)$ spectra for run $R_{13b}^2$ resulting from a balanced driving. Right: time evolution of ${\cal E}^-/{\cal E}^+$ for run $R_{13b}^2$ (red line)  and of ${\cal E}^+/{\cal E}^-$ for run  $R_{13}^2$ (green line). The balanced state is unstable and evolves to a regime where ${\cal E}^-/{\cal E}^+ \simeq 2$.}
    \label{fig:R14b}
\end{figure}

The case  where the $\mu^+$ and $\mu^-$ waves are driven in a balanced way, i.e. with a zero GCH injection rate, is illustrated with run $R_{13b}^2$, for which $\chi_f=0.26$, in Fig. \ref{fig:R14b}. The left panel shows the energy spectra $E^\pm$ of the two counter-propagating waves.  We observe that the balanced regime is unstable. 
The spectra of the counter-propagating waves, which identify at early times, separate and their difference increases with time.   While positive GCH is transferred towards small scales and dissipated, negative GCH is transferred towards large scales, creating a finite global imbalance. As seen on the right panel, after some time, the global imbalance  of run $R_{13b}^2$, as measured by the ratio ${\mathcal E}^+/{\mathcal E}^-$ of the energy of the counter-propagating  waves, saturates to a value close to $0.5$, in contrast with the simulation with imbalanced driving for which this ratio keeps increasing with time. Interestingly, when the energy injection rate $\epsilon_E$ is increased by a factor 256 (run $R_{13sb}^2$), leading to $\chi_f= 1.68$, balanced turbulence becomes stable. Another example where the balanced regime is unstable is provided by $R_{36b}^2$ for which $\chi_f= 0.16$. In this case, we first observe a decay instability before the amplitudes of the counter propagating waves separate (not shown).

When for the same $\beta_e$, and the same energy injection rate, the wavenumber $k_f$ is reduced to $1.3$ (run $R_{1.3b}^2$), thus increasing the value of $\chi_f$ to $0.85$, the instability is suppressed. 
It is recovered when reducing $\beta_e$ to $0.2$ (run $R_{1.3b}^{0.2}$), which has the effect of reducing $\chi_f$  to $0.38$.  When, keeping $\beta_e=0.2$ and $k_f=1.3$, the energy injection rate is increased by a factor $16$ (run $R_{1.3sb}^{0.2}$), $\chi_f = 1.68$ and the instability is again suppressed.  A more quantitative study would be required to determine a precise instability threshold and to confirm the prediction that $\chi_f$ is the only governing parameter that governs the apparition of the instability of the balanced state.

\section{Inverse cascade near the weakly dispersive range} \label{smallkf}

\begin{figure}
    \begin{subfigure}{.5\textwidth}
      \centering
      \includegraphics[width=1\linewidth]
      {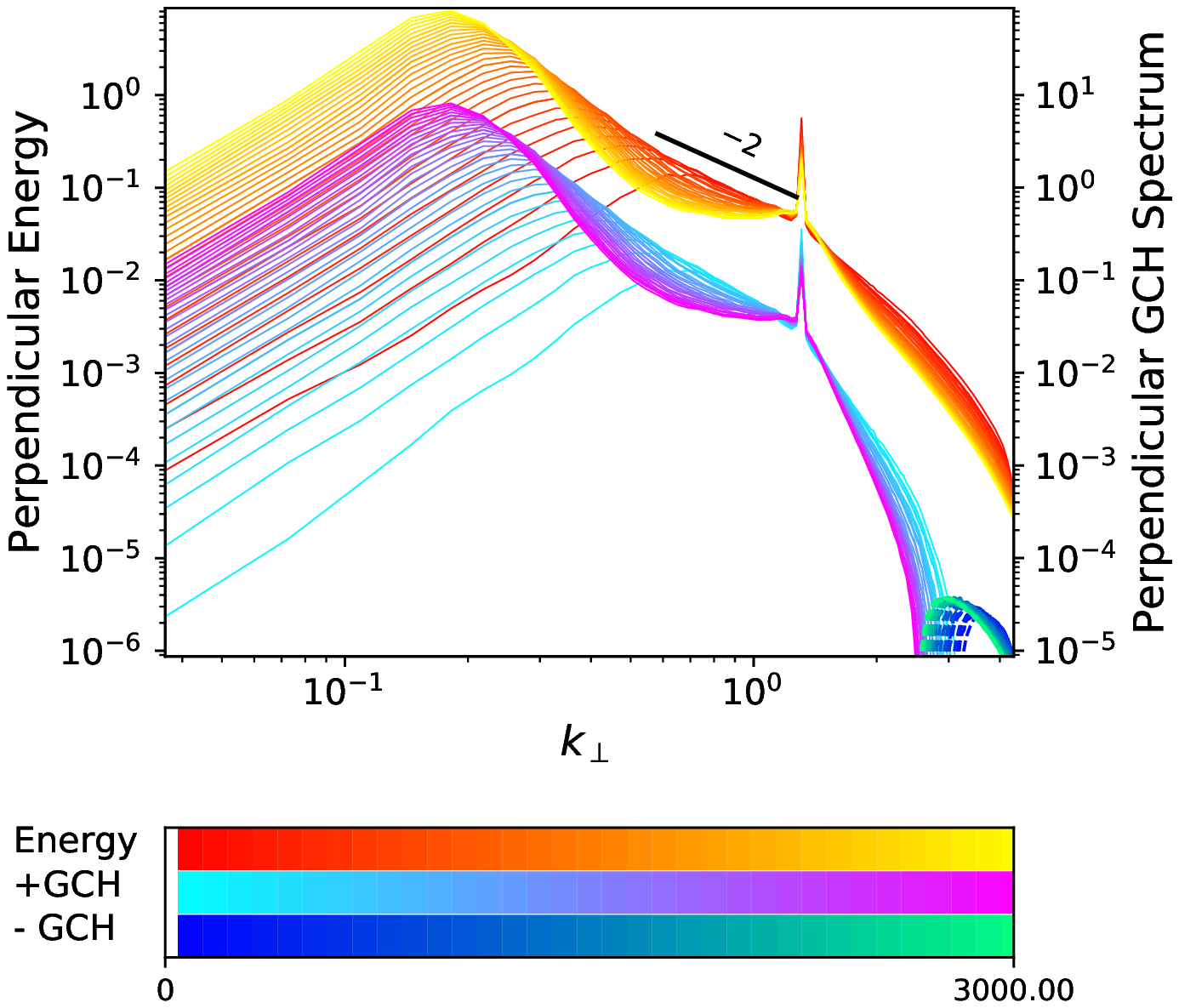} %
      \label{fig:R1PerpSpec}
    \end{subfigure}
    \begin{subfigure}{.5\textwidth}
      \centering
      \includegraphics[width=1\linewidth]
      {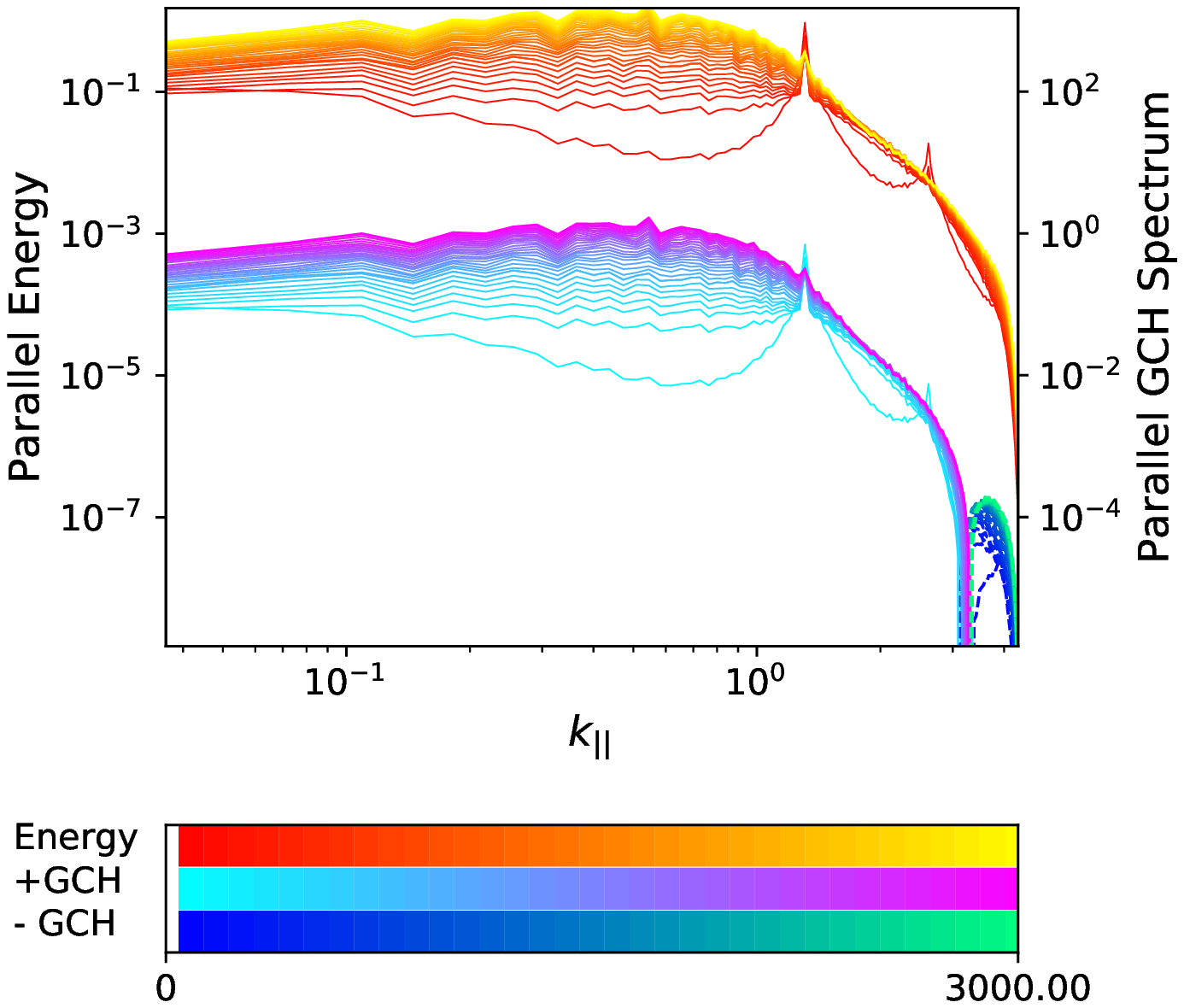} %
      \label{fig:R1ParlSpec}
    \end{subfigure}
    \caption{
    Perpendicular (left) and parallel (right) energy and GCH spectra for run $R^{2}_{1.3}$.}
    \label{fig:R1}
\end{figure}

The goal of this section is to investigate the behavior of the inverse cascade as the energy bump reaches the weakly dispersive range, i.e. for $k_\perp <1$. We thus keep $\beta_e=2$ and the same value of $\epsilon_E$ as in run $R_{13}^2$ but the driving is now centered at $k_f = 1.3$  (run $R_{1.3}^2$).
As a result, the value of $\chi_f^t$ is larger. The  top panels of Fig. \ref{fig:R1} respectively display the  transverse spectra $E(k_\perp)$ and $E_C(k_\perp)$  (left) and the parallel spectra $E(k_\|)$ and $E_C(k_\|)$ (right) at increasing times.  Inspection of these spectra also indicates the existence of an inverse cascade. In contrast with run $R_{13}^2$,  at early time  both  $E(k_\perp)$ and $E_C(k_\perp)$ developa  self-similar range displaying a $k_\perp^{-2}$ power-law. The similarity between the  $E(k_\|)$ and $E_C(k_\|)$ results from the very weak variation of the parallel phase velocity $v_{ph}$  in this spectral range for $\beta_e=2$ (see Fig. \ref{fig:phasevelocities}). This self-similar dynamics does not however proceed  to longer times. After the cascade reaches scales where the variation of $v_{ph}(k_\perp)$ becomes very small, the transfer to larger scales slows down significantly while the amplitude of the spectra still increases under the effect of the persistent driving. As a result, the spectra tend to bend,  flattening  close to the driving wavenumber and developing a spectral bump at the minimal excited wavenumber. 
The slowing down is evident from the fact that the position of the spectral maximum follows two different power laws in time, behaving like $t^{-1/3}$ in the dispersive range and like $t^{-1/2}$ in the the weakly dispersive range. On the contrary, for the  (clear) bump of run $R_{13}^2$ the dependence in the dispersive range is of the form  $t^{-2/3}$. Longer integrations are needed to check whether this behaviour is persistent.

Interestingly, this  cascade depletion is the result of the inner turbulence dynamics, due to a strong decrease of the dispersion, with no need for an external effect such as a hypo-viscosity, as required for example in simulations of the inverse magnetic helicity cascade in incompressible isotropic MHD \citep{Linkmann17} or in EMHD \citep{KimCho15}. In contrast,
in the parallel direction, no inverse cascade is  observed. Both the energy  and GCH spectra become flat, corresponding to an absolute equilibrium regime where the energy is uniformly distributed among the $k_\|$ wavenumbers.

The existence of an inverse cascade of GCH in the transverse spectral plane is also conspicuous when considering the transverse GCH flux (not shown) which displays a flat range between the injection wavenumber and the maximum of the perpendicular spectra. Compared to $R_{13}^2$, in $R_{1.3}^2$ the transverse energy flux is also flat in this spectral range, but it is significantly stronger at scales smaller than the driving, suggesting a split cascade with most of the energy transferred to the small scales.




\section{Impediment to the transverse cascade and formation of a finite-scale condensate} \label{arrest}

\begin{figure}
     \begin{subfigure}{.48\textwidth}
        \centering
        \includegraphics[width=1\textwidth]
        {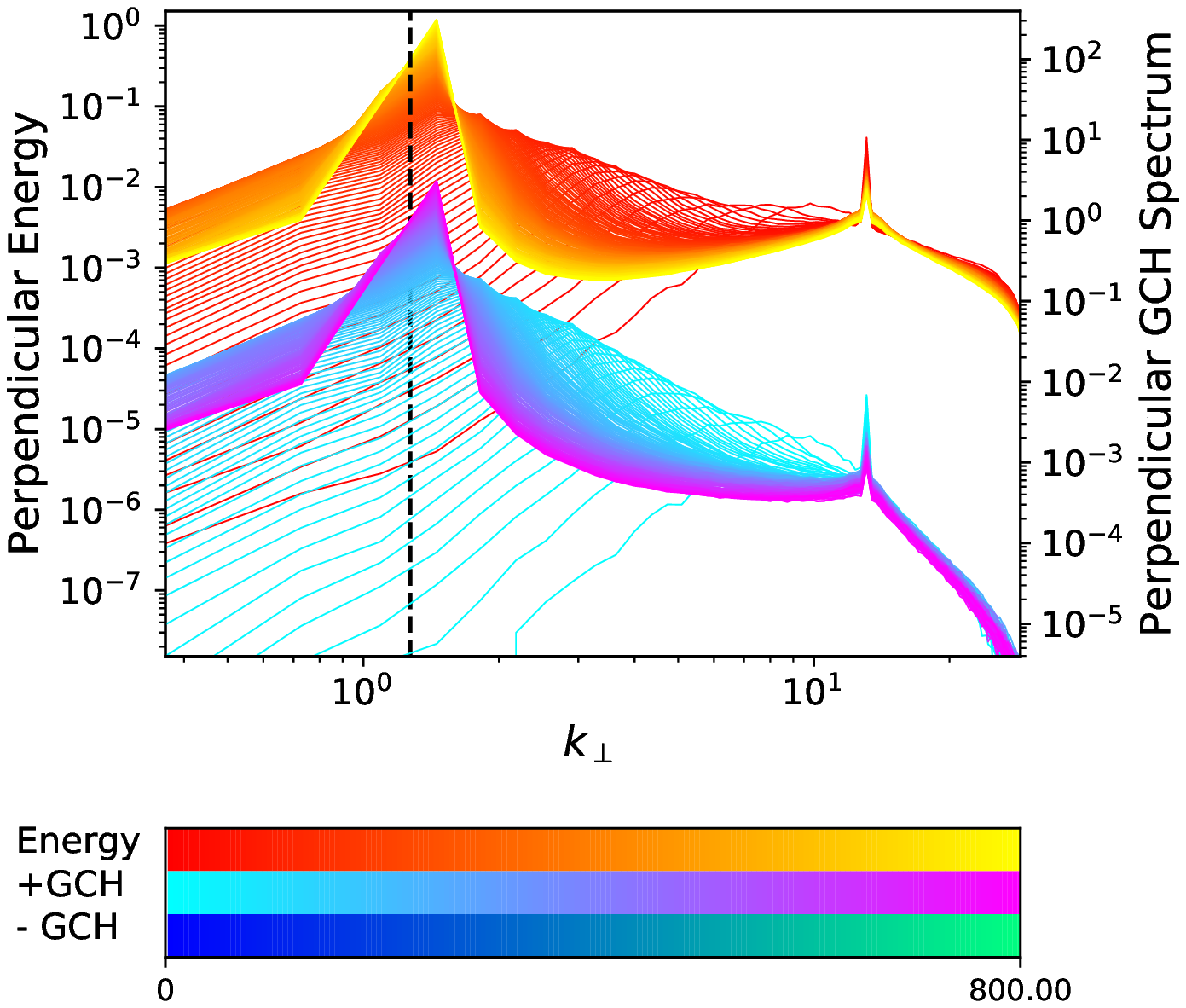} %
    \end{subfigure}
    \begin{subfigure}{.48\textwidth}
      \centering
      \includegraphics[width=1\linewidth]
      {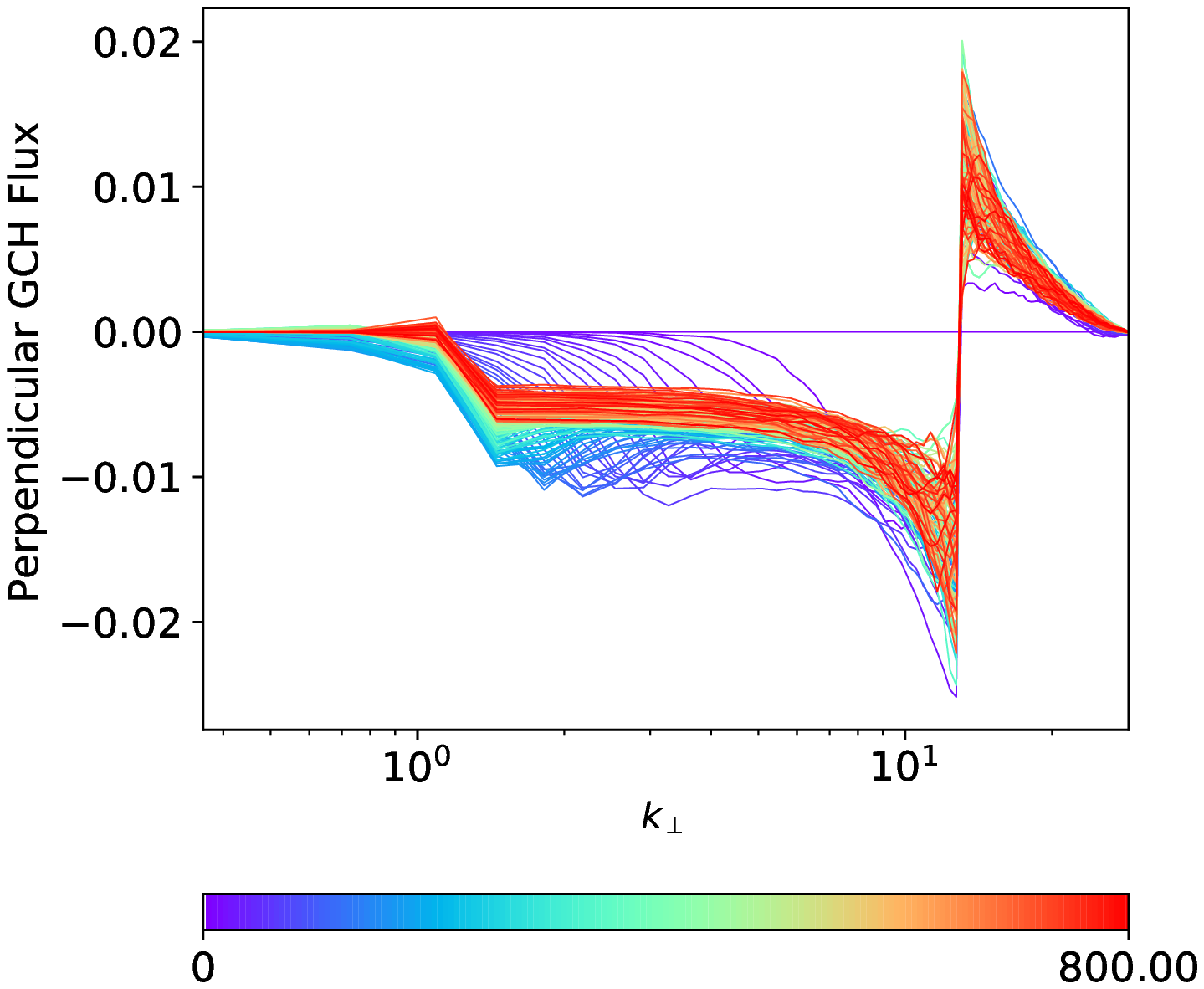} %
    \end{subfigure}
     \begin{subfigure}{.48\textwidth}
        \centering
        \includegraphics[width=1\textwidth]
        {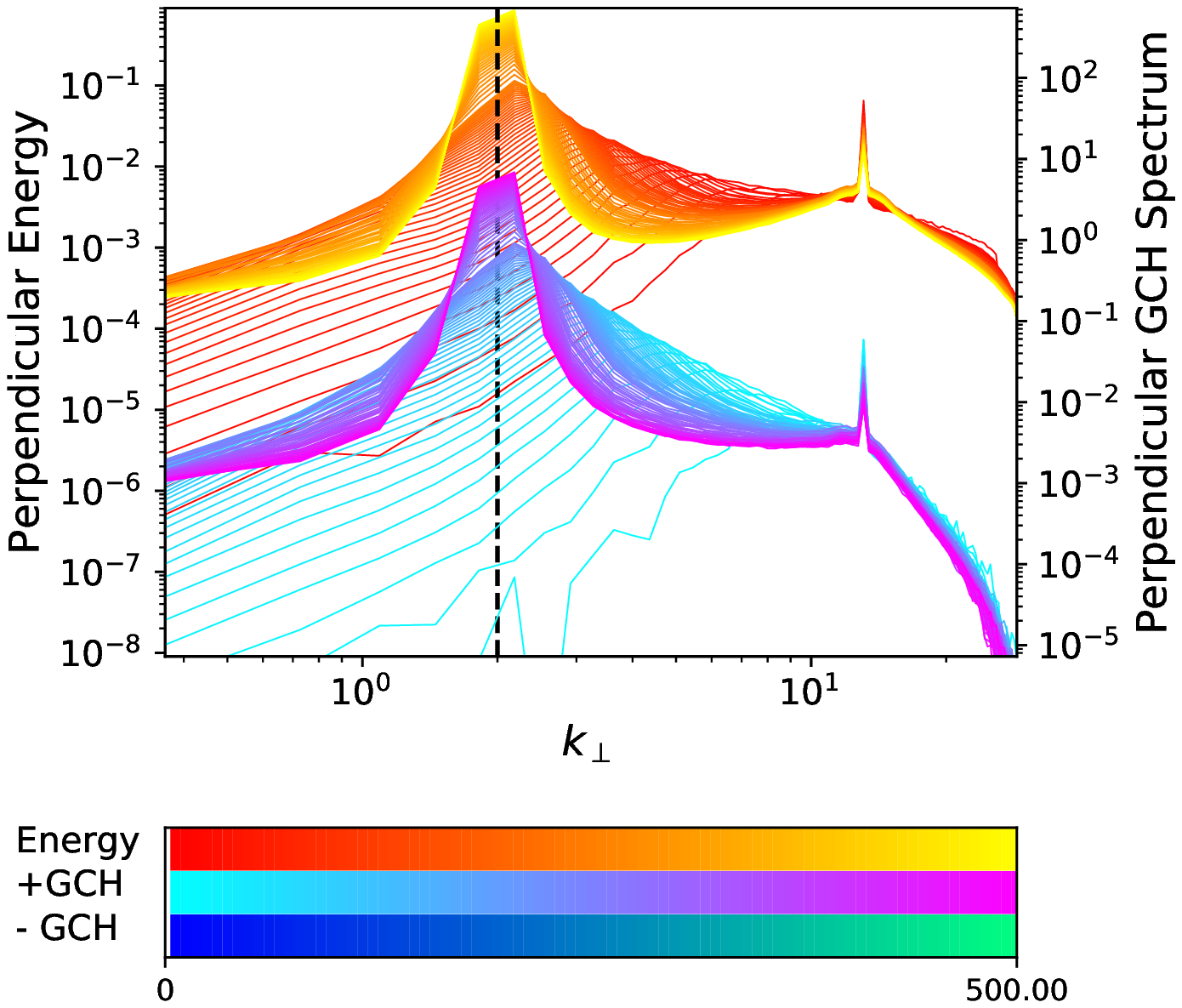} %
    \end{subfigure}
    \begin{subfigure}{.48\textwidth}
      \centering
      \includegraphics[width=1\linewidth]
      {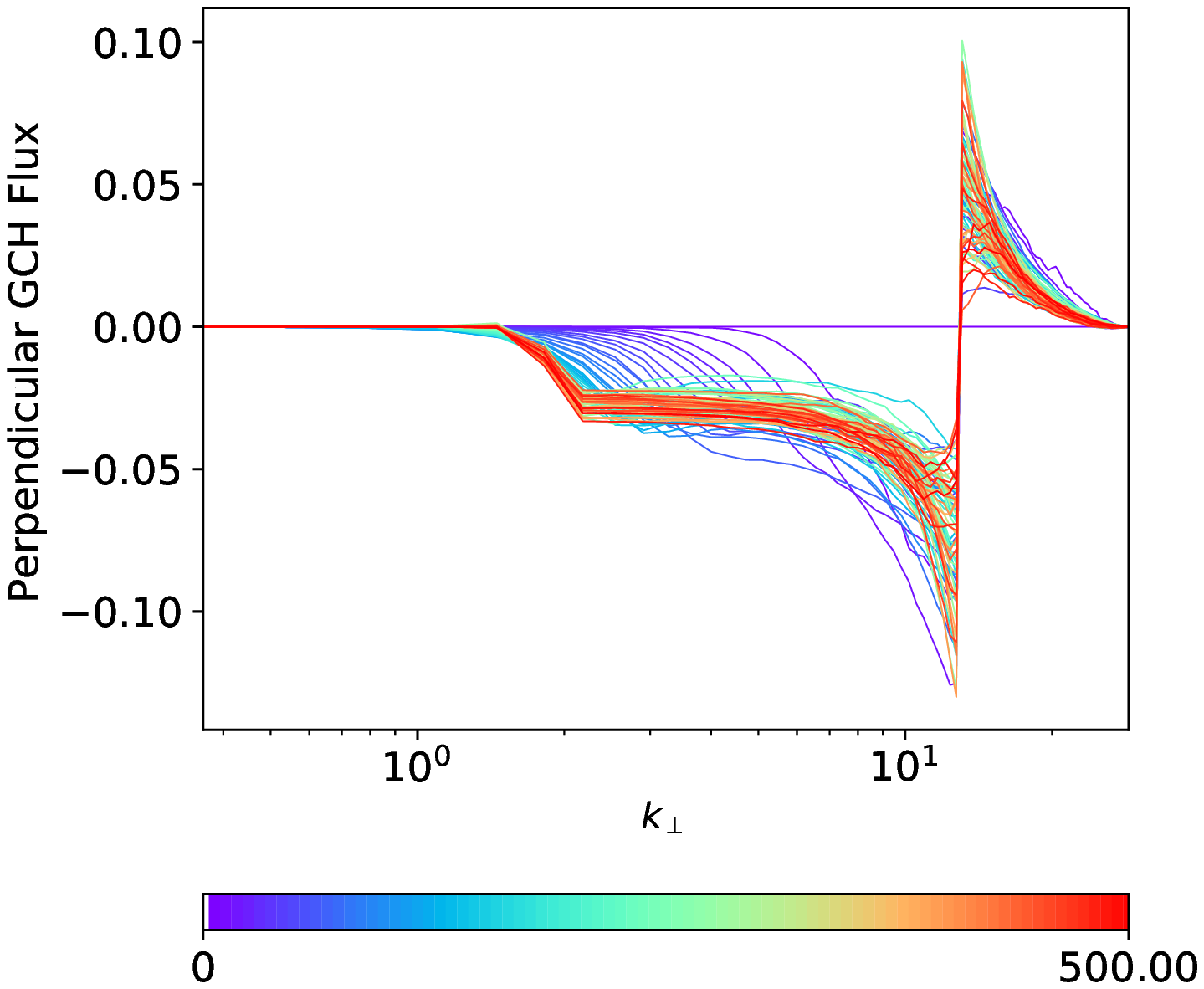} %
    \end{subfigure}
     \caption{Energy and GCH spectra (left) and perpendicular GCH flux (right) for Run $R^{10}_{13}$ (top) and $R^{50}_{13}$ (bottom).  In the former simulation, the maximum of the spectral bump is located at $k_\perp =1.3$ and in the latter at $k_\perp = 2$, corresponding to the respective local minima of the parallel-phase velocity $v_{ph}$ for the respective $\beta_e$, displayed in Fig.  \ref{fig:phasevelocities}.
     }
     \label{fig:R15-and-5}
\end{figure}
As mentioned previously, the shape of the dispersion curve has an important effect on the cascade dynamics. 
In order to confirm that the spectral location of this condensate corresponds to the minimum of the parallel phase velocity, we show 
in Fig. \ref{fig:R15-and-5} two simulations with a driving at $k_f = 13$, where $\beta_e=10$ (top) and $\beta_e=50$ (bottom).
In the case where $\beta_e=10$, the cascade clearly stops at a perpendicular wavenumber slightly larger than unity, where the energy and GCH accumulate at late times. 
  
A similar behavior is observed in the left bottom panel for $\beta_e=50$, but in this case the peak is centered at a slightly larger wavenumber, at a value close to $k_\perp=2$. It turns out that the wavenumbers where the cascade is arrested precisely correspond to the minimum of the parallel phase velocity, as displayed in Fig. \ref{fig:phasevelocities}, and only depend on the parameter $\beta_e$. It is remarkable that such a small variation in the curvature of this quantity (e.g. between the cases $\beta_e=2$ and $\beta_e=10$) can lead to such an important dynamical effect.
 
In both cases, the arrest of the cascade leads to the development of a flat GCH spectrum, together with an energy spectrum growing like $k_\perp$, at scales slightly larger than the injection scales. Such an effect was also observed in simulations of the inverse cascade of magnetic helicity in MHD after the cascade has been arrested by the effect of  an hypo-diffusivity \citep{Linkmann17}. These spectra correspond to those predicted by absolute equilibrium arguments by \citet{Linkmann16} (up to the geometrical factors associated with space dimension). In the case of the present model, the GCH spectrum $E_C(k_\perp)$ and the energy spectrum $E(k_\perp)$, are related, in the strongly imbalanced regime, by $E_C(k_\perp)=E(k_\perp)/v_{\rm ph}(k_\perp)$, which leads us to predict that they will have a similar slope for scales larger than the ion-scale but that 
$E_C(k_\perp)$ will be steeper than $E(k_\perp)$ by a factor of $k_\perp$ at sub-ion scales. Energy equipartition between the modes corresponds to transverse and parallel energy spectra scaling like $k_\perp$ and $k_\|^0$ respectively.

\begin{figure}
    \begin{subfigure}{.51\textwidth}
        \centering
        \includegraphics[width=1\textwidth]
        {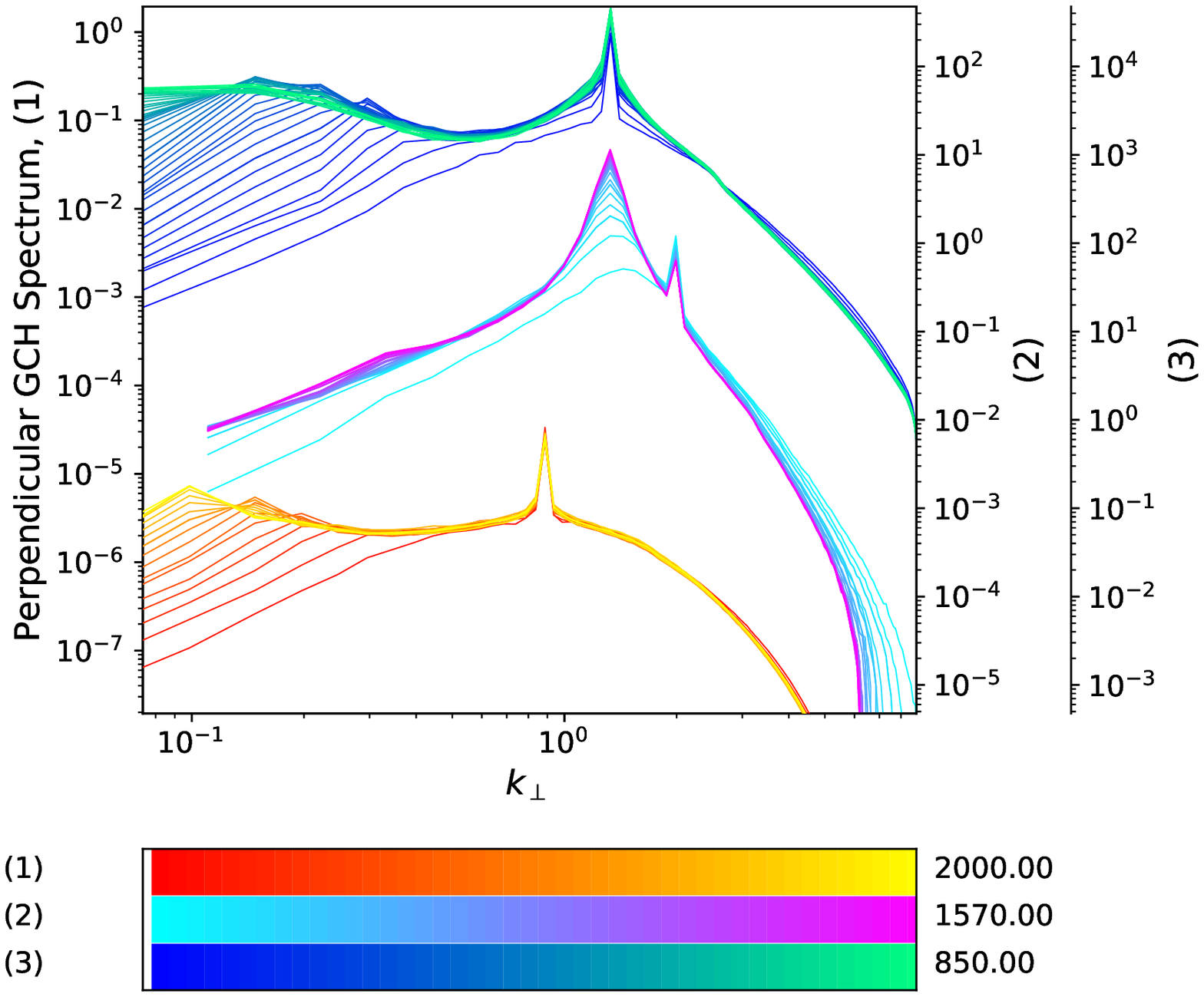} 
    \end{subfigure}
   \begin{subfigure}{.51\textwidth}
        \centering
        \includegraphics[width=1\textwidth]
        {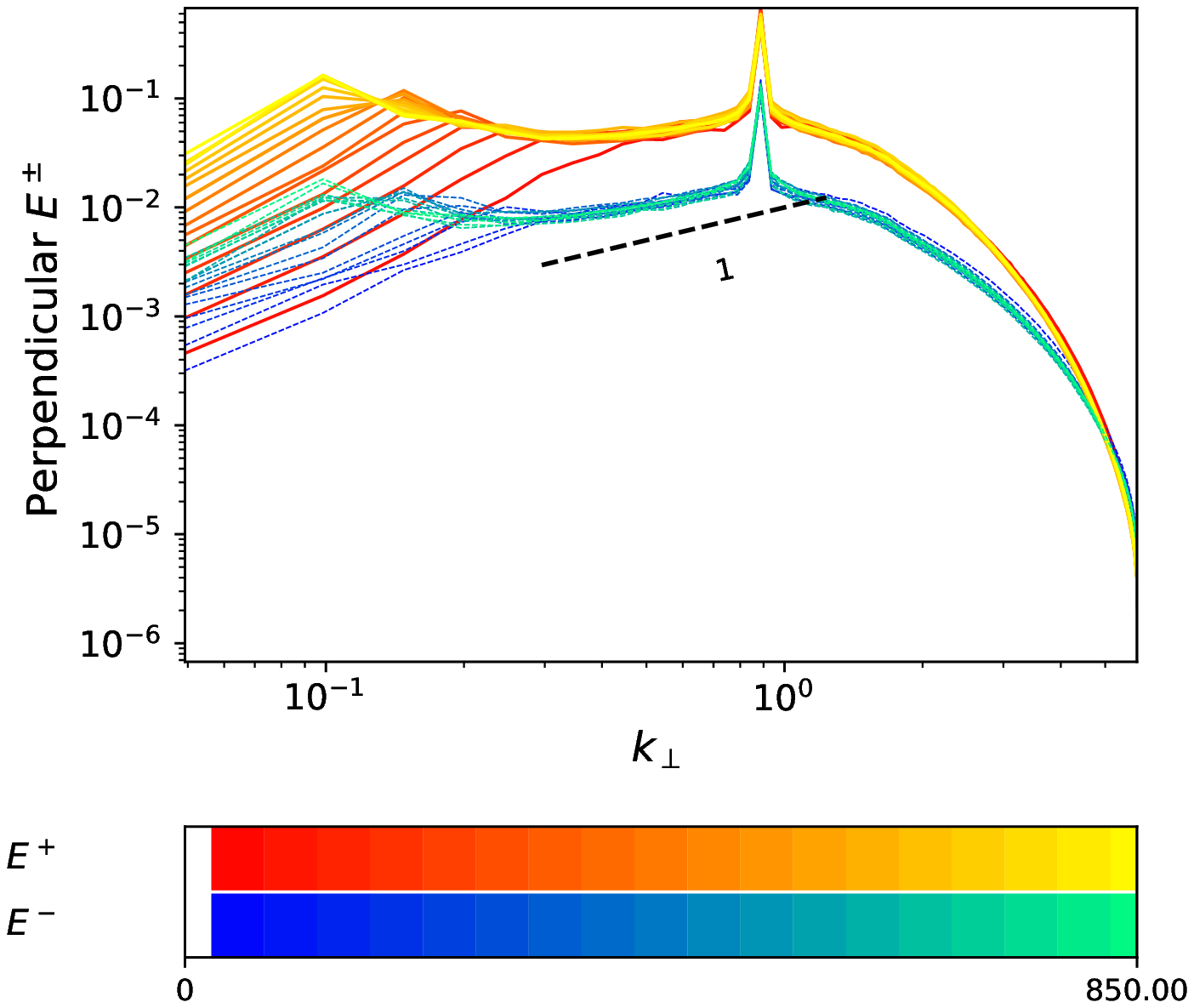} 
    \end{subfigure}
    \caption{Left: perpendicular GCH spectra  for runs $R_{0.89}^{10}$ (1), $R_{2}^{10}$ (2) and $R_{1.3}^{10}$ (3). Right: $E^\pm(k_\perp)$ spectra for run $R_{0.89}^{10}$ (1).}
    \label{fig:R17}
\end{figure}

In order to highlight the sensitivity of the nonlinear dynamics to the shape of the phase velocity, we performed simulations,  keeping $\beta_e =10$,  while the injection wavenumber $k_f$ was taken slightly larger than, equal to or slightly smaller than the wavenumber $k_m= 1.3$ where $v_{ph}$  has a minimum.  The left panel of Fig. \ref{fig:R17} displays the transverse GCH spectra corresponding to these three runs $R_{2}^{10}$, $R_{1.3}^{10}$ and $R_{0.89}^{10}$. For $k_f =2$, thus larger than $k_m$, the cascade is arrested at this latter wavenumber. A zone of negative GCH transverse flux develops in the small spectral range between the driving and the spectral peak, which only forms on the dominant wave (not shown). In contrast, for $k_f = 1.3$ or $0.89$ (thus $k_f \le k_m$), a weak non self-similar inverse transfer of energy and GCH develops at large scales, with a GCH flux  that is essentially zero (not shown).  A tendency for analogous behavior can be inferred from the GCH spectrum of  run $R_{1.3}^{10}$ (cf. the small knee in the spectrum at $k_\perp\approx 0.3$), but the dynamics is much slower than for the two other simulations. Additional information is provided by the Elsasser spectra $E^\pm$ of run $R_{0.89}^{10}$ displayed in the right panel. Transfer to the large scales in run $R_{0.89}^{10}$ takes place in a similar way for  both waves, in contrast with  the simulations  presented in previous  sections in which only the dominant wave cascades.
This large-scale part of the cascade, which is associated with a very small negative flux (not shown) is thus of a different nature and much slower, possibly due to the negative curvature of the parallel phase velocity in this range\footnote{In the context of weak turbulence, resonant three-wave interactions are impossible for negative dispersion (i.e. negative curvature) \citep{Zakharov92} and four-wave processes then have to be taken into account.
} This point deserves a more detailed investigation in future.

\begin{figure}
    \centering
    
    \includegraphics[width=.7\textwidth]{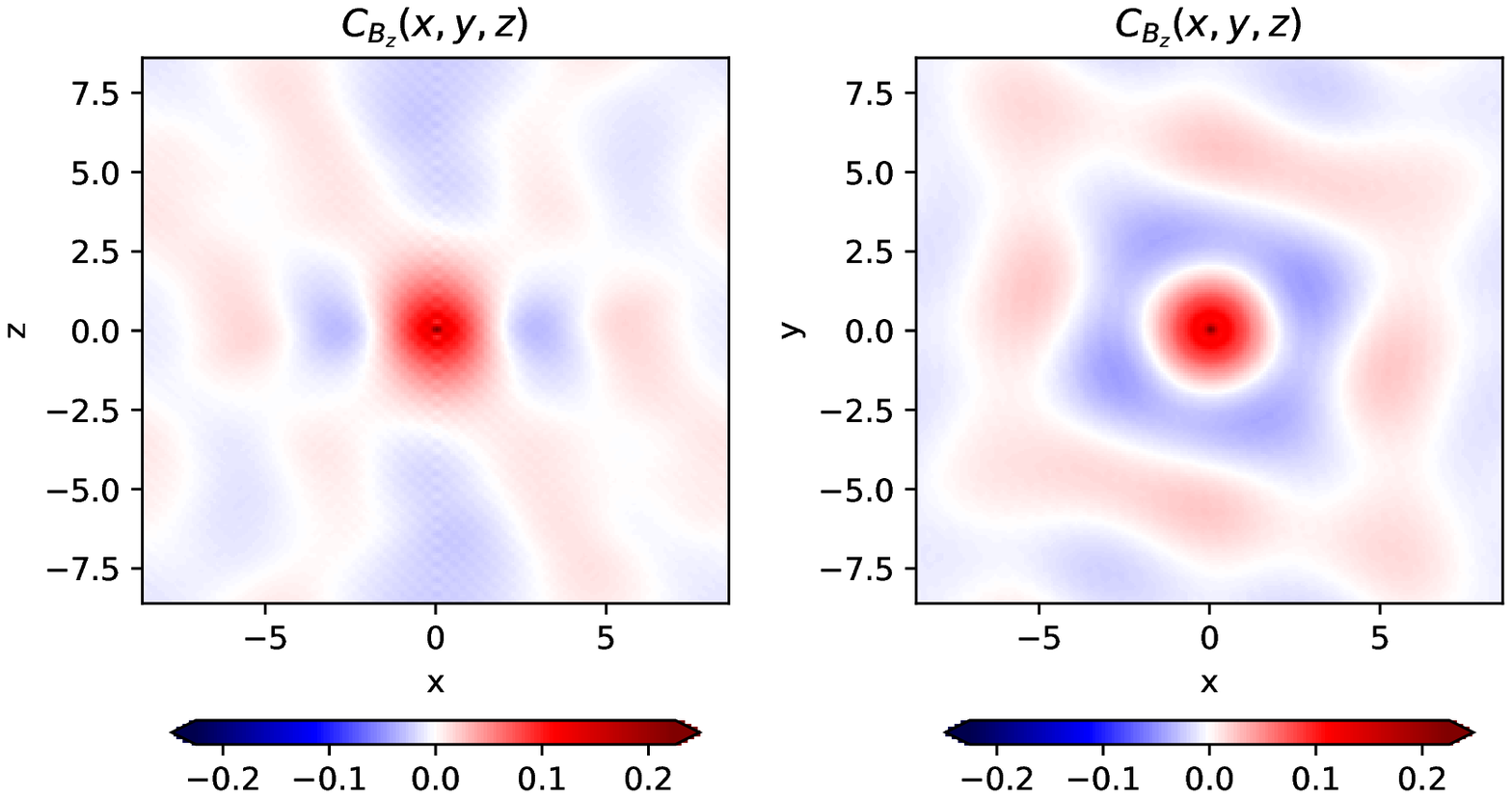}
    \includegraphics[width=1\textwidth]{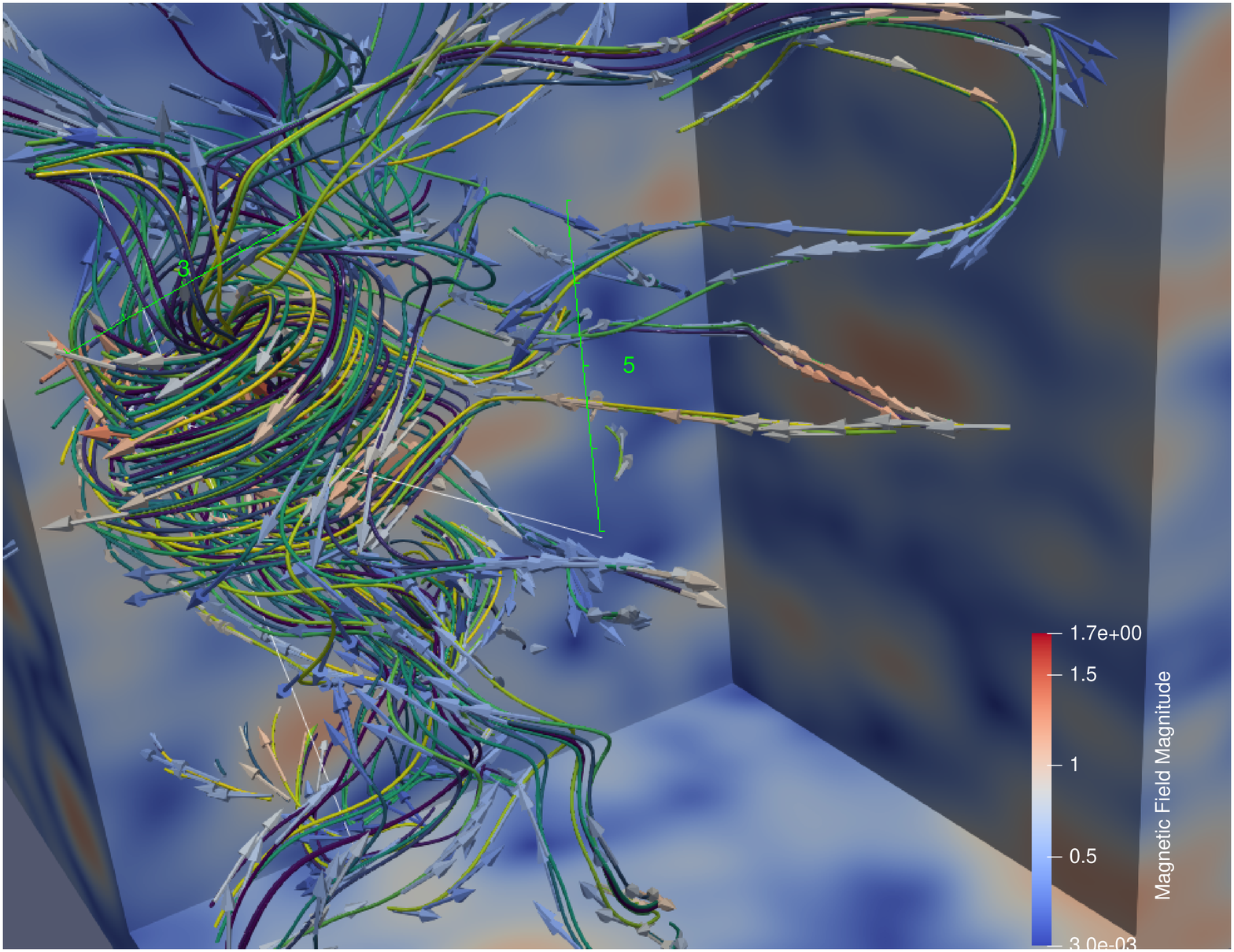}\llap{\raisebox{7.5cm}{\includegraphics[width=0.35\textwidth]{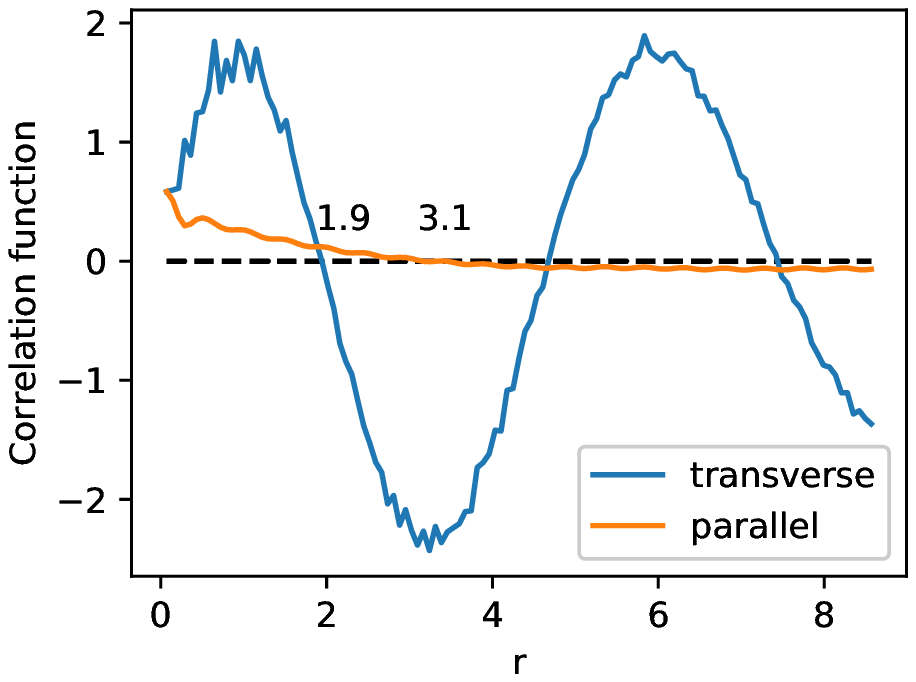}}}\llap{\raisebox{4cm}{\includegraphics[width=0.35\textwidth]{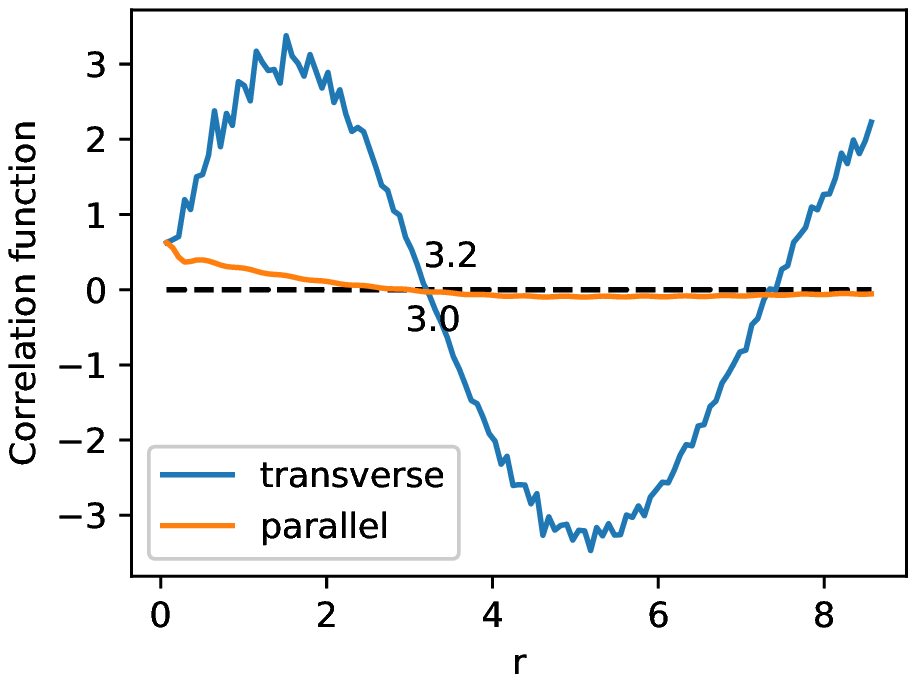}}}
    \caption{Top panels: The parallel and perpendicular cross-sections containing the origin of the correlation function defined in Eq.~\eqref{correlationfun}. Panels below on the right: radial (blue line) and longitudinal (orange line) dependence of the correlation function integrated over the azimuthal angle and restricted to $z = 0$ and $|{\bf r}_\perp| = 0$ respectively. Bottom panel: Magnetic-perturbation field lines for run  $R_{13}^2$  at an intermediate time of the simulation, showing the formation of an elongated vortex. The field has been filtered to remove the imprint of the small scales such as forcing and dissipation. Field lines are differentiated by colors. Arrows indicate the direction of the field. Color map corresponds to the magnitude of the magnetic field both on the arrows and on the horizontal and the vertical cuts. The typical size of the 3D structures is indicated using green rulers. }
    \label{fig:R14Spatial}
\end{figure}

\begin{figure}
    \centering
    \includegraphics[width=0.7\textwidth]{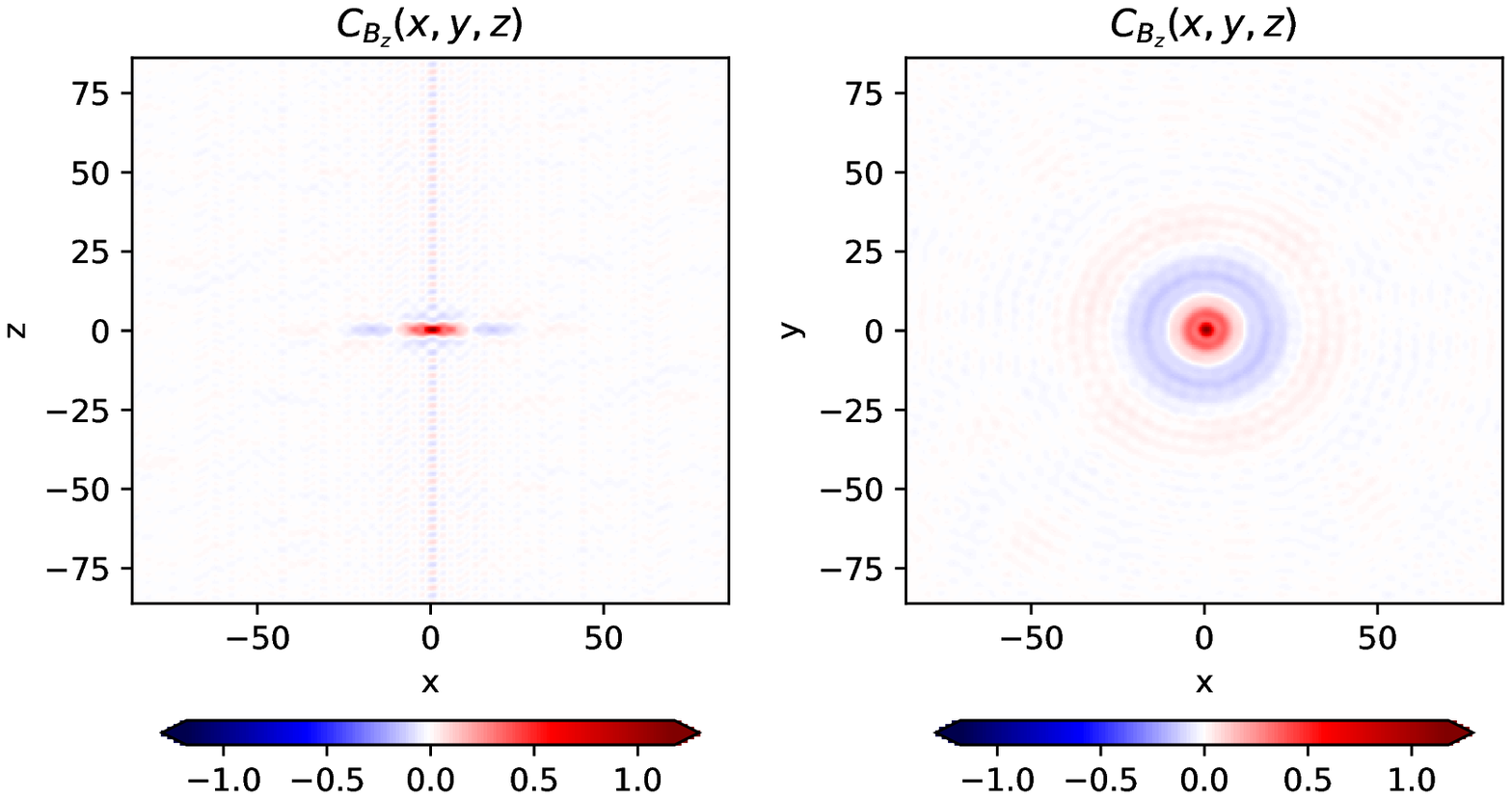}
    \includegraphics[width=1\textwidth]{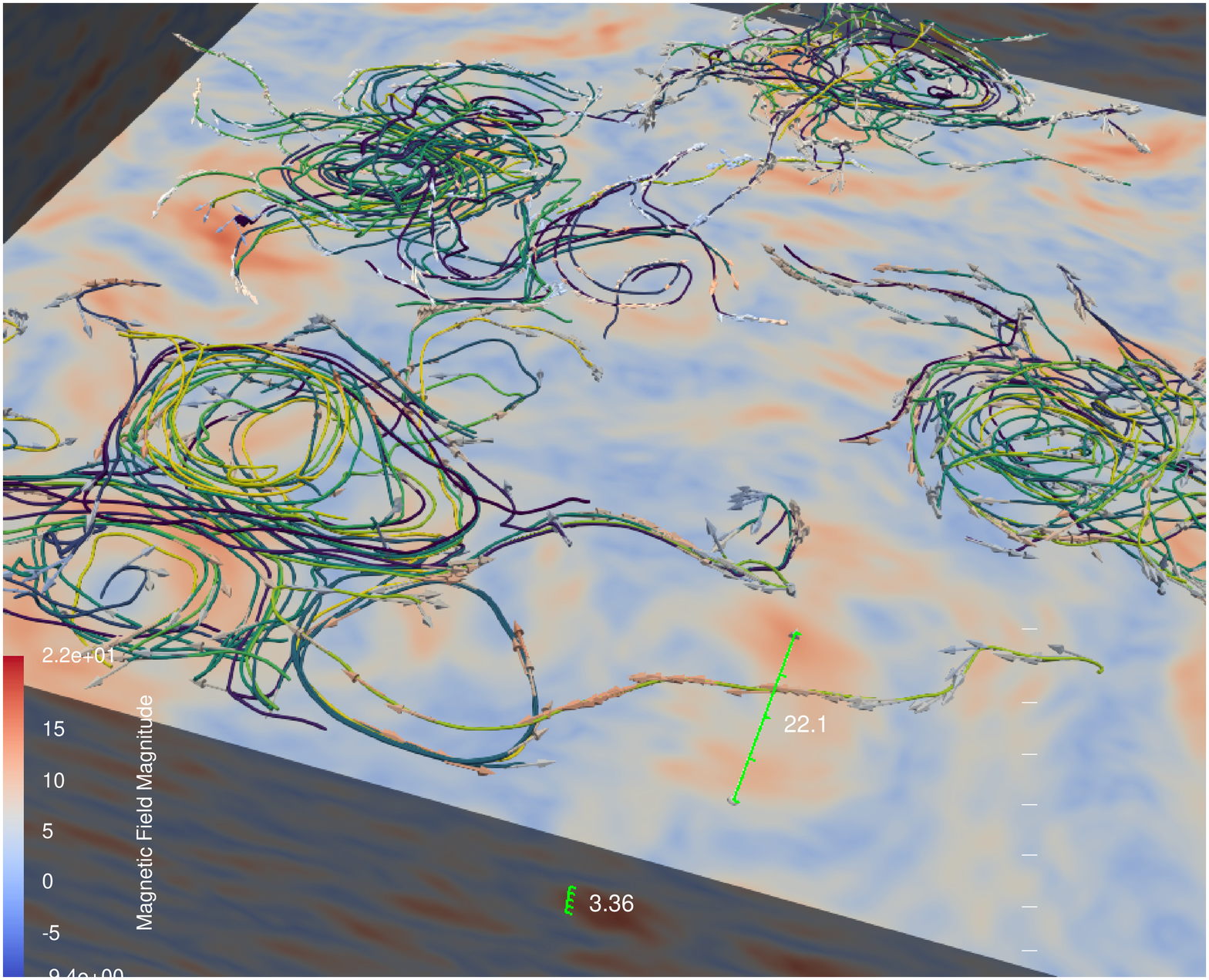}\llap{\raisebox{6cm}{\includegraphics[width=0.55\textwidth]{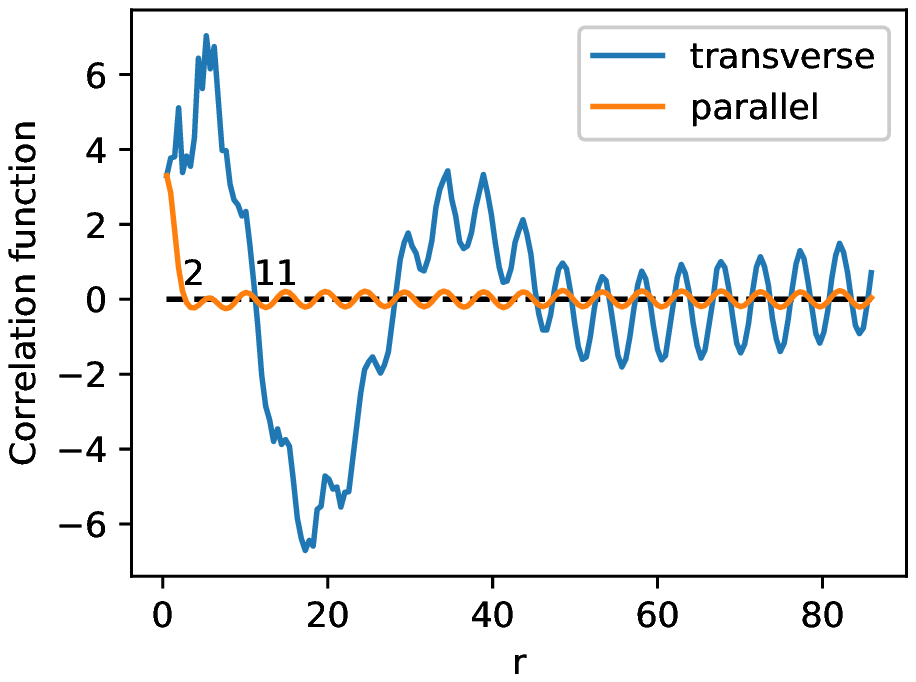}}}
    \caption{Top panels: vertical (left) and horizontal (right) cuts through correlation functions. Panel below on the right: radial (blue line) and longitudinal (orange line) dependence of the correlation function integrated over the azimuthal angle and restricted to $z = 0$ and $|{\bf r}_\perp| = 0$ respectively. Bottom panel: Magnetic-fluctuation field lines  for Run $R_{1.3}^2$, showing a layer of looping filaments at a time corresponding to the late stage of the simulation. Field lines are differentiated by colors. Arrows indicate the direction of the field. Color map corresponds to the magnitude of the magnetic field both on the arrows and on the horizontal and the vertical cuts. The typical size of the 3D structures is indicated using green rulers. }
    \label{fig:R1magfield3d}
\end{figure}

\section{Structures in physical space} \label{vortices}

Another point, important in the context of the formation of coherent structures, concerns the existence of a parallel inverse cascade.
When comparing GCH parallel spectra of the fiducial run displayed in Fig. \ref{fig:R14SpecandFlux} with those of Fig.~\ref{fig:R1}, it becomes evident that there is a transition from a split cascade with a significant inverse parallel component for run $R^2_{13}$ at large $k_f$, to a predominantly forward cascade in run $R^2_{1.3}$ at smaller $k_f$. In the latter case, the beating of  $\pm k_z$ modes can feed the zero mode in the parallel direction and further interactions can then drive the intermediate modes to an absolute equilibrium. We checked that a parallel inverse cascade only develops when $k_f$ is large enough. A precise determination of the condition for the existence of a parallel cascade would require a large number of simulations and is outside the scope of this paper.

These qualitative spectral features have interesting consequences in physical space. As in two-dimensional  hydrodynamics, we can expect the formation of large-scale vortices, but the situation is more complex due to the three-dimensional nature of the dynamics and due to the different types of behavior in the parallel direction. To give a general overview, the transverse magnetic field appears  to form various inter-connected vortices, with a typical diameter that can be appreciated from the wavenumber corresponding to the maximum in the energy or helicity spectrum. Alternatively one may estimate the size of these structures based, for example, on the behavior of the $B_z$-correlation function, defined as
\begin{equation}\label{correlationfun}
    C_{B_z} ({\bf r}):= \int d^3 x\; B_z ({\bf x} + {\bf r})B_z ({\bf x} ).
\end{equation}

As the perpendicular inverse cascade proceeds, the structures, which originate from initial fluctuations at a scale comparable to the forcing,   grow in time.  For $\beta_e > 3.5$, this process continues until  the finite-$k_\perp$ condensate is reached. For smaller $\beta_e$, no condensate forms but the dynamics slows down significantly when approaching the  MHD scales. Therefore several large-scale vortices may exist simultaneously in a large enough domain, even at late times. As a representative case, consider the fiducial run that develops the vortex presented in Fig~\ref{fig:R14Spatial} (bottom panel). The field line filaments displayed in this figure are associated with the magnetic fluctuations ${\boldsymbol B} =\nabla_\perp\times (A_\|\widehat{{\boldsymbol z}}) + B_z \widehat{\boldsymbol z}$. 
We observe a vortex whose diameter can be roughly estimated as 3 to 4 length units, while the vertical extension are  5 to 6 units. The size estimates correspond to the domain of increased field intensity, as displayed by the color of the arrows. The parallel and perpendicular cross-sections containing the origin of the correlation function defined in Eq.~\eqref{correlationfun} are displayed in the left and right top panels respectively. In the inserted panels, we have plotted the radial (blue line) and longitudinal (orange line) dependence of the correlation function integrated over the azimuthal angle and restricted to $z=0$ and $|{\boldsymbol r}_\perp| =0$ respectively. The top right inserted panel, which corresponds to the time at which the vortex is displayed, has two maxima. The typical transverse (parallel) size of the vortices corresponds to the distance at which the radial (longitudinal)  correlation functions reach zero and the distance (in the transverse direction)  between vortices  of the same sign  is given by the distance between the two maxima. The bottom right panel shows the same plots at a later time when the  cascades have reached the largest scales of the domain.  From these figures, we can conclude that the typical vortices that are formed have generally elongated features, with an aspect ratio of order unity (in the rescaled variables of the model). Note that when transforming the rescaled coordinates to the physical ones, the structures are to be stretched along this direction by a quantity given by the inverse expansion parameter, for example the degree of spectral anisotropy.

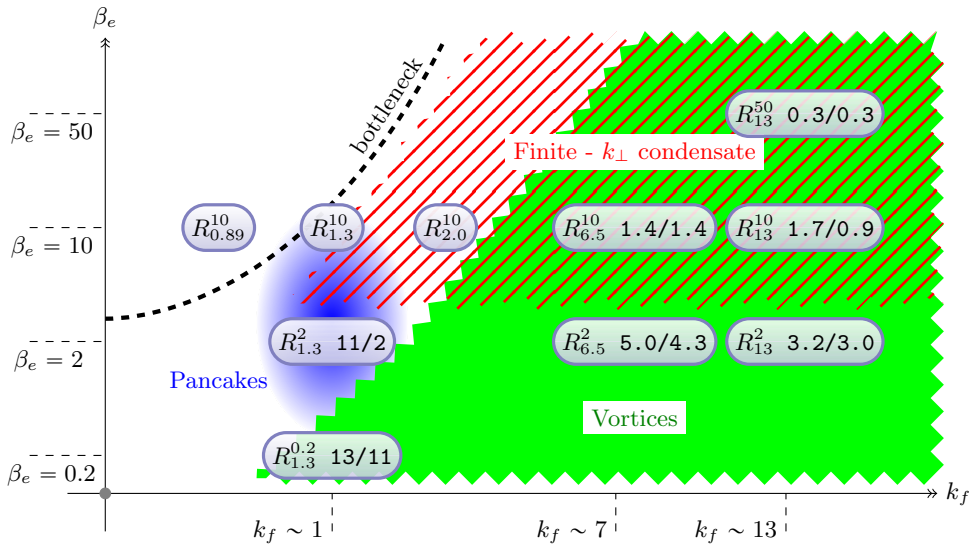
\begin{figure}
\centering
    \begin{tikzpicture}[node distance=5mm,
    nonterminal/.style={
rectangle,
minimum size=6mm,
very thick,
draw=red!50!black!50, 
top color=white, 
bottom color=red!50!black!20, 
font=\itshape
},
terminal/.style={
rectangle,minimum size=6mm,rounded corners=3mm,
very thick,draw=blue!50!black!50,
top color=white,bottom color=blue!50!black!20,
font=\ttfamily}
]
        \draw[->>] (-.5,0) -- (11,0) node[right] {$k_f$}; 
        \draw[->>] (0,-.5) -- (0,6) node[above] {$\beta_e$}; 
        \draw[dashed] (3.0,-0.5)  node[anchor = east] {$k_f \sim 1$} --  (3.0,0) node[anchor = east] {}; 
        \draw[dashed] (6.75,-0.5)  node[anchor = east] {$k_f \sim 7$} --  (6.75,0) node[anchor = east] {}; 
        \draw[dashed] (9,-0.5)  node[anchor = east] {$k_f \sim 13$} --  (9,0) node[anchor = east] {}; 
        \draw[dashed] (-1,.5) -- node[sloped,below, xshift=-.3cm] {$\;\;\beta_e = 0.2$} (0,.5); 
        \draw[dashed] (-1,2) -- node[sloped,below, xshift=-.25cm] {$\beta_e = 2$} (0,2); 
        \draw[dashed] (-1,3.5) -- node[sloped,below, xshift=-.25cm] {$\;\beta_e = 10$} (0,3.5); 
        \draw[dashed] (-1,5) -- node[sloped,below, xshift=-.25cm] {$\;\beta_e = 50$} (0,5); 
        \filldraw [gray] (0,0) circle (2pt); 
        \filldraw[inner color=blue,outer color=white,draw=white] (3,2.25) ellipse (1 and 1.5);
        \node at (1.5,1.5) {\textcolor{blue}{Pancakes}};
        \fill[color=green, decorate, decoration=zigzag]  (2,0.2)  -- (4,2)   -- (7,6) -- (11,6) --(11,0.2) -- cycle;
        \node[fill=white] at (7,1) {\textcolor{green!50!black}{Vortices}};
        \fill[color=red, pattern=north east hatch, hatch distance=3mm, hatch thickness=1pt, pattern color = red, decorate, decoration=zigzag]  (2.5,2.5)  -- (5,6) -- (11,6) --(11,2.5) -- cycle;
        \node[fill=white] at (7,4.5) {\textcolor{red}{Finite - $k_\perp$  condensate}};
        \draw[dashed,ultra thick](0,2.3) .. controls(1,2.3) and (3,3)  .. node[very near end,sloped,above] {bottleneck} (4.5,6) ;
        \node (a) [terminal] at (1.5,3.5) {$R^{10}_{0.89}$};
        \node (b) [terminal] at (3,3.5) {$R^{10}_{1.3}$};
        \node (c) [terminal] at (4.5,3.5) {$R^{10}_{2.0}$};
        \node (d) [terminal] at (3,2) {$R^{2}_{1.3}$ 11/2};
        \node (e) [terminal] at (3,.5) {$R^{0.2}_{1.3}$ 13/11};
        \node (f) [terminal] at (7,2) {$R^{2}_{6.5}$ 5.0/4.3};
        \node (g) [terminal] at (7,3.5) {$R^{10}_{6.5}$ 1.4/1.4};
        \node (h) [terminal] at (9.25,2) {$R^{2}_{13}$ 3.2/3.0};
        \node (i) [terminal] at (9.25,3.5) {$R^{10}_{13}$ 1.7/0.9};
        \node (j) [terminal] at (9.25,5) {$R^{50}_{13}$ 0.3/0.3};
        
    \end{tikzpicture}
    \caption{Phase space diagram  in the $(k_f,\beta_e)$-plane, where the locations of the various types of cascades and structures in physical space are indicated as shaded areas. The relevant runs are also added in the diagram, together with the bottleneck locus associated with the values of $k_\perp$ where the parallel phase velocity has a minimum. The two numbers on the right of the run label provide the estimated correlation lengths in the transverse/parallel format. }
    \label{fig:runs}
\end{figure}

The parallel inverse cascade affects the parallel correlation length of the fluctuating magnetic field of vortices. In particular, in run $R_{1.3}^2$, no inverse cascade takes place in the parallel direction and, consequently, the structures in physical space are found to be flattened in horizontal layers, as shown in Fig. \ref{fig:R1magfield3d}.

Figure \ref{fig:runs}  provides a schematic  graphical summary of the performed simulations, together with the phenomenology observed in different regimes, in a phase diagram whose axes correspond to the driving wavenumber $k_f$ and the electron beta parameter $\beta_e$. The two additional numbers following the  run labels correspond to the transverse and longitudinal correlation lengths determined using the approach described above. We can see from the figure that there are two main intersecting areas at sub-ion scales: the green part of the phase space, which corresponds to finite aspect ratio vortices, and the red part  which corresponds to the runs with $\beta_e > 3.5$ that, due to the existence of a local minimum  in the  phase velocity curve,  produce a finite-$k_\perp$ condensate. The ``bottleneck'' curve is the locus of points on the diagram for which $v_{ph}$ displays such a minimum.
 
 For moderate values of $\beta_e$, a transition is also observed close to the ion scale between situations without a parallel inverse cascade (at smaller values of $k_f$), where pancake structures are observed, and those at larger $k_f$ where a clear inverse cascade develops in the parallel direction, associated with more elongated  vortices. As we cross the bottleneck curve we encounter a critical behavior described in Section~\ref{arrest}, associated with runs at $k_f\sim 1$. Beyond, for scales much larger than the ion sonic Larmor radius, the dynamics is dominated by  the forward cascade of GCH,  that we do not address in this paper. Below the red region, we have runs that do not produce finite-$k_\perp$ condensate, since the minimum in phase velocity disappears. 
It must be born in mind that only a selected choice of runs is available for drawing general conclusions, and therefore, this diagram has to be understood as a qualitative approximation.

\section{Conclusion} \label{conclu}

Direct numerical simulations of a two-field gyrofluid model for  KAW turbulence randomly driven at the sub-ion scales have been presented. A main observation is the development
of an inverse cascade of GCH in the transverse direction and, to a lesser degree, of energy. This cascade is essentially nonlocal and some of its properties are sensitive to the nonlinearity parameter $\chi_f$ (ratio of the period of the waves to the characteristic nonlinear time at the driving scale), and to the shape of the dispersion relation. When $\chi_f$ is large enough, a self-similar cascade develops at early time while, for smaller values of $\chi_f$, a spectral bump is observed to propagate towards small wavenumbers. In the latter case, transient parametric decay instability events are also possible, which enhance the development of the cascade. At later times, we observe a similar evolution in all the simulations, where the spectrum gets distorted, with the establishment of an absolute equilibrium range extending from the forcing wavenumber to larger scales.
When the cascade reaches non-dispersive scales, it slows down while the maxima of the energy and GCH spectra still increase.
For the moderate values of the GCH injection rate we use throughout the paper, one type of wave is usually dominant, while the other develops an equilibrium spectrum. When the injection wavenumber is large or $\beta_e$  small enough (i.e. when $\chi_f$ is small), an inverse GCH cascade is even possible  when the driving is balanced.  
This result suggests that  fast magnetic reconnection events could not only drive small-scale turbulence \citep{CerriNJP17}, but also induce a cascade to larger scales, a dynamics somewhat similar to the island merging proposed in \citet{Franci17}. 
When $\beta_e \gtrapprox 3.5$, the parallel-phase velocity $v_{ph} =\omega/k_\|$ displays a minimum at a perpendicular wavenumber close to the ion scale. In this case, the inverse GCH cascade is arrested, leading to the formation of finite-wavenumber condensates.
In the parallel direction, depending on the parameters, an inverse cascade resulting mostly from local interactions can also develop.

Another  result concerns the capability of the transverse GCH inverse cascade to lead to the formation of coherent vortices with a transverse size comparable to the ion scale. Depending on the plasma parameters, their longitudinal correlation length can be enhanced when an inverse cascade develops in the parallel direction. 
Typically, the structures that form have an aspect ratio of order unity in the anisotropic coordinates used in the model. They will be more elongated in the physical coordinates by a quantity given by the inverse expansion parameter, associated with the amplitude of the fluctuations. These structures are possibly related to those   observed in the solar wind \citep{Perrone17} or in the terrestrial magnetosheath \citep{Wang19}, which in some instances are associated by \citet{Alexandrova08} to the $k_\perp^{-4}$ transition range.
The simulations we have discussed above provide a natural way to form  vortices, and to demonstrate  their dynamical stability.
Futher development would be needed to analyze the process of their formation, i.e. whether it proceeds in a conventional two-dimensional hydrodynamics sense of vortex merger/nucleation, and if one can observe instances of reconnection, as discussed  by \citet{zhou20} in the context of RMHD. Another issue is the analytical characterization of these vortices and the comparison with the Alfv\'en vortices found analytically in \citet{Petviashvili} at MHD scales or in \citet{Jovanovic20} (and references therein) at ion scales and large values of the beta parameter.

The present study can be extended to include the coupling to slow magnetosonic waves, using a four-field gyrofluid model \citep{Tassi20}.  
This description appears especially relevant  in the regions near the Sun, explored by Parker Solar Probe and Solar Orbiter,  where large-scale compressibility effects are important. Such a model allows for the development of the  Alfv\'en wave parametric decay instability, an effect that is viewed as an efficient process for generating counter-propagating waves, requested for the existence of turbulence at the MHD scales \citep{Vinas91,DelZanna01,Shoda18}. 

Another avenue could involve an detailed theoretical understanding of the impediment of the inverse cascade and the formation of a finite-$k_\perp$ condensate, observed in the simulations in the presence of a  small depression in the KAW parallel-phase velocity.

\section*{Acknowledgments}
This work was granted access to the HPC resources of CINES/IDRIS under the allocation A0060407042. Part of the computations have also been done on the "Mesocentre SIGAMM" machine, hosted by Observatoire de la C\^ote d'Azur. 
We are thankful to D. Borgogno for his contribution to the initial version of the numerical code.

\begin{appendix}
	
\section{Decay instability}  \label{App:decay}

Using the same notation for the operators and their Fourier symbols, we define the Fourier components $c_{\boldsymbol k}^{\sigma_k} =
D_e(k_\perp) \mu_{\boldsymbol k}^{\sigma_k}$ of the fields $D \mu^{\pm}$ (where $\sigma_k= \pm$) and,  in the interaction representation,
\begin{equation}
a_{\boldsymbol k}^{\sigma_k} = e^{i\omega_k^{\sigma_k} t} c_{\boldsymbol k} \label{mode-ak-sigma},
\end{equation}
with $\omega_{\boldsymbol k}^{\sigma_k} = \sigma_k v_{ph}(k_\perp) k_z$.  The gyrofluid equations rewrite
\begin{equation}
\partial_t a_{\boldsymbol k}^{\sigma_k}  - \int \sum_{\sigma_p, \sigma_q} e^{i \Omega_{{\boldsymbol k};{\boldsymbol p} {\boldsymbol q}}^{\sigma_k \sigma_p \sigma_q}t}
V_{{\boldsymbol k}{\boldsymbol p}{\boldsymbol q}}^{\sigma_k \sigma_p \sigma_q} a^{\sigma_p}_{\boldsymbol p}a^{\sigma_q}_{\boldsymbol q} \delta({\boldsymbol p}+{\boldsymbol q} - {\boldsymbol k} )
d{\boldsymbol p} d{\boldsymbol q}=0, \label{eq:ak}
\end{equation}
where we define 
\begin{equation}
\Omega_{{\boldsymbol k};{\boldsymbol p} {\boldsymbol q}}^{\sigma_k \sigma_p \sigma_q} = \omega_{\boldsymbol k}^{\sigma_k} - \omega_{\boldsymbol p}^{\sigma_p} - \omega_{\boldsymbol q}^{\sigma_q}= \sigma_k v_{ph}(k_\perp) k_\| - \sigma_p v_{ph}(p_\perp) p_\| - \sigma_q v_{ph}(q_\perp) q_\|.  \label{Omega}.
\end{equation}
When neglecting electron inertia,
\begin{equation}
V_{{\boldsymbol k}{\boldsymbol p}{\boldsymbol q}}^{\sigma_k \sigma_p \sigma_q}= \frac{1}{8}  \frac{\sigma_p\sigma_q({\widehat {\boldsymbol z}}\bcdot({\boldsymbol p}\times {\boldsymbol q}))}{k_\perp p_\perp q_\perp}
\left ( \frac{\sigma_p}{\xi(p_\perp)} - 
\frac{\sigma_q}{\xi(q_\perp)}\right )\left ( \sigma_k k_\perp^2 \xi(k_\perp) 
+ \sigma_p p_\perp^2 \xi(p_\perp) + \sigma_q q_\perp^2 \xi(q_\perp) \right )
\end{equation}
with $\xi =s/v_{ph}$,
consistent with the vertex given in Eq. (6.4) of \citet{Voitenko98a}, directly derived from the Vlasov-Maxwell equations.

At this point it is easy to estimate the growth rate of the decay parametric instability of a KAW into two other KAWs.
For this purpose, one considers a pump of type $\sigma_k$, with wavenumber ${\boldsymbol k}$, frequency $\omega_{\boldsymbol k}$ and complex amplitude $a_{\boldsymbol k}^{\sigma_k}= D_e \left (\Lambda \varphi(\boldsymbol k, t)+ \sigma_k A_\|(\boldsymbol k,t)\right )$ (see~Eq. (\eqref{mupmequation}). It can interact with KAWs of wavevectors 
${\boldsymbol p}$ and ${\boldsymbol q}$ that satisfy the resonance conditions 
\begin{eqnarray}
&&{\boldsymbol k}={\boldsymbol p}+ {\boldsymbol q}\\
&&\omega_{\boldsymbol k}= \omega_{\boldsymbol p}+ \omega_{\boldsymbol q},
\end{eqnarray}
in the way that 
\begin{eqnarray}
&&\partial_t a_{\boldsymbol p}^{\sigma_p} = V_{{\boldsymbol p}\,{\boldsymbol k}\,{-{\boldsymbol q}}}^{\sigma_p \sigma_k \sigma_q} a_{\boldsymbol k}^{\sigma_k}
{a_{\boldsymbol q}^{\sigma_q}}^*\\
&&\partial_t{a_{\boldsymbol q}^{\sigma_q}}^* = V_{-{\boldsymbol q}\,-{\boldsymbol k}\,{\boldsymbol p}}^{\sigma_q \sigma_k \sigma_p} {a_{\boldsymbol k}^{\sigma_k}}^*
a_{\boldsymbol p}^{\sigma_p}.
\end{eqnarray}
This results in a growth rate $\gamma$ for the modes $a_{\boldsymbol p}^{\sigma_p}$ and $a_{\boldsymbol q}^{\sigma_q}$ given by 
\begin{eqnarray}
\gamma^2 &=& V_{{\boldsymbol p}\,{\boldsymbol k}\,{-{\boldsymbol q}}}^{\sigma_p \sigma_k \sigma_q}V_{-{\boldsymbol q}\,-{\boldsymbol k}\,{\boldsymbol p}}^{\sigma_p \sigma_k \sigma_q} |a_{\boldsymbol k}^{\sigma_k}|^2 \\
&=& \frac{1}{64} \frac{({\widehat {\boldsymbol z}}\bcdot({\boldsymbol p}\times {\boldsymbol q}))^2}{\xi(p_\perp)\xi(q_\perp)}\frac{1}{k_\perp^2 p_\perp^2 q_\perp^2}
\left ( \frac{\sigma_k}{\xi(q_\perp)}-\frac{\sigma_p}{\xi(k_\perp)}\right)
\left(\frac{\sigma_p}{\xi(k_\perp)}-\frac{\sigma_k}{\xi(p_\perp)}\right)
\nonumber\\
&& \times \left (\sigma_k k_\perp^2 \xi(k_\perp) + \sigma_p p_\perp^2 \xi(p_\perp) + \sigma_q q_\perp^2 \xi(q_\perp)\right )^2 |a_{\boldsymbol k}^{\sigma_k}|^2, 
\end{eqnarray}
where the eigenmode $a_{\boldsymbol k}^{\sigma_k}$ satisfies  $|a_{\boldsymbol k}^{\sigma_k}|^2 = (8/\beta_e)|B_\perp(\boldsymbol k)|^2$ \citep{PS19}.
Instability thus requires
\begin{equation}
\left ( \frac{\sigma_k}{\xi(q_\perp)}-\frac{\sigma_p}{\xi(k_\perp)}\right)
\left(\frac{\sigma_p}{\xi(k_\perp)}-\frac{\sigma_k}{\xi(p_\perp)}\right)>0.
\end{equation}
Equation (7.5) of \citet{Voitenko98a} (see also \citet{ZhaoJGR10}) is reproduced when noting 
that in the latter equation the length unit is not $\rho_s$ but $\rho_i$ defined as $\sqrt{\tau}\rho_s$ and that $\xi$, defined as the ratio of the Alfv\'en to the KAW frequency, is independent of the length unit.
In the framework of the present paper, the growth rate of the parametric decay instability scales, in the ERMHD regime, as $\gamma\sim k_\perp^2 \beta_e^{-1/2} |B_k|$ when $\beta_e\lesssim 1$ and $\gamma\sim k_\perp^2 \beta_e^{-1}|B_k|$ when $\beta_e\gg 1$ (recall that for $\beta_e\gg 1$, $\xi(k_\perp) =s/v_{ph}\sim \frac{1}{sk_\perp}$).
Note that, if we use $d_i$ as length unit, thus writing $k_\perp = ({\overline k}_\perp d_i)/s$, where ${\overline k}_\perp$  denotes the corresponding dimensional wavenumber, the growth rate $\gamma$ does not scale with $\beta_e$ in the large $\beta_e$ limit, as it is the case in EMHD for whistler waves \citep{ZhaoApJ10}.

\section{Energy and GCH shell-to-shell transfers}\label{App:transfers}
Useful information on the cascade dynamics is provided by analyzing the energy and GCH shell-to-shell transfers in Fourier space, when proceeding as in \cite{Alexakis07} and~\cite{mininni07}. As in the calculation of the transverse and parallel energy and GCH spectra,  we perform a partition of the Fourier space either  in cylindrical shells $S_{K_\perp}$ 
including all the wavevectors whose transverse component obeys $K_\perp<|\boldsymbol{k}_\perp| \leqslant K_\perp+ dk_x$, or in slabs $S_{K_\|}$ including all the wavevectors whose parallel component $k_z$ obeys $K_z<|k_z| \leqslant K_z+dk_z$, with the goal to study the
perpendicular and parallel transfer respectively. Here, $dk_x$ and $dk_z$ refer to the spectral mesh along the transverse axes and along the parallel axis, respectively.

 Our aim is to estimate the energy  and GCH  transfer rates $T_E(K,P)$ and $T_C(K,P)$), respectively,  between a shell $K$ that receives and a shell $P$ that gives,  such that it is possible to write ${\partial_t E_K = \sum_P T_E(K,P)}$ and ${\partial_t C_K = \sum_P T_C(K,P)}$, where $E_K$  and $C_K$  are the contributions of a cylindrical or of a slab-shaped shell $S_K$ to the energy ${\mathcal E}$ and to the GCH ${\mathcal C}$ respectively (${{\mathcal E} = \sum_K E_K}$ and ${{\mathcal C} = \sum_K C_K}$).
Due its physical meaning, we expect that $T_E(P,K) = - T_E(K,P)$ and $T_{C}(P,K) = - T_{C}(K,P)$  are antisymmetric, since the amount of the invariant that the $K$-shell gives to $P$ equals the amount of the same invariant that the $P$-shell receives. 
Finally,  the total fluxes of energy $\Pi_E(K)$ and GCH $\Pi_C(K)$ through a wavenumber $K$ 
\begin{equation}
\Pi_E(K) =-\sum_{K^{\prime}=0}^{K} \sum_{P=0}^{\infty}  T_E\left(K^{\prime}, P\right), \qquad \Pi_C(K) =-\sum_{K^{\prime}=0}^{K} \sum_{P=0}^{\infty}  T_C\left(K^{\prime}, P\right)  \label{eq:fluxes}
\end{equation}
are obtained from the shell-to shell transfer $T_E(K',P)$  or  $T_C(K',P)$  by the  summing  over $P$ and over $K'\le K$. 

A scalar field $A$ restricted to a shell $S_{K}$ (here and in the following $K$ holds for either $K_\perp$ or $K_\|$) can be represented in terms of its Fourier modes $\hat{A_{\boldsymbol{k}}}$ by
\begin{equation}
A_{K}(\boldsymbol{x})=\sum_{\boldsymbol{k} \in S_{K}} \widehat{A}_{\boldsymbol{k}} e^{i \boldsymbol{k} \cdot \boldsymbol{x}},
\end{equation}
so that 
\begin{equation}
\sum_{K} A_{K}(\boldsymbol{x}) = A(\boldsymbol{x}).
\end{equation}

When considering two scalar fields $A$ and $B$, one has,  
\begin{equation}
	\int d^3 x  A_K(\mathbf{x}) B(\mathbf{x}) = \sum_{\mathbf{k} \in S_{K}^{\pm}}\int d^3 x {\widehat A}_{\mathbf{k}} e^{i \mathbf{k} \cdot \mathbf{x}} B(\mathbf{x}) = \dfrac{1}{(2\pi)^3}  \sum_{\mathbf{k} \in S_{K}^{\pm}}  {\widehat A}_{\mathbf{k}}  {\widehat B}^\star_{\mathbf{k}}
\end{equation}
and 
\begin{equation}
\int d^3 x  A_K(\mathbf{x}) B_K(\mathbf{x}) = \sum_{\mathbf{k} \in S_{K}^{\pm}}\sum_{\mathbf{k}^\prime \in S_{K}^{\pm}}\int d^3 x {\widehat A}_{\mathbf{k}} e^{i (\mathbf{k} + \mathbf{k}^\prime )\cdot \mathbf{x}} {\widehat B}_{\mathbf{k}^\prime}  = \dfrac{1}{(2\pi)^3}  \sum_{\mathbf{k} \in S_{K}^{\pm}}  {\widehat A}_{\mathbf{k}}  {\widehat B}^\star_{\mathbf{k}},
\end{equation}
where again  $K$ holds for either $K_\perp$ or $K_\|$, depending if perpendicular or parallel transfers are considered.
In practice, the latter expression is used to evaluate various integrals arising in the expression of the  transfer $T_{E}(K,P)$ and  $T_C(K,P)$.

For the two-fluid gyrofluid, the two-fluid Lie-dragging formulation~\eqref{Gpmequation} can be conveniently used when  electron inertia is retained. The total GCH reads ${\mathcal C} = \int C d^3x$, where we define the GCH density as
\begin{equation}
C= -\frac{1}{4\delta} \left \{ (G^+)^2 - (G^-)^2 \right\}.
\end{equation}
It follows that 
\begin{eqnarray}
\partial_t C &=& \frac{1}{2\delta}\Big  \{ G^+ [\varphi^+, G^+] - G^-[\varphi^- , G^-]  -\frac{1}{2\delta}\partial_z ((G^+)^2 + (G^-)^2) \nonumber \\
&+&\frac{2}{\delta} L_e A_\| \partial_z A_\| + 2\delta N_e \partial_z (1-M_1)\varphi \Big  \}.
\end{eqnarray}
To simplify the notation, we will use 
$$
\sum_\pm \pm G^\pm[\varphi^\pm, G^\pm]    \equiv G^+ [\varphi^+, G^+] - G^-[\varphi^- , G^-].
$$
In this case, the contribution $C_K= \int_{S_K} C \,d^3x$ of shell $K$ to  GCH ($\displaystyle{{\mathcal C} = \sum_K C_K}$) obeys
\begin{equation}
\dot{C}_{K}=\frac{1}{2 \delta} \int d^{3} x \sum_{P} \sum_{\pm} \pm G_{K}^{\pm}\left[\varphi^{\pm}, G_P^{\pm}\right],\label{dotCdef}
\end{equation}
where the contribution of the terms involving a $z$-derivative cancel out by space integration, when using that $N_e = -M_2\varphi$ and the hermiticity of the operators $M_1$, $M_2$ and $-\Delta_\perp$.
The formulation  given by Eq. (\ref{dotCdef}) is interesting, since it is as though the transfer is achieved via Lie-drag over the flow corresponding to $\varphi^\pm$. 
In terms of primitive fields, the expression evaluates to
\begin{equation}\begin{aligned}
\dot{C}_{K} &=\frac{1}{2 \delta} \int d^{3} x \sum_{P} \left( L_{e} A_\| \right)_K \left[\frac{A_\|}{\delta}, \left(L_{e} A_\| \right)_P\right]-\left( L_{e} A_\| \right)_K\left[\left(M_3-M_{2}\right) \varphi,\right.\\
\left.\delta (M_{2} \varphi)_{P}\right] &-\delta (M_{2} \varphi)_{K}\left[\left(M_{3}-M_{2}\right) \varphi,\left( L_{e} A_\| \right)_P\right]+\delta (M_{2} \varphi)_{K}\left[\frac{A_\|}{\delta}, \delta (M_{2} \varphi)_{P}\right].
\end{aligned}\end{equation}
The first summand, which involves a term $A_{\| K} [A_{\|}, A_{\|P}]/\delta$  has a singular behavior as $\delta\rightarrow 0$, but it does not contribute to the flux and is thus physically irrelevant. We will ignore it
in the estimate of the shell-to-shell transfer. 


Turning to the energy, we write $\displaystyle{{\mathcal E}= \sum_K E_K}$ where 
$E_K$, given by 
\begin{equation}
    E_K= \frac{1}{2} \int \left \{ (M_2 \varphi_K)((1-M_1) \varphi_K) + (M_2\varphi_K)^2- \frac{2}{\beta_e} (L_e A_K) (\Delta_\perp A_K) \right\} \, d^3x,
\end{equation}
is the contribution of shell K to the energy.
The operator arising in the above equation being hermitian, we can write
\begin{equation}
\partial_t E_K=  \int \left \{ ((1-M_1) \varphi_K)\partial_t(M_2 \varphi_K) 
+ (M_2 \varphi_K)\partial_t(M_2\varphi_K)- \frac{2}{\beta_e}  (\Delta_\perp A_K) \partial_t (L_e A_K)\right\} \, d^3x
\end{equation}
or 
\begin{eqnarray}
&&\partial_t E_K=  \int \left \{ ((1-M_1) \varphi_K)\partial_t(M_2 \varphi) 
+ (M_2 \varphi_K)\partial_t(M_2\varphi)- \frac{2}{\beta_e}  (\Delta_\perp A_K) \partial_t (L_e A_)\right\} \, d^3x ,\nonumber \\
&&\partial_t E_K=  \frac{1}{4}\int \Big \{ -(\varphi^+_K + \varphi^-_K) \partial_t \frac{G^+ -G^-}{\delta}  + \frac{G^+_K -G^-_K}{\delta} \partial_t \frac{G^+ -G^-}{\delta} +  \frac{G^+_K +G^-_K}{\delta}\partial_t \frac{G^+ +G^-}{\delta} \nonumber \\ 
&&-(\varphi^+_K -\varphi^-_K)\partial_t \frac{G^+ +G^-}{\delta} 
\Big \} \, d^3x
\end{eqnarray}
where we have replaced $L_e$ by its  definition. Equivalently, after a simple algebraic rearrangement, 
\begin{equation}
\partial_t E_K = \frac{1}{2\delta} \int \left \{\left (\frac{G^+_K}{\delta} -\varphi^+_K \right ) \partial_t G^+ + \left (\frac{G^-_K}{\delta} +\varphi^-_K\right) \partial_t G^-\right\} \, d^3x,
\end{equation}
that we rewrite in the more compact form
\begin{equation}
\partial_t E_K = \frac{1}{2\delta} \int \sum_\pm \left (\frac{G^\pm_K}{\delta} \mp \varphi^\pm_K \right ) \partial_t G^\pm \, d^3x.
\end{equation}
We then use  the model equations  in the form
\begin{equation}
 \partial_t G^\pm + [\varphi^\pm, G^\pm \mp \delta \varphi^\pm ] + \partial_z(\varphi^\pm \mp \frac{1}{\delta} G^\pm  ) =0 
\end{equation}
where we include a additional term in the second argument of the bracket which does not contributes, and get
\begin{equation}
\partial_t{E}_{K}=-\frac{1}{2} \int  \sum_{\pm}\left(\frac{G_{K}^{\pm}}{\delta} \mp \varphi_{K}^{\pm}\right)\left[\varphi^{\pm}, \frac{G^{\pm}}{\delta} \mp \varphi^{\pm}\right] \, d^{3} x.
\end{equation}
This enables defining the shell-to-shell energy transfer $T(K,P)$ such that 
\begin{equation}
    \partial_t E_K = \sum_P T(K,P)
\end{equation}
via
\begin{equation}
    T(K,P) = \frac{1}{2} \int  \sum_{\pm}\left(\frac{G_{K}^{\pm}}{\delta} \mp \varphi_{K}^{\pm}\right)\left[\varphi^{\pm}, \frac{G_P^{\pm}}{\delta} \mp \varphi_P^{\pm}\right] \, d^{3} x,
\end{equation}
which is antisymmetric in $K$ and $P$.

In terms of the primitive variables ($M_3= 1+M_2-M_1)$, 
\begin{eqnarray}
 &&   \left(\frac{G_{K}^{\pm}}{\delta} \mp \varphi_{K}^{\pm}\right)\left[\varphi^{\pm}, \frac{G_P^{\pm}}{\delta} \mp \varphi_P^{\pm}\right] = \nonumber \\
 &&\left (\mp(M_3\varphi)_K - \frac{2\delta}{\beta_e} \Delta_\perp A_K \right)  \Big \{ \left [M_3\varphi \pm \frac{1}{\delta} A_\| , \mp (M_3\varphi)_P - \frac{2\delta}{\beta_e} \Delta_\perp A_P \right ] \nonumber \\
 && -\left [M_2\varphi \pm \frac{1}{\delta} A_\| , \mp (M_3\varphi)_P - \frac{2\delta}{\beta_e} \Delta_\perp A_P \right ] \Big \}
\end{eqnarray}
includes a term $(M_3\varphi)_K [M_3\varphi, (M_3 \varphi)_P]$ that does not contribute to the flux, and that we thus suppress.
As a result, the renormalized transfers we report are calculated according to the formulas
\begin{equation}
T_C(K,P)= \int d^{3} x \left( \frac{1}{\delta}A_{\| K} [A_{\|}, A_{\|P}]  - \frac{1}{2 \delta}\sum_{\pm} \pm G_{K}^{\pm}\left[\varphi^{\pm}, G_P^{\pm}\right] \right)\label{CTransfer}
\end{equation}
and
\begin{equation}
{T}_{E}(K,P)= \int d^{3} x\left(-(M_3\varphi)_{K} [(M_3 \varphi), (M_3 \varphi)_P] -\frac{1}{2} \sum_{\pm}\left(\frac{G_{K}^{\pm}}{\delta} \mp \varphi_{K}^{\pm}\right)\left[\varphi^{\pm}, \frac{G_{P}^{\pm}}{\delta} \mp \varphi_{P}^{\pm}\right]\right).
\label{ETransfer}
\end{equation}



\end{appendix}
\bibliographystyle{jpp}
\bibliography{biblio}

\end{document}